\documentclass{aa} 
\usepackage[table]{xcolor}
\usepackage{graphicx}
\usepackage{txfonts}
\usepackage{natbib}
\usepackage{longtable}
\usepackage{siunitx}
\usepackage{hyperref}
\usepackage[normalem]{ulem}
\usepackage{multirow}

\usepackage{array,booktabs}
\usepackage{siunitx}
\hypersetup{
 colorlinks= true,  
 urlcolor = blue,   
 linkcolor = blue, 
 citecolor = blue
}

\makeatletter
\renewcommand*\aa@pageof{, page \thepage{} of \pageref*{LastPage}}
\makeatother

\bibpunct{(}{)}{;}{a}{}{,}  %% natbib cite format used by A&A and ApJ

\defcitealias{Prusinski2021}{PR21}
\defcitealias{Rubin2014}{RU14}

\newcommand\MgII{\ion{Mg}{II} $\lambda\lambda$2796, 2803}
\newcommand\MgI{\ion{Mg}{I} $\lambda$2852}
\newcommand\FeIIa{\ion{Fe}{II} $\lambda\lambda$2374, 82}
\newcommand\FeIIb{\ion{Fe}{II} $\lambda\lambda$2586, 2600}
\newcommand\FeIIc{\ion{Fe}{II} $\lambda$2344}

\usepackage{orcidlink}

\begin{document}

  \title{The Lockman--SpReSO project}

  \subtitle{Galactic flows in a sample of far-infrared galaxies}

\author{Mauro González-Otero \inst{1,2} \orcidlink{0000-0002-4837-1615}   
   \and
   Carmen P. Padilla-Torres \inst{1,2,3,4} \orcidlink{0000-0001-5475-165X}
   \and
   J. Ignacio González-Serrano \inst{4,5} \orcidlink{0000-0003-0795-3026}
   \and
   Jordi Cepa \inst{1,2,4} \orcidlink{0000-0002-6566-724X}
   \and
   Ana María Pérez García \inst{4,6} \orcidlink{0000-0003-1634-3588}
   \and
   J. Jesús González \inst{7}\orcidlink{0000-0002-3724-1583}
   \and
   Erika Benítez\inst{7}\orcidlink{0000-0003-1018-2613}
   \and
   Ángel Bongiovanni\inst{4,8}\orcidlink{0000-0002-3557-3234}
   \and
   Miguel Cerviño \inst{6}\orcidlink{0000-0001-8009-231X}
   \and
   Irene Cruz-González \inst{7}\orcidlink{0000-0002-2653-1120}
   \and
   Jesús Gallego\inst{9}\orcidlink{0000-0003-1439-7697}
   \and
   Martín Herrera-Endoqui \inst{7}\orcidlink{0000-0002-8653-020X}
   \and
   Héctor J. Ibarra-Medel\inst{7}\orcidlink{0000-0002-9790-6313}
   \and
   Yair Krongold \inst{7}\orcidlink{0000-0001-6291-5239}
   \and
   Maritza A. Lara-López \inst{9}\orcidlink{0000-0001-7327-3489}
   \and
   Jakub Nadolny \inst{11}\orcidlink{0000-0003-1440-9061}
   \and
   C. Alenka Negrete\inst{7}\orcidlink{0000-0002-1656-827X}
   \and
   Ricardo Pérez-Martínez\inst{4,10}\orcidlink{0000-0002-9127-5522}
   \and
   Mirjana Povi\'c\inst{12,13,14}\orcidlink{0000-0002-9766-6110}
   \and
   Miguel Sánchez-Portal\inst{4,8}\orcidlink{0000-0003-0981-9664}
   \and
   Bernabé Cedrés\inst{4,8}
   \and
   José A. de Diego \inst{7}
   \and
   Héctor Hernández-Toledo \inst{7}\orcidlink{0000-0001-9601-7779}
   \and
   Rocío Navarro Martínez \inst{4} 
   }

\institute{Instituto de Astrofísica de Canarias, 
    E-38205 La Laguna, 
    Tenerife, Spain %1
  \and Departamento de Astrofísica, Universidad de La Laguna (ULL), 
    E-38205 La Laguna, Tenerife, 
    Spain %2
  \and Fundación Galileo Galilei-INAF Rambla José Ana Fernández Pérez, 7, 
    E-38712 Breña Baja, 
    Tenerife, Spain %3
  \and Asociación Astrofísica para la Promoción de la Investigación, Instrumentación y su Desarrollo, ASPID, 
    E-38205 La Laguna, 
    Tenerife, Spain %4
  \and Instituto de Física de Cantabria (CSIC-Universidad de Cantabria), 
    E-39005, 
    Santander, Spain %5
  \and Centro de Astrobiología (CSIC/INTA), 
    E-28692 ESAC Campus, Villanueva de la Cañada, 
    Madrid, Spain %6   
  \and Instituto de Astronomía, Universidad Nacional Autónoma de México, 
    Apdo. Postal 70-264, 04510 
    Ciudad de México, Mexico %7
  \and Institut de Radioastronomie Millimétrique (IRAM), Av. Divina Pastora 7, Núcleo Central 
    E-18012, 
    Granada, Spain %8
  \and Departamento de Física de la Tierra y Astrofísica, Instituto de Física de Partículas y del Cosmos, IPARCOS.    Universidad Complutense de Madrid (UCM), 
    E-28040, 
    Madrid, Spain. %9
  \and ISDEFE for European Space Astronomy Centre (ESAC)/ESA, P.O. Box 78, 
    E-28690 Villanueva de la Cañada, 
    Madrid, Spain %10
  \and Astronomical Observatory Institute, Faculty of Physics, Adam Mickiewicz University, ul.~S{\l}oneczna 36, 
    60-286 Pozna{\'n}, Poland %11
  \and Space Science and Geospatial Institute (SSGI), Entoto Observatory and Research Center (EORC), Astronomy and    Astrophysics Research Division, 
    PO Box 33679, 
    Addis Abbaba, Ethiopia %12
  \and Instituto de Astrofísica de Andalucía (CSIC), 
    E-18080, 
    Granada, Spain %13
  \and Physics Department, Mbarara University of Science and Technology (MUST), 
    Mbarara, Uganda %14
\\
        \email{mauro.gonzalez-ext@iac.es, mauromarago@gmail.com}
}   
\date{Received ---; accepted ---}

% \abstract{}{}{}{}{} 
% 5 {} token are mandatory
 
 \abstract
  {The Lockman--SpReSO project is an optical spectroscopic survey of 956 far-infrared (FIR) objects within the Lockman Hole field limited by magnitude $R_{\rm C}(AB)<24.5$. \ion{Fe}{II} and \ion{Mg}{II} absorption lines have been detected in 21 out of 456 objects with a determined spectroscopic redshift in the catalogue. The redshift of these objects are in the range $0.5 \lesssim z \lesssim 1.44$.}
  {This study aims to investigate material ejection from star-forming regions and material infalling into galaxies by analysing the \ion{Fe}{II} and \ion{Mg}{II} absorption lines. Additionally, we explore whether the correlations found in previous studies between these galactic wind velocities, line equivalent widths (EWs), and galaxy properties such as stellar mass ($M_{*}$), star formation rate (SFR), and specific star formation rate (sSFR) are valid for a sample with FIR-selected objects. The objects analysed span an $M_{*}$ range of $9.89 < \log\left( M_{*} /M_\odot\right) < 11.50$ and an SFR range of $1.01 < \log\left(\mathrm{SFR}\right) < 2.70$.}
  {We performed measurements of the \MgII, \MgI, \FeIIa, \FeIIb, and \FeIIc\ spectral lines present in the spectra of the selected sample to determine the EW and velocity of the flows observed in the star-forming galaxies. Subsequently, we conducted $10^7$ bootstrap simulations using Spearman's rank correlation coefficient ($\rho_s$) to explore correlations with galaxy properties. Furthermore, we calculated the covering factor, gas density, and optical depth for the measured \ion{Fe}{II} doublets.}
  {Our analysis revealed strong correlations between the EW of \ion{Mg}{II} lines and both $M_{*}$ ($\rho_s=0.43$, 4.5$\sigma$) and SFR ($\rho_s=0.42$, 4.4$\sigma$). For the \ion{Fe}{II} lines, we observed strong correlations between the EW and SFR ($\rho_s\sim0.65$, $>3.9\sigma$), with a weaker correlation for $M_{*}$ ($\rho_s\sim0.35$, $>1.9\sigma$). No notable correlations were found between velocity measurements of \ion{Mg}{II} line and $M_{*}$, SFR, or sSFR of the objects ($\rho_s\sim0.1)$. However, a negative strong correlation was found between the velocity of the \ion{Fe}{II} lines and the SFR of the galaxies ($\rho_s\sim-0.45$, $\sim3\sigma$). Our results align with previous studies but studying FIR-selected objects. Finally, we detected a candidate \textit{loitering outflow}, a recently discovered subtype of FeLoBAL quasar, at redshift of $z=1.4399$, exhibiting emission in \ion{C}{III}] and low line velocities ($|v|\lesssim$ 200 km/s).}
  {}


  \maketitle
%
%-------------------------------------------------------------------

\section{Introduction}

To understand how galaxies evolve, we need to study the processes that galaxies undergo over time. Galactic winds are one of the mechanisms that play a fundamental role in their evolution. When giant stars explode as supernovae, they eject material enriched in heavy elements into the galactic halo, enriching the intergalactic medium through a mechanism known as galactic winds \citep{Veilleux2005}. Subsequently, this ejected material could be re-accreted by the gravitational potential of the host galaxy, leading to a re-injection, an inflow, of enriched material to re-fuel the galaxy and trigger the birth of the new generation of stars. The study of this material cycle is a major challenge in cosmology \citep{Peroux2020}. Investigation of the frequency with which this cycle occurs, its characteristics, and its dependence on host galaxy properties such as stellar mass ($M_{*}$) or star formation rate (SFR) will help us to better understand the dynamical and chemical evolution of the galaxies, the circumgalactic medium, and the intergalactic medium \citep{Tumlinson2017,Veilleux2020}.

Studies of galactic outflows have shown that this phenomenon is a  common feature of star-forming galaxies (SFGs) across cosmic time.  In the local Universe, \cite{Chen2010} showed the ubiquity of  outflows in galaxies with a high SFR. \cite{Martin2012} found winds  in about half of their sample of $\sim$200 objects with redshifts  between 0.4 and 1.4. They found no relationship between the rate of  detection of outflows and the physical properties of galaxies,  but they reported a strong relationship with the viewing angle, thus strengthening the idea of the presence of bi-conical outflows in SFGs. 

The kinematics of galactic winds allow them to be studied in the spectrum of their host galaxy. The spectral lines produced by the flow itself are blueshifted (redshifted) in cases where the material is ejected (captured) by the galaxy. Studies were therefore carried out on individual objects to look for properties of the flows and their possible dispersion \citep[among others]{Martin2012,Erb2012,Rubin2014,Chisholm2015,Finley2017a,Finley2017b,Prusinski2021,Xu2022,Davis2023}, and on co-added spectra of similar objects to look for general properties of galactic flows \citep[among others]{Weiner2009,Rubin2010,Erb2012,Zhu2015,Prusinski2021}, a useful approach for distant objects where the signal-to-noise ratio is poor, as it enables the acquisition of generalised information from the whole sample, although unique properties of each individual object may be forfeited.

Many studies have focused on the analysis of low-ionization resonant absorption spectral lines. Optical features such as \ion{Ca}{II} $\lambda\lambda$3933, 69 or \ion{Na}{I} $\lambda\lambda$5890, 96 \citep{Martin2005,Chen2010} are useful because of their presence in low-redshift SFGs. There are other interesting low-ionization absorption lines in the UV range that are becoming available to ground observations for objects at redshifts $z\gtrsim0.3$. The \MgII\ doublet is one of the most widely used \citep{Weiner2009,Erb2012,Rubin2014,Zhu2015,Finley2017a,Prusinski2021} because it is the most common ionized state of magnesium under a wide range of environmental conditions and also has a large oscillator strength. At bluer wavelengths, the \ion{Fe}{II} lines (\FeIIc, \FeIIb, \FeIIa) are found that have the advantage of being less sensitive to emission filling \citep{Erb2012,Martin2012,Rubin2014,Zhu2015,Finley2017a,Prusinski2021}. For the study of more distant objects, far-UV transitions such as \ion{Si}{II}, \ion{Al}{II}, \ion{C}{II}, \ion{C}{IV}, and Ly$\alpha$ are commonly used to the study of outflows \citep[among others]{Shapley2003,Steidel2010,Jones2012,Leclercq2020}.

One of the properties of the outflows derived from the analysis of the spectral lines is the velocity difference of the outflow with respect to the host galaxy. Some papers have found that the velocity of the outflows tends to increase with $M_{*}$ and the SFR of the host galaxy \citep{Martin2005,Rupke2005a,Weiner2009,Chisholm2015}, implying that galaxies with higher SFRs have more energy to eject material via supernovae and also a greater amount of material. However, this correlation has not been found in other studies, where the outflow velocity is found to be independent of $M_{*}$ or the SFR of the host galaxy \citep{Rupke2005b,Chen2010,Martin2012,Rubin2014,Prusinski2021}. These correlations have shown a significant intrinsic scatter \citep[$\sigma\sim0.2$ dex,][]{Chisholm2015,Heckman2015}, which requires to study large samples of objects with a wide range of analysed parameters. \cite{Davis2023} discovered a scatter that is half of what was obtained by those previous studies, resulting in a large sample that spans almost three orders of magnitude in $M_*$ and SFR. In their study of a sample with a wide range of $M_*$ ($\log M_* \sim 6-10\ M_\odot$) and SFR ($\log$ SFR$ \sim 0.01-100\ M_\odot$yr$^{-1}$), \cite{Xu2022} found a correlation of high significance between outflow velocity and both SFR and $M_*$.

Most studies that investigate the correlations between galaxy properties and outflow properties tend to focus on normal SFGs. Furthermore, for distant objects ($z>0.5$), they are selected based on the UV emission observed in the optical range, which introduces a bias towards a specific type of object. Therefore, it is crucial to conduct studies on objects selected using different criteria. In the paper of \cite{Banerji2011}, they analysed a sample of 19 submillimetre galaxies and 21 submillimetre-faint radio galaxies with an average redshift of $z\sim1.3$. A correlation was found between the velocity of the outflows and the SFR, which is consistent with the results seen for lower redshift ULIRGS.

In this paper we identify and analyse the galactic flows of objects in the Lockman--SpReSO project \cite{Gonzalez2022}. We use a sample of IR selected SFGs with \ion{Fe}{II} and \ion{Mg}{II} absorption lines in their spectra. We fit these lines to determine the EW and the velocity of the flow relative to that of the system. We explore the correlation between these line properties and the physical parameters of the host galaxies, using the flow data from Lockman--SpReSO, \citet[hereafter RU14]{Rubin2014} and \citet[hereafter PR21]{Prusinski2021} together. In addition to creating a statistically significant set of objects ($\sim200$), we include SFGs chosen for their far-infrared (FIR) brightness, which complements the sample. This is important, as studies of flow in IR-selected distant objects are limited, and such work is usually carried out on a composite spectrum that lacks the individual properties of the objects. We also determine the covering factor, optical depth and ion densities for the \ion{Fe}{II} doublets.

This is the second paper in the Lockman--SpReSO project series and is structured as follows. In Sect.\ \ref{sec:sample}, we describe the sample, the selection criteria and the comparison samples. In Sect.\ \ref{sec:flow_study}, we describe the flow properties and Sect. \ref{sec:fit_line} shows the line-fitting process. Sect.\ \ref{sec:analysis} describes the analysis of the correlations between flow and galaxy properties. The Sect.\ \ref{sec:ew} analyses the EW of the flows, Sect.\ \ref{sec:vel} analyses the velocity of the flows, and Sect.\ \ref{sec:covering} studies the local covering factor, optical depths and ion densities. The results and conclusions are summarised in Sect. \ref{sec:conclusions}. Throughout the paper, magnitudes in the AB system \citep{Oke1983} are used. The cosmological parameters adopted in this work are: $\Omega_\mathrm{M} = 0.3$, $\Omega_\mathrm{\Lambda} = 0.7$, and $H_\mathrm{0} = 70$ km s$^{-1}$ Mpc$^{-1}$. Both the SFR and $M_{*}$ assume a \cite{Chabrier2003} initial mass function (IMF).

%--------------------------------------------------------------------
\section{Sample selection and description} \label{sec:sample}
\subsection{Lockman--SpReSO data}
\begin{table*}
  \centering
    \caption{Basic information on the objects selected from the Lockman--SpReSO catalogue, ordered by increasing redshift.}
    \label{tab:object_info}
  \begin{tabular}{ccccccccccc}
  \hline \hline
   ID  & RA  & DEC & $z_\mathrm{spec}$ & $R_\mathrm{C}$ & $A_V$ &$\log\left(M_{*}\right)$ & $\log\left(L_\mathrm{TIR}\right)$ & $\log\left(\rm{SFR}\right)$ & 12 + $\log\left(\rm{O/H}\right)$ \\
      & (deg) & (deg) &           & (mag) & (mag) &($M_\odot$)      & ($L_\odot$)           & ($M_\odot$ yr$^{-1}$)   &                 \\
  (1) & (2) & (3) & (4) & (5) & (6) & (7) & (8) & (9) & (10)\\
  \hline
  123207       & 163.15606 & 57.58581 & 0.4914 & 20.39 &  1.8 &  10.38 $\pm$ 0.12 & 10.98 $\pm$ 0.06 & 1.15 $\pm$ 0.06 & $8.77^{+0.06}_{-0.07}$\\[0.5ex]
   96864       & 163.51695 & 57.43780 & 0.5870 & 20.83 &  1.9 &  10.11 $\pm$ 0.10 & 11.15 $\pm$ 0.02 & 1.32 $\pm$ 0.02 & $8.94^{+0.06}_{-0.08}$\\[0.5ex]
  101926       & 163.18977 & 57.46801 & 0.6049 & 21.20 &  1.2 &   9.89 $\pm$ 0.07 & 10.84 $\pm$ 0.03 & 1.01 $\pm$ 0.03 & - \\[0.5ex]
  120080       & 163.08320 & 57.56991 & 0.6108 & 21.22 &  2.7 &  10.29 $\pm$ 0.08 & 11.30 $\pm$ 0.05 & 1.47 $\pm$ 0.05 & $9.03^{+0.04}_{-0.05}$\\[0.5ex]
  118338       & 163.51918 & 57.55793 & 0.6153 & 21.01 &  2.7 &  10.78 $\pm$ 0.07 & 11.35 $\pm$ 0.05 & 1.52 $\pm$ 0.05 & - \\[0.5ex]
   95738       & 162.89970 & 57.43148 & 0.6165 & 21.15 &  2.1 &  10.25 $\pm$ 0.14 & 11.15 $\pm$ 0.05 & 1.32 $\pm$ 0.05 & $8.85^{+0.07}_{-0.10}$\\[0.5ex]
  109219       & 163.54524 & 57.50826 & 0.6443 & 21.09 &  2.8 &  10.79 $\pm$ 0.06 & 11.40 $\pm$ 0.03 & 1.58 $\pm$ 0.03 & - \\[0.5ex]
   94458       & 162.78952 & 57.42330 & 0.6708 & 21.29 &  3.6 &  10.77 $\pm$ 0.08 & 11.73 $\pm$ 0.02 & 1.91 $\pm$ 0.02 & - \\[0.5ex]
   92467       & 162.96713 & 57.41357 & 0.6908 & 20.90 &  2.7 &  10.57 $\pm$ 0.06 & 11.59 $\pm$ 0.06 & 1.76 $\pm$ 0.06 & - \\[0.5ex]
  120257       & 163.33868 & 57.56994 & 0.7204 & 21.41 &  2.0 &  10.12 $\pm$ 0.12 & 11.26 $\pm$ 0.06 & 1.44 $\pm$ 0.06 & - \\[0.5ex]
   95958       & 163.43787 & 57.43224 & 0.7800 & 21.56 &  2.1 &  10.51 $\pm$ 0.12 & 11.42 $\pm$ 0.06 & 1.59 $\pm$ 0.06 & - \\[0.5ex]
  116662       & 162.97753 & 57.54978 & 0.8053 & 22.29 &  1.9 &   9.95 $\pm$ 0.18 & 10.99 $\pm$ 0.08 & 1.17 $\pm$ 0.08 & - \\[0.5ex]
  133957       & 162.94574 & 57.64587 & 0.8116 & 21.74 &  3.0 &  10.49 $\pm$ 0.17 & 11.56 $\pm$ 0.05 & 1.73 $\pm$ 0.05 & - \\[0.5ex]
  186820       & 163.25419 & 57.67225 & 0.8146 & 21.86 &  3.2 &  10.41 $\pm$ 0.04 & 11.54 $\pm$ 0.02 & 1.71 $\pm$ 0.02 & - \\[0.5ex]
   97778       & 163.41833 & 57.44325 & 0.8194 & 22.11 &  3.1 &  10.64 $\pm$ 0.13 & 11.36 $\pm$ 0.09 & 1.53 $\pm$ 0.09 & - \\[0.5ex]
   77155       & 162.95000 & 57.32068 & 0.8683 & 21.33 &  2.5 &  10.43 $\pm$ 0.17 & 11.85 $\pm$ 0.04 & 2.03 $\pm$ 0.04 & - \\[0.5ex]
  102473       & 163.50675 & 57.47030 & 0.8881 & 22.42 &  2.9 &  10.25 $\pm$ 0.10 & 11.56 $\pm$ 0.06 & 1.73 $\pm$ 0.06 & - \\[0.5ex]
  120237$^{a}$ & 163.44230 & 57.56971 & 1.0810 & 21.05 &  1.2 &  10.19 $\pm$ 0.13 & 11.75 $\pm$ 0.04 & 1.92 $\pm$ 0.04 & - \\[0.5ex]
  206641$^{a}$ & 163.31984 & 57.59742 & 1.2023 & 18.96 &  1.2 &  11.50 $\pm$ 0.04 & 11.81 $\pm$ 0.03 & - & - \\[0.5ex]
   78911       & 162.92261 & 57.33095 & 1.2124 & 22.13 &  4.1 &  10.81 $\pm$ 0.06 & 12.53 $\pm$ 0.04 & 2.70 $\pm$ 0.04 & - \\[0.5ex]
  206679$^{a}$ & 163.12537 & 57.65379 & 1.4399 & 20.54 &  2.9 &  11.29 $\pm$ 0.04 & 12.52 $\pm$ 0.03 & - & - \\[0.5ex]
  \hline
  \end{tabular}
  \tablefoot{Column (1) is the unique identification number for each object in the Lockman--SpReSO main catalogue \citep{Gonzalez2022}. Columns (2) and (3) give the optical coordinates (J2000) of the object. Column (4) is the spectroscopic redshift of the objects determined by \cite{Gonzalez2022}. Column (5) is the AB magnitude in the $R_\mathrm{C}$ band. Column (6) is the extinction obtained from the SED fits using the CIGALE and the \cite{Calzetti2000} extinction law. Column (7) is the stellar mass obtained from the SED fits using CIGALE. Column (8) is the total infrared luminosity derived from the SED fits using CIGALE. Column (9) is the star formation rate derived using the IR calibration from \cite{Kennicutt2012} and the IMF from \cite{Chabrier2003}. Column (10) is the gas phase metallicity derived using the calibration from \cite{Tremonti2004}}.
  \raggedright\small{$^{a}$ Objects classified as AGN by González-Otero et al. (in prep). }\\
\end{table*}

\begin{figure}
  \begin{center}
    \includegraphics[width=0.48\textwidth]{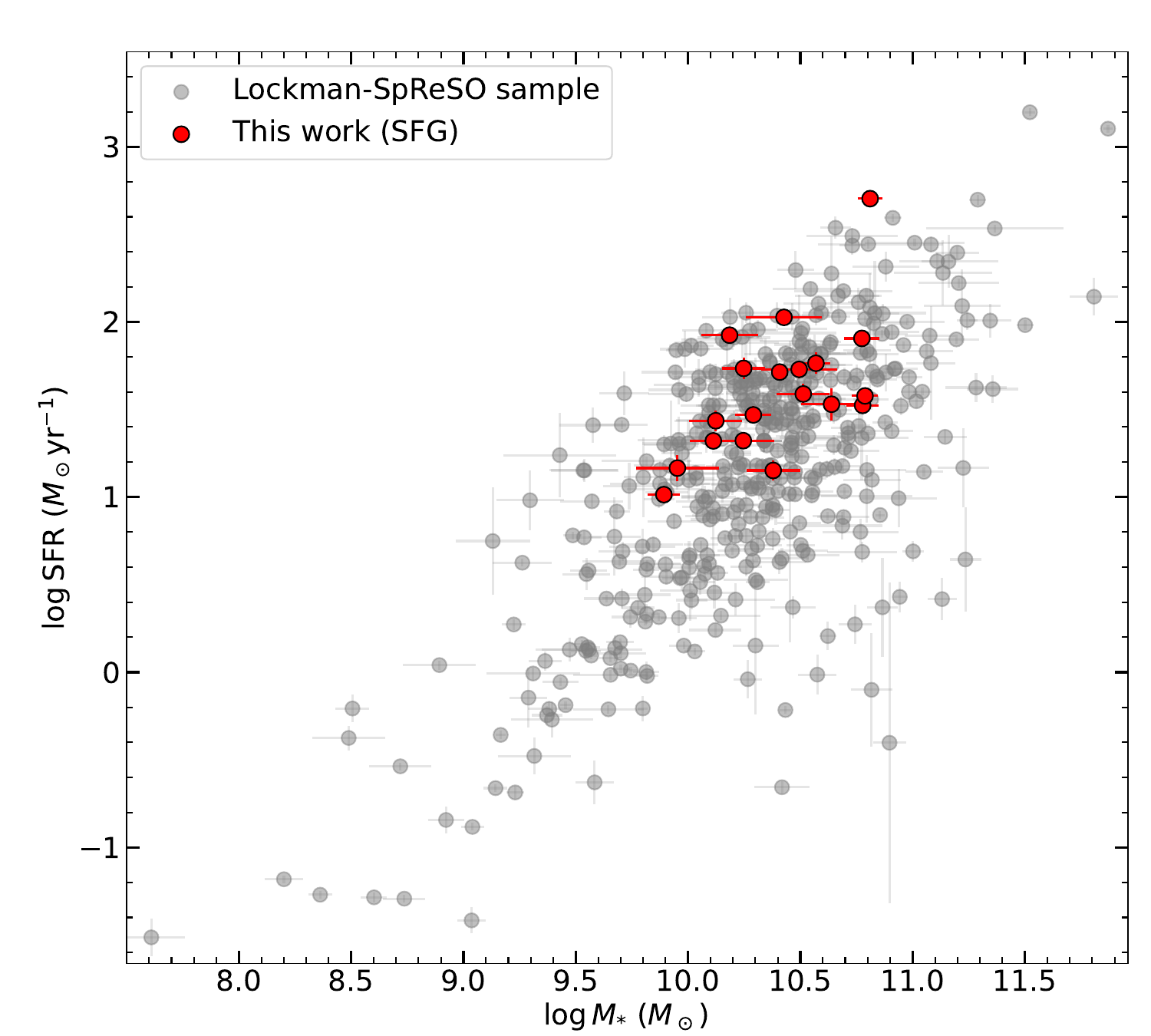}
  \end{center}
  \caption{SFR against $M_*$ of the SFGs selected for this study (red) compared to the Lockman--SpReSO sample (grey). The galaxies exhibiting \ion{Mg}{II} and \ion{Fe}{II} absorption lines populate the region with the highest density of objects in the parent sample.}
  \label{fig:comparison}
\end{figure}

\begin{figure}
  \begin{center}
    \includegraphics[width=0.48\textwidth]{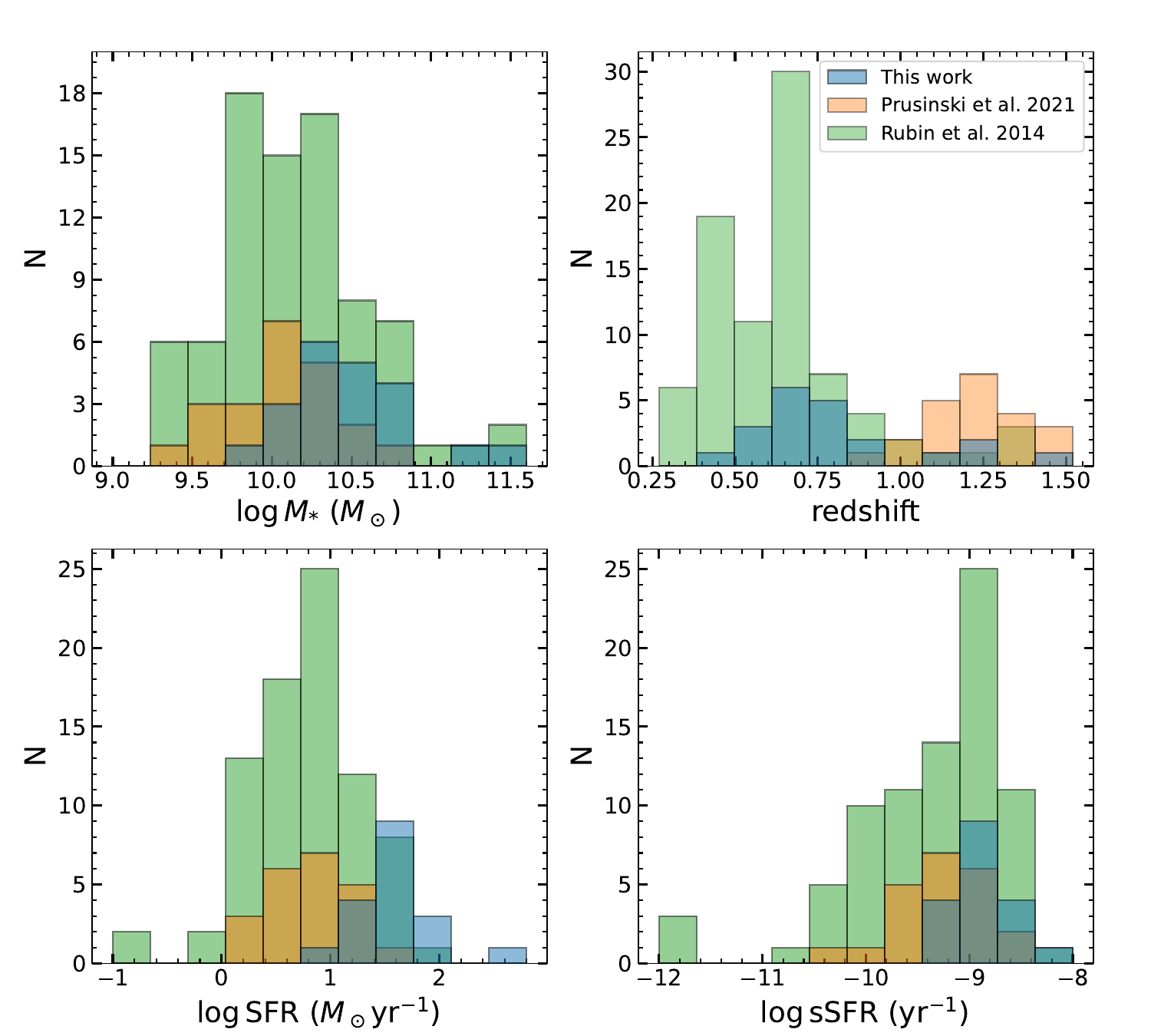}
  \end{center}
  \caption{Main properties of the studied objects. The blue bars represent the objects from Lockman--SpReSO survey, the orange bars represent the objects from \citetalias{Prusinski2021} and the green bars are the objects from \citetalias{Rubin2014}. The top left panel shows the $M_{*}$ of the objects, the top right shows the spectroscopic redshift, the bottom left shows the SFR and, the bottom right shows the sSFR.}
  \label{fig:main_properties}
\end{figure}

The galaxies studied in this paper have been selected from the Lockman--SpReSO project object catalogue. Detailed description of the observations, reduction and catalogue compilation can be found in the presentation paper \cite{Gonzalez2022}. In summary, the Lockman--SpReSO project focuses on a spectroscopic follow-up of 956 objects selected from FIR observations of the Lockman Hole field with the \textit{Herschel Space Observatory}, plus a sample of 188 interesting objects in the field, with a sample limiting magnitude in the Cousins $R$ band of $R_{\rm C}<24.5$. The spectroscopic observations were made with the WHT/A2F-WYFFOS\footnote{\url{https://www.ing.iac.es/Astronomy/instruments/af2}}\citep{wyffos2014} and WYIN/HYDRA\footnote{\url{https://www.wiyn.org/Instruments/wiynhydra.html}} instruments for the bright subset of the catalogue ($R_{\rm C}<20.6$ mag) and with the GTC/OSIRIS\footnote{\url{http://www.gtc.iac.es/instruments/osiris/osiris.php}} instrument \citep{JCepa2000} for the faint subset ($R_{\rm C}>20$ mag). The objects studied here belong to the faint subset, where the resolving power used ($R\equiv\lambda/\delta\lambda$, $\delta\lambda$ being the spectral resolution at wavelength $\lambda$) was $R=500$ ($\sim4$ $\AA\,\mathrm{pix}^{-1}$) for the blue grism with a FWHM resolution at the central wavelength of 560 km/s, covering the electromagnetic spectra from 3600 $\AA$ to 7200 $\AA$. Two different red grisms were used with $R=500$ and $R=1000$ ($\sim3$ $\AA\,\mathrm{pix}^{-1}$) and a FWHM resolution at the central wavelength of 511 km/s and 267 km/s, respectively, covering from $\sim$5000 $\AA$ to 10000 $\AA$. The spectral analysis made it possible to determine the spectroscopic redshift for 456 objects.  

The objects for this work were selected by visual inspection among those with a spectroscopic redshift obtained in the framework of the Lockman--SpReSO project, and which also had the \ion{Mg}{II} $\lambda\lambda$2796, 2803 doublet in emission or absorption along with some of the \ion{Fe}{II} lines in the near-UV range. The observations cover a wavelength that establishes the minimum redshift of the objects we could study. Objects having these properties can only be studied at redshifts greater than 0.4. A total of 21 objects were selected in which both the \MgII\ doublet and the \ion{Fe}{II} lines were found. In all of them, the absorption \ion{Fe}{II} $\lambda\lambda$2586, 2600 doublet was detected; in 19 of them, the \ion{Mg}{II} $\lambda\lambda$2796, 2803 doublet was found in absorption, and in the remaining two, one (ID 206641), showed a total emission component and the other (ID 120237) both emission and absorption components. Other UV absorption lines were also measured in this study; the \ion{Mg}{I} $\lambda$2852 was detected in 15 objects, the \ion{Fe}{II} $\lambda\lambda$2374, 82 doublet was detected in 11 objects, and the \ion{Fe}{II} $\lambda$2344 line was detected in seven of them. 

The sample spans the spectroscopic redshift range between 0.5 and 1.44 (see Table \ref{tab:object_info}). For the determination and study of the galactic flows, it is important to have a good determination of the velocity of the object or, in essence, a good determination of the redshift. All the selected objects show strong emission in the [\ion{O}{II}] $\lambda\lambda$3726, 29 doublet. This emission is normally produced in the photoionised gas near the star-forming regions, so this is a good determinant of the velocity of the system. Other emission lines, such as the most intense Balmer lines (H$\alpha$, H$\beta$ and H$\gamma$), available for the lower redshift objects, or the [\ion{O}{III}] $\lambda\lambda$4959, 5007, were also used to determine the systemic velocity (or redshift). 

Using the selection criteria described above, the selected sample could be contaminated by AGNs since this type of object can also show emission or absorption in both \ion{Mg}{II} and \ion{Fe}{II} lines. We used the classification performed by Gonzalez-Otero et al. (in prep.), in which the objects in the Lockman--SpReSO project catalogue were divided into SFGs and AGNs, using different photometric and spectroscopic criteria. 
For the former, the X-ray-to-optical flux ratio \citep{Szokoly2004}, the X-ray total luminosity \citep{Luo2017}, near-infrared (NIR) and FIR data \citep{Donley2012,Messias2012} were used. For the spectroscopic ones, the BPT \citep{BPT1981}, the \cite{Cid-Fernandes2011} diagrams, and visual inspection were used. With these criteria, objects 120237, 206641, and 206679 were classified as AGNs.

Basic properties of the galaxies were retrieved from the work of \cite{Gonzalez2022} and compiled here in Table \ref{tab:object_info}. They used their spectroscopic redshift determinations to perform SED fits of the UV to FIR photometric data using the CIGALE software (Code Investigating GALaxy Emission, \citealt{Bugarella2005}, \citealt{Boquien2019}), providing more accurate measurements of $M_{*}$ and the total infrared luminosity ($L_{\rm TIR}$) of the objects than that provided by conventional photometric redshifts. The CIGALE parameter configuration is shown in the Appendix B of \cite{Gonzalez2022}. The $M_{*}$ of the objects samples the range $9.89<\log\left(M_{*}/M_\odot\right)<11.50$ and the $L_{\rm TIR}$ values are in a range value between $10.84<\log\left(L_{\rm TIR}/L_\odot\right)<12.53$, with 18 objects compatible with Luminous Infrared Galaxies (LIRGs) and two others compatible with Ultra-Luminous Infrared Galaxies (ULIRGs). The individual cutouts of the objects and their SED fittings obtained in the framework of \cite{Gonzalez2022} are compiled in Appendix \ref{sec:appendix}.

The SFR and metallicity of these objects were also obtained from the work of Gonzalez-Otero et al.\ (in prep.). The SFR of the Lockman--SpReSO project galaxies were studied using the flux of the Balmer lines, $L_{\rm TIR}$, and the [\ion{O}{II}] doublet flux as tracers. 
In Fig. \ref{fig:comparison} we have compared the complete Lockman--SpReSO sample with the spectroscopic redshift determined by \cite{Gonzalez2022} with the sample of objects selected in this paper with absorption in \ion{Mg}{II} and \ion{Fe}{II} and classified as SFG.

The metallicity was analysed following several criteria, although because of the redshift range of the objects studied in this paper only four of them have a metallicity measurement available. Studies such as \cite{Pettini2004} and \cite{Pilyugin2016}, which use spectral lines over the whole optical range, can be applied to only one of the objects in the sample (ID 123207). Studies based on the R23 method, such as \cite{Tremonti2004} and \cite{Kobulnicky2004}, which use the blue lines in the optical spectrum, can be applied to four of the objects studied. The SFRs collected in Table \ref{tab:object_info} were obtained using $L_{\rm TIR}$, the calibration of \cite{Kennicutt2012} and the IMF of \cite{Chabrier2003}. The metallicities given in Table \ref{tab:object_info} were obtained using the calibration of \cite{Tremonti2004} and the errors are estimated using Monte Carlo simulations.

In Fig. \ref{fig:main_properties} we show the basic properties ($M_{*}$, redshift, SFR, and sSFR) of the objects from Lockman--SpReSO (blue data), together with data from papers that have also carried outflow studies in SFG (see Sect. \ref{sec:comparison} for a detailed description).

\subsection{The comparison samples}\label{sec:comparison}

In order to have a frame of reference with which to compare, we have included in our work samples from other papers that have studied the galaxy winds produced in SFG.

The numerous sample of 105 galaxies from \citetalias{Rubin2014} was used. They selected objects from existing spectroscopic catalogues for which HST deep observations are available in order to study the morphology, orientation, and spatial distribution of the star formation. They selected objects with redshifts $z>0.3$ to ensure \ion{Mg}{II} coverage and with B-band magnitude $<23$. The $M_{*}$ and the SFR were obtained from a SED fitting procedure in the wavelength range between 2400 $\AA$ and 24 $\mu$m assuming the IMF of \cite{Chabrier2003}. 
The EW of the lines were derived using the feature-finding algorithm outlined in \cite{Cooksey2008}. To normalize the line doublets for the study, the authors used continuum windows located in proximity to the lines region.
To determine the flow velocity, two models were used to fit the lines: one with a single component per line and another with two components per line to separate the systemic component at zero velocity from the flow component. The profile of the lines was fitted using a Voigt shape. Additionally, the maximum flow velocity was also determined. To compare with the data obtained from the Lockman--SpReSO objects, we used the results from their one-component fit for consistency. (see Sect. \ref{sec:fit_line}). The distributions of the main properties of the sample are shown in Fig. \ref{fig:main_properties} in green.

In addition, we also used the data from \citetalias{Prusinski2021}, a sample of 22 galaxies from the \cite{Skelton2014} data of the CANDELS and COSMOS surveys. These were selected based on their $\mathrm{SFR}$ ($>1$ $M_\odot$ yr$^{-1}$), with photometric redshift range $0.7\lesssim z_\mathrm{phot} \lesssim 1.5$ at 99\% confidence and magnitude $R_\mathrm{C}\leq24$. The SFR was computed from the \textit{Hubble Space Telescope} (HST) H$\alpha$ emission-line maps using the \cite{Kennicutt1998} recipe and the $M_*$ was retrieved from the 3D-HST catalogue \citep{Skelton2014}. This SFR has been multiplied by the correction factor (0.68) given in \cite{Kennicutt2012} to make the used IMF consistent. To study the outflows, they examined the bluest line of the \MgII\ doublet, along with the \ion{Fe}{II} absorption lines present in the spectra of the sample. The outflow velocity was estimated from the centroids of the lines, and the EW was measured by directly integrating the lines. Weighted averages of the velocity and EW for the \ion{Fe}{II} lines of each object were obtained due to their saturation. The distributions of the main properties of the sample are shown in Fig. \ref{fig:main_properties} in orange.

\begin{figure}
  \begin{center}
    \includegraphics[width=0.48\textwidth]{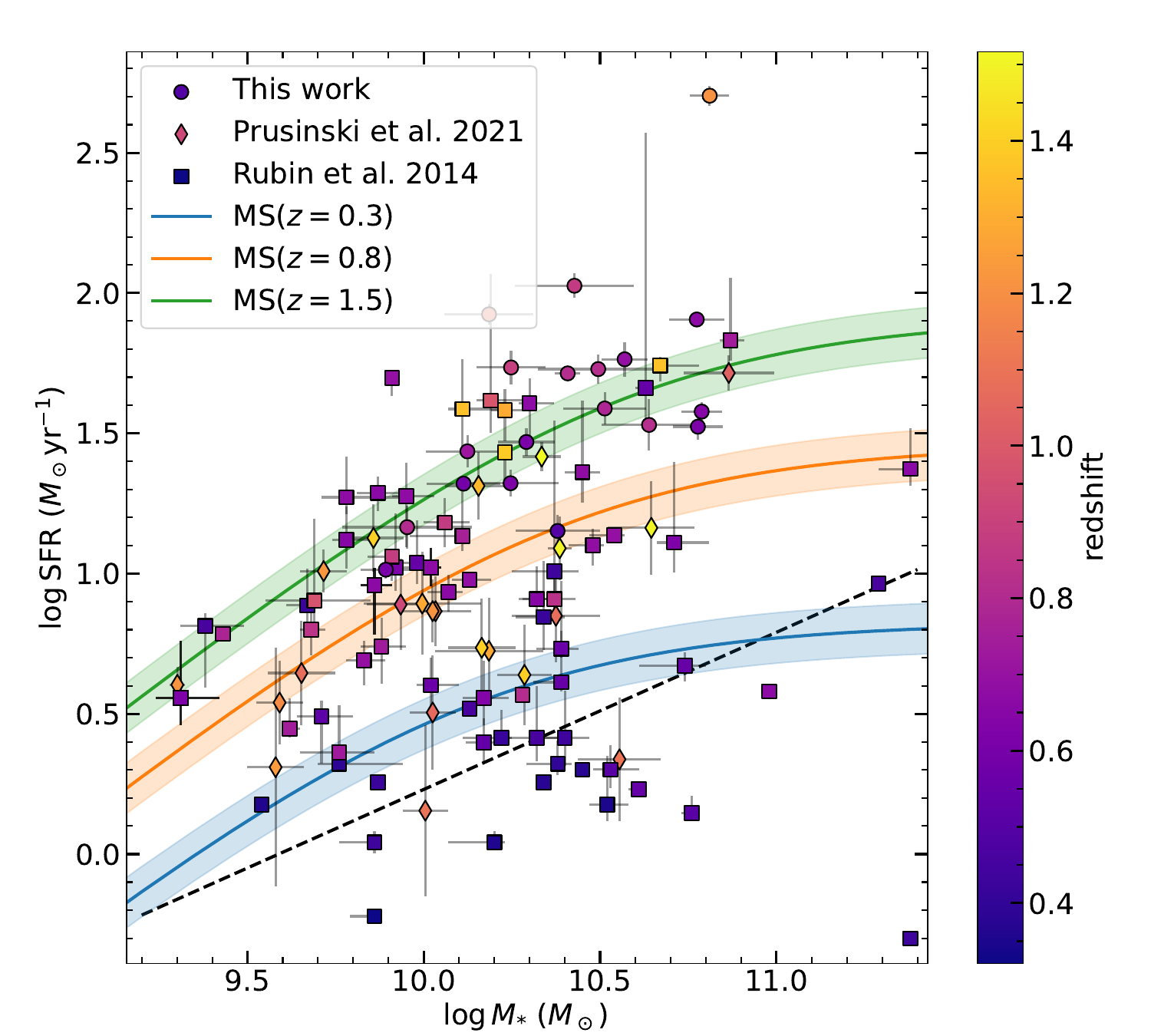}
  \end{center}
  \caption{Diagramn of $M_*$ against SFR of the sample studied color coded by the spectroscopic redshift. The data from Lockman--SpReSO are represented by circles, \citetalias{Prusinski2021} data are represented by diamonds, and \citetalias{Rubin2014} data are represented by squares. The dotted line represents the \cite{Murray2011} limit for producing winds in galaxies. The blue, red, and green solid lines and shaded areas are the \cite{Popesso2023} fit for the main sequence and its scatter of 0.09 dex calculated at the minimum, median, and maximum redshift of the sample, respectively.}
  \label{fig:SFR_M}
\end{figure}

The $M_{*}$ distribution is similar for three samples, unlike the SFR distribution where a larger difference is found, as can be seen in Fig. \ref{fig:main_properties}. The Lockman--SpReSO sample populates the highest SFR region, which is understandable given the selection criterion for objects with FIR emission. The range of SFRs between the samples presented allows us to cover a much wider range of parameters. 
In Fig. \ref{fig:SFR_M} we can see the relationship between SFR and $M_{*}$ for the objects in the three data sets. The dashed line represents the \cite{Murray2011} limit for the production of winds in galaxies, adapted from the Eq. 3 of \citetalias{Rubin2014}. It can be seen that the full sample agrees well with the theoretical limit, filling the upper right region in the diagram. The solid lines represents the main sequence developed by \cite{Popesso2023} for the minimum, average, and maximum redshifts of the sample ($z\sim0.3,\ 0.8,\ 1.5$, respectively) and the shaded area represents a scatter of 0.09 dex. Almost 85\% of the sample is located over the MS, that is the region of starburst galaxies.

\section{Inflow and outflow study} \label{sec:flow_study}
To determine whether our objects have inflows or outflows we need to know the velocity difference, if any, between the selected absorption lines and the velocity of the system. To do this, the redshift of the object and the centre of the lines are required. The redshift was retrieved from \cite{Gonzalez2022} (see the paper for details) and is given in Table \ref{tab:object_info}. The properties of the \ion{Mg}{II} and \ion{Fe}{II} lines were derived for this paper by fitting the spectral lines (see the Section \ref{sec:fit_line} for details).

As we have already seen, the \ion{Fe}{II} and \ion{Mg}{II} lines are useful for studying the galactic flows. 
Some studies have described the UV lines and their main properties in detail (\citealp[and references therein]{Erb2012,Martin2012,Zhu2015}). As with UV lines, their analysis with ground-based telescopes is limited to objects with redshifts $\gtrsim0.4$ owing to the attenuation of the UV component by the atmosphere; otherwise, it is necessarily to observe with space telescopes. Another major problem with these lines is the resonant emission that occurs in some of them, which fills in the absorption and makes it difficult to accurately measure the line. In fig.\ 5 of \cite{Zhu2015} and their Appendix A are gathered the energy-level diagrams of the \ion{Fe}{II}, \ion{Mg}{II}, and \ion{Mg}{I} atoms for the transitions that we studied for this paper. Lines that do not produce fluorescent emission (those that can only be de-excited to the ground state) are more affected by emission filling. When this happens, the line profile may be modified. 

Absorption is produced by the gas between the observer and the galaxy while emission can come from other parts of the galaxy, leading to a difference between the centres of the absorption and emission components. Among the lines used in our study, the one most affected by this effect is the \ion{Fe}{II} $\lambda$2382 line, since the only way to de-excite it is by resonant emission. In addition, the \ion{Fe}{II} $\lambda$2344 line produces a fluorescent emission at 2381.49 \AA, which also affects the \ion{Fe}{II} $\lambda$2382 line. Even the \ion{Fe}{II} $\lambda$2600 line could suffer from emission filling as the fluorescent emission it produces at 2626 \AA\ has a very low probability of occurring. Lines such as \ion{Fe}{II} $\lambda$2373 and \ion{Fe}{II} $\lambda$2586 suffer very little from this effect because the fluorescent emission de-excitation channels have high probabilities of occurrence (high Einstein A coefficient), so the lines suffer very little from emission filling. 

However, when analysing the spectral lines of our sample, we found no emission filling effect. Figure \ref{fig:doublet_fit} shows examples where both \ion{Fe}{II} and \ion{Mg}{II} lines are detected with no apparent emission filling effect. This was also seen in analyses of nearby galaxies, where the \ion{Na}{I} $\lambda\lambda$ 5890,96 lines showed little emission filling \citep{Heckman2000,Martin2005,Martin2006,Chen2010}. This has been attributed to the presence of regions of high gas density and high neutral sodium concentration where the long path length of the scattered photons leads to a high probability of absorption by dust. The Lockman--SpReSO survey objects are selected for their IR emission, in other words they are dusty, as indicated by a mean extinction of $A_V\sim2.7$ mag. Furthermore, as we have discussed previously (see Table \ref{tab:object_info}), we see that most of the objects are LIRGs or even ULIRGs, which favours the non-production of emission filling according to the above reasoning. 
In the work of \cite{Prochaska2011}, they find that the \ion{Fe}{II} fluorescence lines scale with the amount of emission filling. Since they are not detected in the spectra of our sample, this supports the fact that no emission filling effect is found in the spectral lines.

\subsection{Fitting of absorption lines} \label{sec:fit_line}

\begin{figure*}
  \begin{center}
    \includegraphics[width=0.96\textwidth]{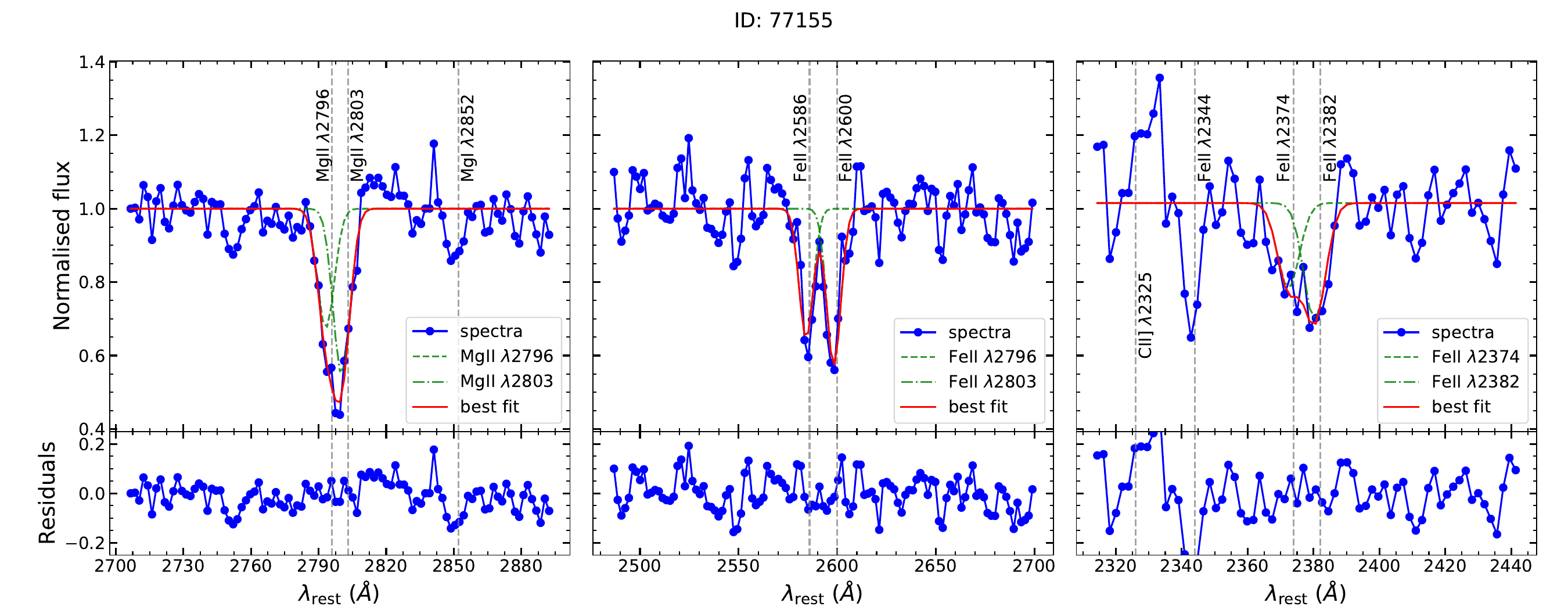}
  \end{center}
  \caption{An example of the fitting for the doublets studied in this study for source 77155. On the left is the \MgII\ fit with the residuals at the bottom. The \MgI\ line is also visible in this slice of the spectrum. In the centre, the fit for \FeIIb, also with the residuals of the fit at the bottom. On the right, using the same schedule, the case of \FeIIa\ with a low S/N ratio. You can also see the \ion{C}{II}] $\lambda$2325 in emission and \ion{Fe}{II} $\lambda$2344 in absorption.}
  \label{fig:doublet_fit}
\end{figure*}

The lines were measured using the same procedure as that used by \cite{Gonzalez2022} where the Python package \textit{LMFIT\footnote{\url{https://lmfit.github.io/lmfit-py/index.html}}} \citep{LMFIT2014} was used for the measurement of the most intense spectral lines. \textit{LMFIT} allows us to perform a non-linear least-squares minimisation line fitting routine with the desired model. The lines were fitted using a Gaussian model for the absorption component and a linear zero-slope continuum model. 
As we have seen, in the work of \cite{Rubin2014} and others \citep[for example]{Chen2010,Xu2022}, each line is also fitted with a two component model: one fixed zero velocity component representing the ISM of the galactic region and a variable velocity component representing the galactic flow. However, the resolution of our study does not allow this type of component decomposition.
The normalisation of the spectra was performed only in the analysis windows of each line or doublet line. The flat continuum model obtained in the fitting was used to normalise by the continuum.

\MgII, \FeIIa, and \FeIIb\ are doublets of spectral lines whose components are very close in wavelength. This implies that at the dispersion used ($\sim$3--4 $\AA\,\mathrm{pix}^{-1}$), depending also on the redshift, these doublets can appear blended. Therefore, in the case of these doublets, we fitted the two lines together with the continuum at the same time, that is two Gaussian components (one for each line) plus a linear model. In order to reduce the computational time, we have used the known relations of the lines as initial parameters of the models, but allowed them to fluctuate. Since each doublet line is produced in the same region and under the same conditions in the galaxy, in the absence of emission filling or other phenomena, the width should be the same, that is, the same initial sigma value for the two Gaussian components. The same reasoning applies to the central wavelength of the line, where the length between peaks is known, although it is a free parameter in the fit. None of the fit parameters were fixed, so they were all allowed to vary during the fitting process. Finally, since the redshift is known, the spectra were transferred to rest-frame to measure the lines. 

Figure \ref{fig:doublet_fit} shows an example of the line fit for ID 77155. The different degrees of blending found in this particular source are representative for our sample. Each panel shows the fit of one of the doublets studied in this paper. Upper panels represent the fit performed and lower panels represent the residuals of the fit itself. The fit results of \MgII, \FeIIb\ and \FeIIa\ (with the lowest S/N ratio) are shown from left to right.

In addition, from the line-fitting procedure we determined the rest-frame equivalent width (EW) of the lines to be analysed in relation to the physical properties of the galaxies. The values obtained and its errors are collected in Table \ref{tab:ew}.  
The error in the EW was calculated by propagating the error obtained for the Gaussian component and the continuum of each line when the EW was calculated (both errors were calculated in the line fitting process). As can be seen, the values of the EW are biased towards higher values, mainly due to the resolution of the Lockman--SpReSO spectra.

Finally, using the centre of the Gaussian components obtained in the fit, we determined the velocity of the lines and hence the velocity difference between the material wind and the system. 
The errors in velocity have been propagated from the central wavelength error obtained in the fit of each line. In Table \ref{tab:velocities} we list the velocities obtained and its errors for the lines analysed.

\begin{table*}
  \centering
  \caption{Rest-frame equivalent width of the absorption lines measured for the objects from Lockman--SpReSO catalogue.}
  \label{tab:ew}
  \begin{tabular}{cccccccccccc}
  \hline \hline
   ${\rm ID}$ &\ion{Fe}{II} $\lambda$2344 &\ion{Fe}{II} $\lambda$2374 &  \ion{Fe}{II} $\lambda$2382 & \ion{Fe}{II} $\lambda$2586 & \ion{Fe}{II} $\lambda$2600 & \ion{Mg}{II} $\lambda$2796 & \ion{Mg}{II} $\lambda$2803 & \ion{Mg}{I} $\lambda$2852 \\
      &       ($\AA$) &   ($\AA$)    &      ($\AA$)  &     ($\AA$)   &    ($\AA$)    &       ($\AA$) &      ($\AA$) &   ($\AA$)    \\
  \hline 
  123207 &   -      &   -      &  -      & 3.9 $\pm$ 1.2 & 3.5 $\pm$ 1.1 &  6.1 $\pm$ 1.1 &  4.5 $\pm$ 1.2 & 0.6 $\pm$ 0.8 \\
   96864 &   -      &   -      &  -      & 5.7 $\pm$ 1.7 & 5.2 $\pm$ 1.7 &  6.5 $\pm$ 1.8 &  6.0 $\pm$ 1.9 & 1.9 $\pm$ 0.5 \\
  101926 &   -      &  1.8 $\pm$ 1.4 & 1.8 $\pm$ 1.4 & 9.1 $\pm$ 2.2 & 1.8 $\pm$ 1.4 &  4.4 $\pm$ 1.2 &  4.2 $\pm$ 1.3 & 0.8 $\pm$ 0.6 \\
  120080 &   -      &   -      &  -      & 2.6 $\pm$ 1.4 & 3.6 $\pm$ 1.4 &  4.3 $\pm$ 1.2 &  2.9 $\pm$ 1.4 & 0.8 $\pm$ 1.3 \\
  118338 &   -      &   -      &  -      & 3.5 $\pm$ 1.3 & 4.3 $\pm$ 1.4 &  6.8 $\pm$ 1.6 &  5.9 $\pm$ 1.7 & 1.2 $\pm$ 0.6 \\
   95738 &   -      &   -      &  -      & 4.6 $\pm$ 1.2 & 4.6 $\pm$ 1.2 &  5.1 $\pm$ 1.3 &  6.3 $\pm$ 1.1 & 1.1 $\pm$ 0.7 \\
  109219 &   -      &   -      &  -      & 3.5 $\pm$ 1.2 & 1.3 $\pm$ 1.2 &  3.9 $\pm$ 1.7 &  7.3 $\pm$ 1.3 &  -      \\
   94458 &   -      &   -      &  -      & 1.6 $\pm$ 1.2 & 4.6 $\pm$ 1.0 &  3.5 $\pm$ 1.4 &  3.4 $\pm$ 1.4 & 1.4 $\pm$ 1.1 \\
   92467 &   -      &   -      &  -      & 2.2 $\pm$ 1.8 & 5.3 $\pm$ 1.1 &  3.2 $\pm$ 1.4 &  2.8 $\pm$ 1.4 & 0.5 $\pm$ 1.4 \\
  120257 &  3.0 $\pm$ 1.2 &  1.5 $\pm$ 1.9 & 6.3 $\pm$ 1.5 & 5.7 $\pm$ 1.7 & 5.6 $\pm$ 1.2 &  7.9 $\pm$ 1.3 &  4.4 $\pm$ 2.0 & 1.7 $\pm$ 0.7 \\
   95958 &   -      &  2.3 $\pm$ 0.9 & 2.1 $\pm$ 0.9 & 2.2 $\pm$ 1.4 & 4.2 $\pm$ 1.0 &  5.1 $\pm$ 1.6 &  6.5 $\pm$ 1.2 & 1.7 $\pm$ 0.9 \\
  116662 &   -      &  1.8 $\pm$ 1.3 & 2.2 $\pm$ 1.3 & 2.8 $\pm$ 1.3 & 5.6 $\pm$ 1.1 &  3.6 $\pm$ 1.4 &  5.0 $\pm$ 1.4 & 2.1 $\pm$ 1.0 \\
  133957 &   -      &   -      &  -      & 9.3 $\pm$ 1.3 & 5.1 $\pm$ 1.4 &  7.6 $\pm$ 1.4 &  3.3 $\pm$ 1.9 &  -      \\
  186820 &   -      &   -      &  -      & 2.7 $\pm$ 1.1 & 1.7 $\pm$ 1.1 &  3.7 $\pm$ 1.3 &  3.9 $\pm$ 1.2 &  -      \\
   97778 &   -      &  1.3 $\pm$ 1.2 & 1.6 $\pm$ 1.1 & 2.5 $\pm$ 1.4 & 4.8 $\pm$ 1.2 &  3.5 $\pm$ 1.2 &  4.5 $\pm$ 1.1 & 1.5 $\pm$ 1.1 \\
   77155 &  2.6 $\pm$ 1.8 &  3.5 $\pm$ 1.6 & 5.1 $\pm$ 1.5 & 5.3 $\pm$ 1.6 & 7.5 $\pm$ 1.4 &  6.7 $\pm$ 1.2 &  6.8 $\pm$ 1.2 & 2.2 $\pm$ 0.6 \\
  102473 &  1.6 $\pm$ 1.3 &  3.5 $\pm$ 1.2 & 3.5 $\pm$ 1.2 & 6.3 $\pm$ 1.2 & 6.3 $\pm$ 1.0 &  5.5 $\pm$ 1.3 &  5.1 $\pm$ 1.3 & 4.3 $\pm$ 0.6 \\
  120237$^{a}$ &  4.0 $\pm$ 1.0 &  2.6 $\pm$ 1.1 & 2.3 $\pm$ 1.1 & 3.7 $\pm$ 0.9 & 5.1 $\pm$ 0.7 &   -      &   -      &  -      \\
  206641$^{a}$ &  0.3 $\pm$ 0.1 &  0.2 $\pm$ 0.2 & 0.9 $\pm$ 0.1 & 0.4 $\pm$ 0.3 & 1.8 $\pm$ 0.2 &   -      &   -      &  -      \\
   78911 &  5.6 $\pm$ 1.3 &  2.1 $\pm$ 1.5 & 5.2 $\pm$ 1.3 & 9.6 $\pm$ 1.9 & 9.0 $\pm$ 1.9 &  9.4 $\pm$ 2.0 &  9.9 $\pm$ 1.9 & 6.5 $\pm$ 1.3 \\
  206679$^{a}$ &  5.1 $\pm$ 0.4 &  4.4 $\pm$ 1.5 & 7.1 $\pm$ 0.9 & 5.9 $\pm$ 1.0 & 7.1 $\pm$ 1.0 &  7.4 $\pm$ 1.4 &  7.0 $\pm$ 1.5 &  -      \\
  \hline
  \end{tabular}\\[0.2cm]
  \raggedright\small{$^{a}$ Objects classified as AGN by González-Otero et al. (in prep). }\\
\end{table*}

\begin{table*}
  \centering
  \caption{Absorption line velocities measured for the objects from Lockman--SpReSO catalogue.}
  \label{tab:velocities}
  \begin{tabular}{
  c
  S[table-format=4(3)]
  S[table-format=4(3)]
  S[table-format=4(3)]
  S[table-format=4(3)]
  S[table-format=4(3)]
  S[table-format=4(3)]
  S[table-format=4(3)]
  S[table-format=4(3)]
   }
  \hline \hline
   {${\rm ID}$} &{\ion{Fe}{II} $\lambda$2344} &{\ion{Fe}{II} $\lambda$2374} &  {\ion{Fe}{II} $\lambda$2382} & {\ion{Fe}{II} $\lambda$2586} & {\ion{Fe}{II} $\lambda$2600} & {\ion{Mg}{II} $\lambda$2796} & {\ion{Mg}{II} $\lambda$2803} & {\ion{Mg}{I} $\lambda$2852} \\
   {}& {(km s$^{-1}$)} &{(km s$^{-1}$)} & {(km s$^{-1}$)} & {(km s$^{-1}$)} & {(km s$^{-1}$)} & {(km s$^{-1}$)} & {(km s$^{-1}$)} & {(km s$^{-1}$)} \\
  \hline
        123207 & \textendash & \textendash & \textendash & -177(144) &  2 ( 85) & -131 (293) & -226 (296) & -161 (364) \\
        96864 & \textendash & \textendash & \textendash & -192(110) & -122 (119) &  18 ( 85) &  131 ( 92) &  -46 ( 33) \\
        101926 & \textendash & 131(100)  & -18(89)  &  59(170) & -148 ( 84) & -322 ( 5) & -269 ( 68) & -210 ( 88) \\
        120080 & \textendash & \textendash & \textendash & 107(53) & -43 ( 95) & -199 ( 70) & -118 (109) &  67 (159) \\
        118338 & \textendash & \textendash & \textendash &  61(59) &  1 (165) &  -24 ( 53) &  25 ( 61) &   9 (127) \\
        95738 & \textendash & \textendash & \textendash & -232(93) & -147 (108) & -305 (235) & -220 (219) & -210 ( 34) \\
        109219 & \textendash & \textendash & \textendash & 463(207) & -134 (150) & -322 ( 39) & -316 (723) & \textendash \\
        94458 & \textendash & \textendash & \textendash &  23(169) & -382 (173) &  -62 (149) &  -16 (155) & -217 ( 78) \\
        92467 & \textendash & \textendash & \textendash & -467(70) & -202 ( 76) & -114 ( 92) & -166 (105) & -159 ( 496) \\
        120257 & -222(92)  & 379(511)  & 321(111)  & -224(249) & -31 ( 73) &  89 ( 63) &  286 (119) & 151 ( 134) \\
        95958 & \textendash & -233(148)  & -54(164)  & -317(139) & -238 ( 83) & -279 (116) & -232 ( 98) &  78 ( 110) \\
        116662 & \textendash & -499(215)  & -303(173)  & -178(154) & -199 (118) & -322 (416) &  -98 (390) & 331 ( 143) \\
        133957 & \textendash & \textendash & \textendash & -171(289) & -228 ( 99) &  -90 ( 87) &  97 (213) & \textendash \\
        186820 & \textendash & \textendash & \textendash & 315(227) & -23 (124) & -214 ( 75) &  -34 ( 88) & \textendash \\
        97778 & \textendash & -478(252)  & -295(208)  & -346(132) & -430 (213) & -214 (225) &  -36 (172) & -420 (247) \\
        77155 & -131(117)  & -363(222)  & -186(147)  & -99(46) & -296 ( 50) & -201 ( 98) & -214 ( 18) & -170 ( 93) \\
        102473 &  63(100)  & 306(117)  & 378(60)  & -232(12) &  68 (154) & -141 ( 63) &  -78 ( 67) & -210 ( 42) \\
    120237$^{ a}$ & -343(133)  & -130(234)  & -378(182)  & -297(95) & -312 ( 60) & \textendash & \textendash & \textendash \\
    206641$^{a}$ & -384(55 )  & -379(147)  & -327(30)  & -313(51) & -170 ( 52) & \textendash & \textendash & \textendash \\
        78911 & -256(27 )  & -13(190)  & 258(76)  & 116(17) &  89 ( 95) &  12 (63) &   93 (59) &   15(47) \\
    206679$^{a}$ & -54(33 )  & -75( 39)  &  35(23)  &  3(18) & -166 ( 23) &  -44 (32) &  -66 (34) & \textendash \\
  \hline
  \end{tabular}\\[0.2cm]
  \raggedright\small{$^{a}$ Objects classified as AGNs by González-Otero et al. (in prep).}\\
\end{table*}

\section{Galactic flows properties analysis}\label{sec:analysis}
\subsection{Equivalent width analysis}\label{sec:ew}

\begin{figure*}
  \begin{center}
    \includegraphics[width=0.96\textwidth]{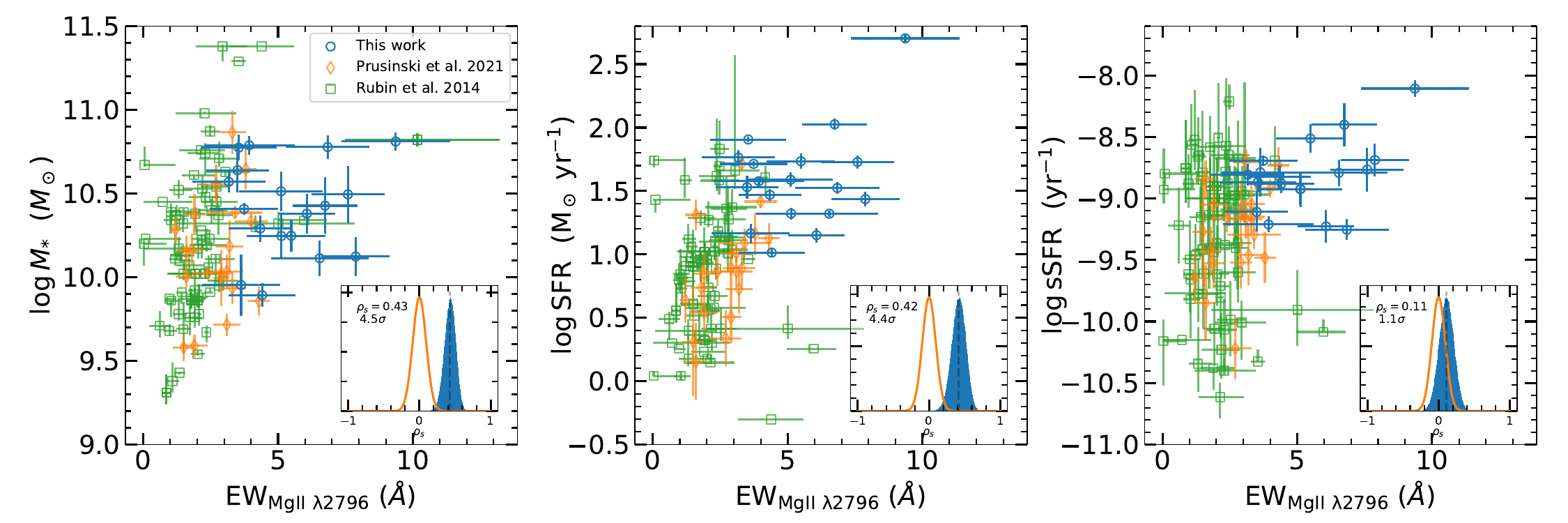}
  \end{center}
  \caption{Plot of the \ion{Mg}{II} $\lambda2796$ equivalent width against, from left to right, $M_{*}$, SFR and sSFR. The empty blue circles are the results obtained here, the empty orange diamonds are the results of \citetalias{Prusinski2021} and the green empty squares are the results of \citetalias{Rubin2014}. The inset graph in each panel contains the probability distribution of the Spearman parameter from the bootstrap procedure in blue (see the text for more details) and the expected distribution in the uncorrelated case is shown in orange. The value of the mode of the obtained distribution and the significance in sigmas for the obtained correlation are also included. See the Sect. \ref{sec:ew} for a discussion of correlation.}
  \label{fig:ew_analysis_MgIIa}
\end{figure*}

\begin{figure*}
  \begin{center}
    \includegraphics[width=0.96\textwidth]{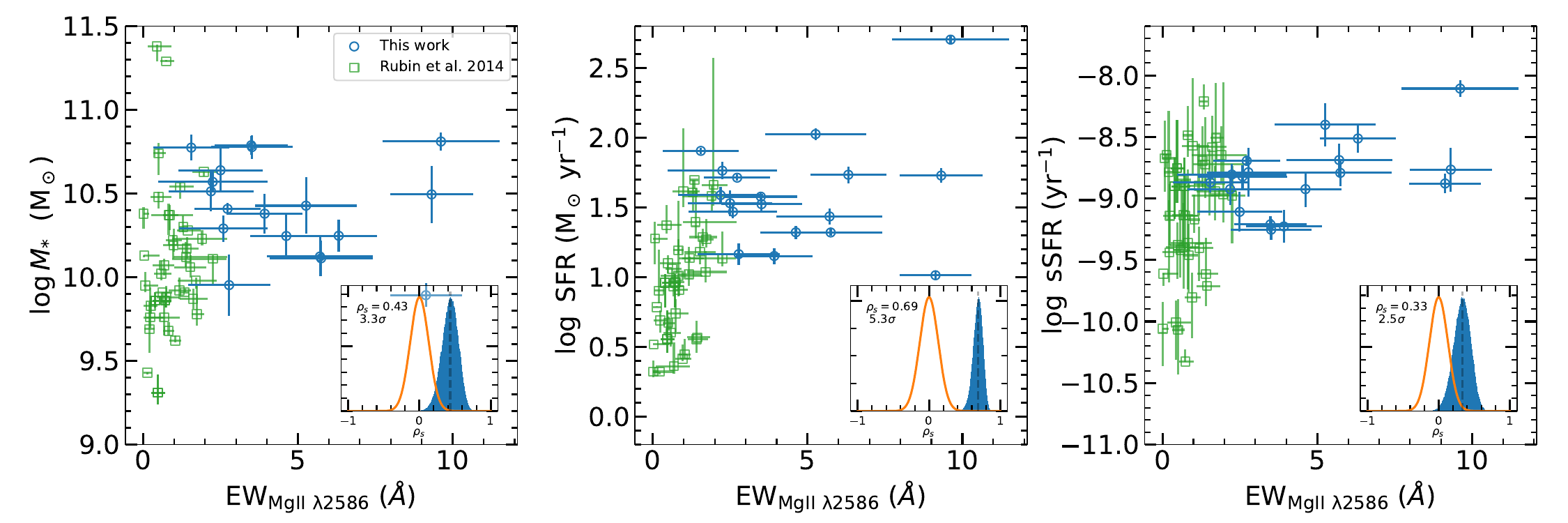}
  \end{center}
  \caption{Plot of the \ion{Fe}{II} $\lambda2586$ equivalent width against, from left to right, $M_{*}$, SFR and sSFR. The empty blue circles are the results of this paper and the empty green squares are the results of \citetalias{Rubin2014}. The inset plots represent the same idea as in Fig. \ref{fig:ew_analysis_MgIIa}. See the Sect. \ref{sec:ew} for a discussion of correlation.}
  \label{fig:ew_analysis_FeIIa_rub}
\end{figure*}

\begin{figure*}
  \begin{center}
    \includegraphics[width=0.96 \textwidth]{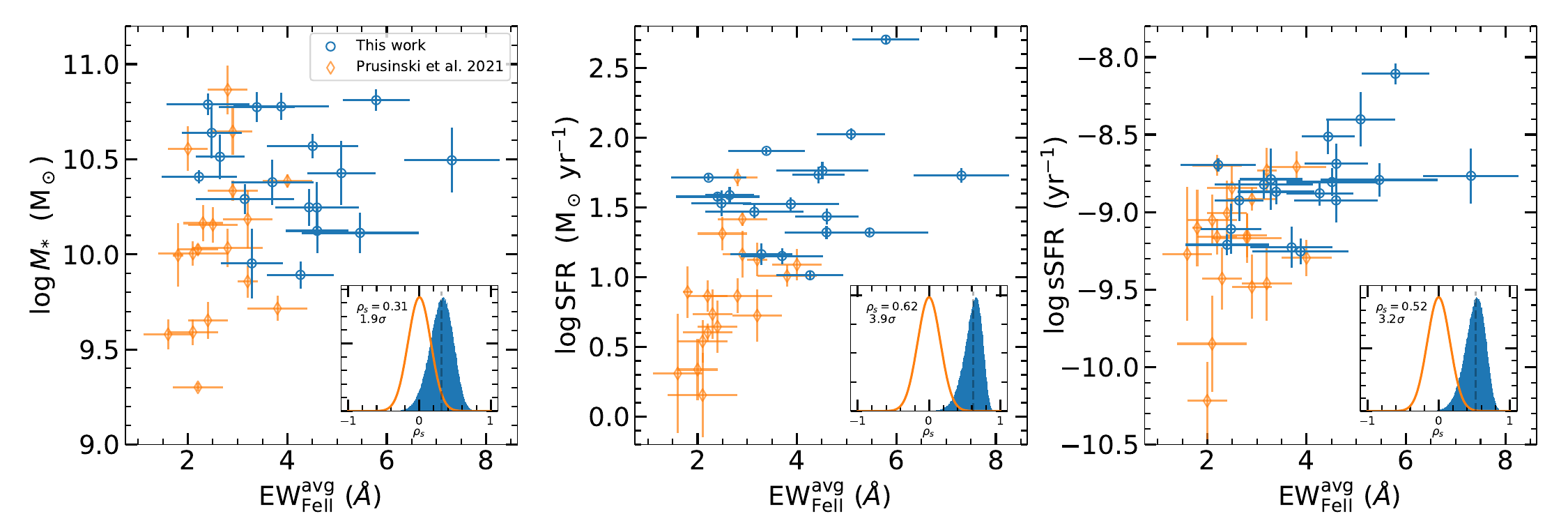}
  \end{center}
  \caption{Plot of the weighted EW average of the \ion{Fe}{II} lines against, from left to right, $M_{*}$, SFR and sSFR. The empty blue circles are the results of this paper and the empty orange diamonds are the results of \citetalias{Prusinski2021}. The inset plots represent the same idea as in Fig. \ref{fig:ew_analysis_MgIIa}. See the Sect. \ref{sec:ew} for a discussion of correlation.}
  \label{fig:ew_analysis_FeII_pru}
\end{figure*}

There are studies in the literature that has led to claims of correlations between EW and certain galaxy properties. \citetalias{Rubin2014} found that the EW of the absorption of \ion{Mg}{II} is correlated with $M_{*}$ with a significance of $3.2\sigma$ under a parameter of the Spearman's rank correlation coefficient\footnote{Spearman's correlation coefficient is a statistical method used to measure the strength and direction of the relationship between two variables. It assesses if there is a consistent monotonic relationship between the ranks of the variables, without assuming a linear association. The correlation coefficient ranges from $-1$ to 1, with positive values indicating a direct relationship, negative values indicating an inverse relationship, and 0 indicating no monotonic relationship. The test provides a p-value to determine the statistical significance of the observed correlation. It is commonly used when dealing with non-linear or non-normally distributed data.} 
($\rho_s$) of $\rho_s=0.44$ and with the SFR at a significance of $3.5\sigma$ ($\rho_s=0.48$). For the EW of \ion{Fe}{II} they found a correlation with SFR at a significance of $2.4\sigma$ ($\rho_s=0.46$) while for $M_{*}$ they found no correlation, claiming that this may be because of the small $M_{*}$ range of their sample. \citetalias{Prusinski2021} performed a similar analysis with other data and found a similar correlation between the EW of \ion{Mg}{II} $\lambda$2796 ($\rho_s=0.67$ at $3.2\sigma$) and \ion{Fe}{II} ($\rho_s=0.65$ at $2.9\sigma$) with the SFR. However, they found no significant correlation between EW and $M_{*}$. 

As we have seen, our objects are dusty and the effect of emission filling is not very important. Nevertheless, we focus our analysis on the \ion{Mg}{II} $\lambda$2796, the blue line of the doublet, which is less affected by this phenomenon. In Fig. \ref{fig:ew_analysis_MgIIa}, from left to right, we have plotted the EW of the \ion{Mg}{II} $\lambda$2796 line against $M_{*}$, SFR, and specific SFR (sSFR) for the objects together from the Lockman--SpReSO, \citetalias{Prusinski2021} and \citetalias{Rubin2014} samples. 
It can be seen that the Lockman--SpReSO galaxies populate regions of the plots that the Rubin and Prusinski objects barely do. The Lockman objects have larger EWs than the other samples, while the properties of the galaxies are comparable (see Fig. \ref{fig:main_properties}). The EW of the Lockman objects is biased towards higher values due to the low resolution of the observations, with values close to $10\ \AA$ and minimum values around $2\ \AA$. However, this allows us to sample other regions of the parameter space and to study in more detail how the galaxy properties relate to the EW of the lines.

To test whether there is a correlation between the EW and the properties of the galaxies, Spearman tests were performed by grouping the three data sets. To estimate the confidence intervals of the correlation, the bootstrap method was applied 10$^7$ times. The blue histograms in the inset of each panel of Fig. \ref{fig:ew_analysis_MgIIa} represent the probability distribution function (PDF) of the Spearman parameter obtained in the bootstrap simulations and the dashed lines mark the mode of the PDF. The expected distribution for the case where there is no correlation between the variables\footnote{For large samples ($n\geq30$), the expected distribution of the Spearman parameter tends to be a normal one $N\left(0,1/\sqrt{n-1}\right)$.} is plotted in orange and the mode of Spearman parameter distribution and the significance are typed. 
It can be seen that for $M_{*}$ the data suggest a positive correlation with the EW, a mode of $\rho_s=0.43$ being obtained at a significance of 4.5$\sigma$, indicating that the greater the mass of the object, the greater the capacity to generate material winds. This result differs from that obtained in the framework of \citetalias{Prusinski2021}, but confirms that found by \citetalias{Rubin2014}, mainly owing to the joint study of all the data, which adds more statistics to the entire range of values, both EW and $M_{*}$. A positive correlation was also found when examining the SFR with a mode of $\rho_s=0.42$ at a significance of 4.4$\sigma$. As we have already discussed, Lockman data add more statistic significance in the high SFR area of the graph, thus better defining the behaviour in this area. The correlation found is not as strong as the one discovered in the work of \citetalias{Prusinski2021} and \citetalias{Rubin2014}, but the significance is higher owing to the larger data set. In the case of sSFR, no correlation was found with the EW of \ion{Mg}{II}. The distribution obtained from the bootstrap simulations and that expected in the case of non-correlation are very close with a mode of $\rho_s=0.11$ at a significance of 1.1$\sigma$. 

In Fig. \ref{fig:ew_analysis_FeIIa_rub} we plot the EW of \ion{Fe}{II} $\lambda$2586 with respect to the properties of the galaxies in our study, together with the \citetalias{Rubin2014} data, using the same procedure and symbols as in Figure \ref{fig:ew_analysis_MgIIa}. It is observed that the EW correlates with $M_{*}$ in a similar way to that found for \ion{Mg}{II}, with $\rho_s=0.43$ at 3.3$\sigma$ of significance. This result differs from that of \citetalias{Rubin2014}, who found no correlation between the parameters ($\rho_s=0.07$ at 0.4$\sigma$). The significant difference in the EW between the data from \citetalias{Rubin2014} and Lockman--SpReSO, mainly due to the different nature of the objects according to the selection criteria, coupled with the increase in the statistics, means that we can further populate the diagram and recover the correlation found between the EW of \ion{Mg}{II} and $M_{*}$. The correlation between the SFR and the EW of \ion{Fe}{II} is the strongest found in this study, with a positive correlation where a mode of $\rho_s=0.69$ was found in the bootstrap simulations at 5.3$\sigma$ of significance. Once more, galaxies with higher SFRs can produce stronger winds. For the sSFR, there is still no significant correlation with the EW ($\rho_s=0.33$ at 2.5$\sigma$).

\citetalias{Prusinski2021} carried out the study of the \ion{Fe}{II} lines using an EW-weighted average because they found the \ion{Fe}{II} lines to be saturated in their spectra. In order to compare our data with their results, we calculated an EW-weighted average of the \ion{Fe}{II} lines present in our objects and we show the results found in Fig. \ref{fig:ew_analysis_FeII_pru}. The correlation with $M_{*}$ is very weak ($\rho_s=0.31$) with a significance of 1.9$\sigma$. In this comparison, the bootstrap distribution and the expected distribution are wider than in the previous cases, mainly because of the smaller data set used. For the SFR we find a significant correlation ($\rho_s=0.62$) at 3.9$\sigma$. This result is consistent with the results of \citetalias{Prusinski2021} and with the previous comparison with what \citetalias{Rubin2014} found. This correlation supports the hypothesis that \ion{Fe}{II} lines are more intense at higher SFRs of galaxies. The sSFR also shows a slight correlation ($\rho_s=0.52$) at 3.2$\sigma$ with the average EW, which we did not find in the previous comparisons. This relation may be due to the smaller statistics we have for $\log\left(\mathrm{sSFR}\right)<-9.5$, with only two sources in this region.

\subsection{Velocity analysis} \label{sec:vel}

\begin{figure*}
  \begin{center}
    \includegraphics[width=\textwidth]{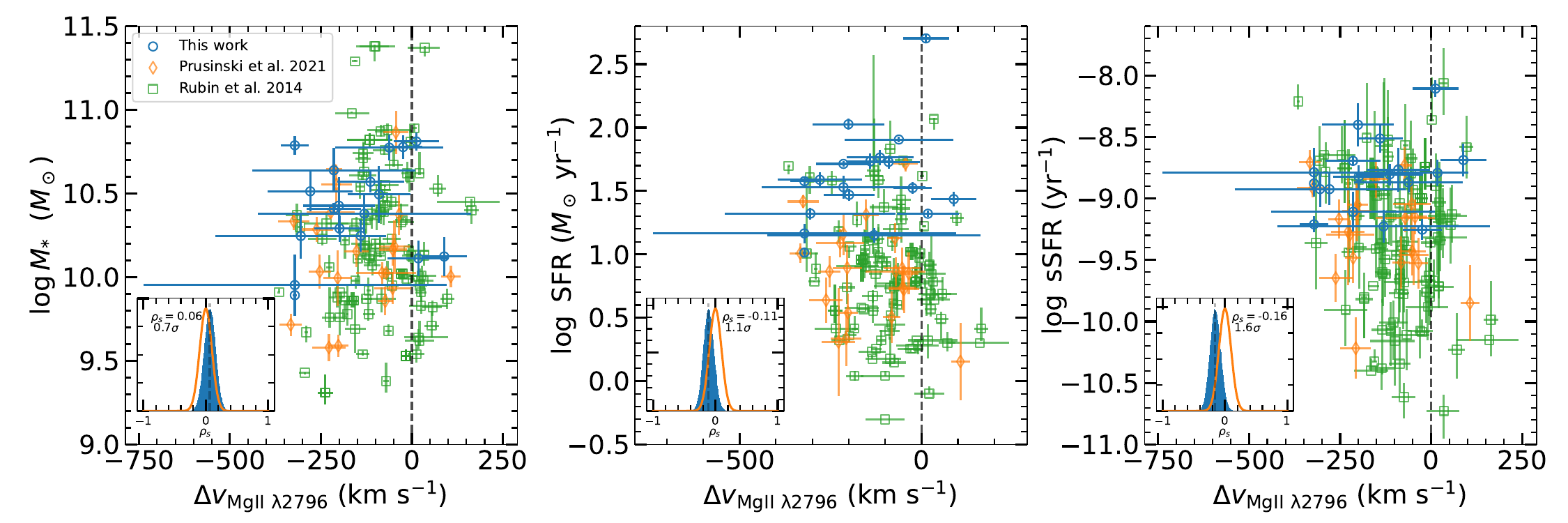}
  \end{center}
  \caption{\ion{Mg}{II} $\lambda$2796 velocity against, from left to right, $M_{*}$, SFR and sSFR. The symbols and colour coding are the same as for Fig. \ref{fig:ew_analysis_MgIIa}. See Sect. \ref{sec:vel} for a discussion of correlation analysis.}
  \label{fig:vel_analysis_MgIIa}
\end{figure*}

\begin{figure*}
  \begin{center}
    \includegraphics[width=\textwidth]{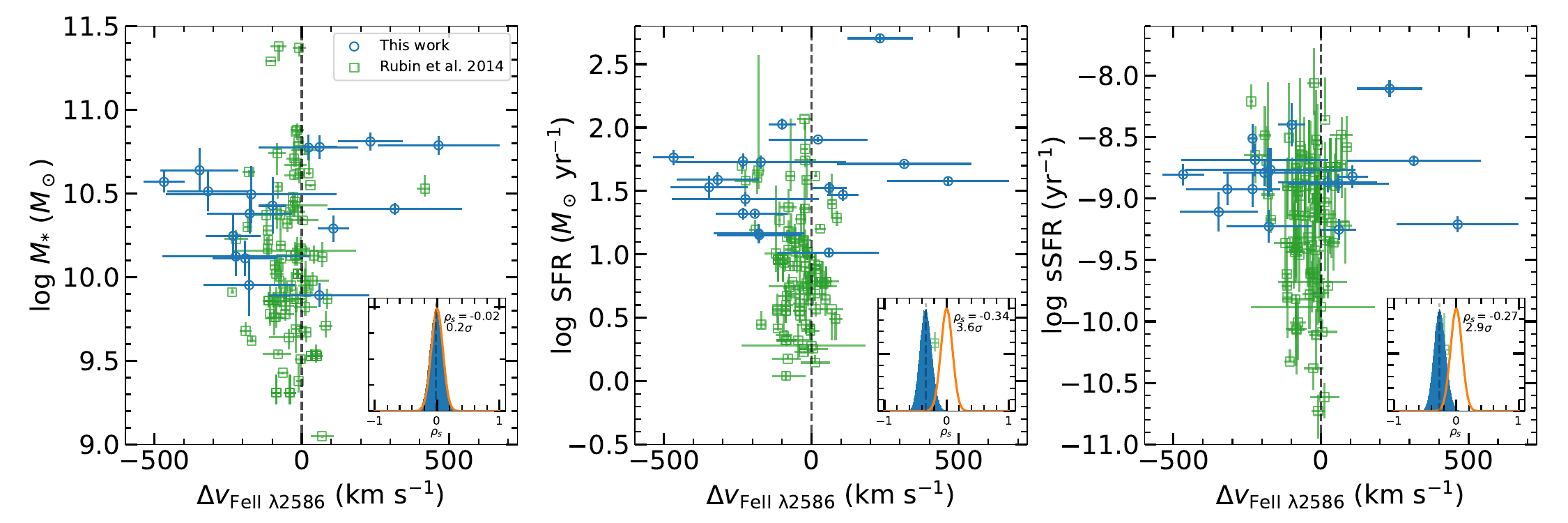}
  \end{center}
  \caption{\ion{Fe}{II} $\lambda$2586 velocity against, from left to right, $M_{*}$, SFR and sSFR. The symbols and colours are as in Fig. \ref{fig:ew_analysis_MgIIa}. See Sect. \ref{sec:vel} for a discussion of correlation.}
  \label{fig:vel_analysis_rub_FeIIa}
\end{figure*}

\begin{figure*}
  \begin{center}
    \includegraphics[width=\textwidth]{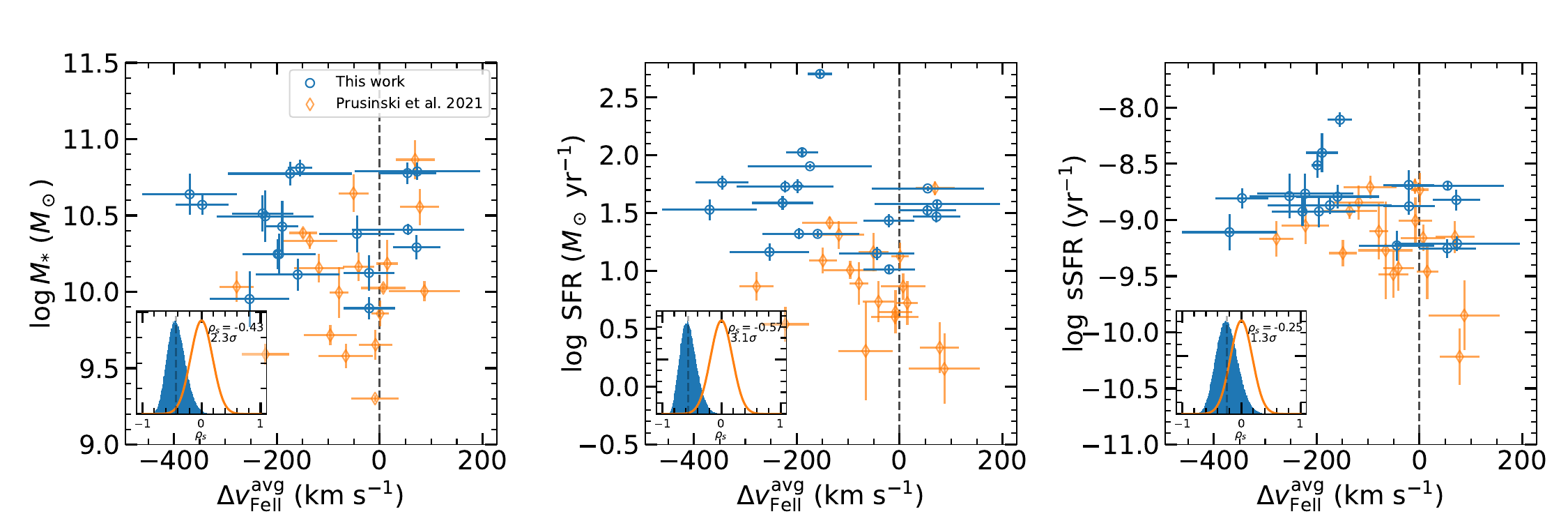}
  \end{center}
  \caption{Average \ion{Fe}{II} velocity against, from left to right, $M_{*}$, SFR and sSFR. The symbols and colours are as in Fig. \ref{fig:ew_analysis_MgIIa}. See the Sect. \ref{sec:vel} for a discussion of correlation.}
  \label{fig:vel_analysis_pru_FeII}
\end{figure*}

As we have seen, the spectral resolution of the Lockman--SpReSO spectra is not the one normally used for the analysis of outflows. It can lead to large errors in the measurement of properties derived from spectral lines, such as the velocity of the ejected material, as can be seen from Table \ref{tab:velocities}.
Despite this severe limitation our sample increases the number of objects that can be studied in this still investigated topic. The initial results provide a fundamental basis for future research with higher resolution, which will help to reduce the margin of error and provide more accurate knowledge about the physical conditions of our sample. Using the velocities obtained we have estimated which objects are receiving an inflow of material, which are ejecting material from the galaxy star-forming regions, and which have velocities compatible with zero (no wind). 

Following the criteria used for the EW analysis, we studied the \ion{Fe}{II} $\lambda$2586 and \ion{Mg}{II} $\lambda$2796 lines to distinguish between galaxies with outflows, inflows, and gas at the systemic velocity. Analysing the \ion{Fe}{II} $\lambda$2586 line, of the 18 objects retrieved from Lockman--SpReSO, nine show material outflows, five show material inflows, and four show velocities compatible with zero, namely absorption only. In the analysis of the \ion{Mg}{II} $\lambda$2796 line, ten objects show outflows, one shows an inflow, and seven have velocities compatible with zero. Considering the three objects classified as AGN, two (120237 and 206641) show an outflows in the \ion{Fe}{II} $\lambda$2586 line and object 206679 has velocities compatible with zero. The \ion{Mg}{II} $\lambda$2796 line is available only in pure absorption for the object 206679, and has an outflow-compatible velocity. The discrepancy between the \ion{Fe}{II} and \ion{Mg}{II} results can be attributed to the high degree of blending of the \ion{Mg}{II} doublet lines with respect to the \ion{Fe}{II} doublet lines, mainly because of the small wavelength separation between the centres of the \ion{Mg}{II} lines ($\sim 7 \, \AA$) and the resolution at which they are observed in Lockman--SpReSO. Furthermore, object 206679, at redshift $\sim$1.44, could be a \ion{Fe}{II} low-ionization broad absorption-line quasar (FeLoBALQ) candidate that shows an AGN-type nature by the presence of \ion{C}{III}] emission, also noticed by \cite{Schmidt1998}, and might be an example of a \textit{loitering outflow} representing a new class of FeLoBALQs described by \cite{Choi2022}. The peculiarity of these FeLoBALQs is that they have low flow velocities and high column density winds located at $\log\left(R_\mathrm{d}\right) \lesssim 1$ pc where $R_\mathrm{d}$ is the distance to the centre.

In Fig.\ \ref{fig:vel_analysis_MgIIa} we have plotted the \ion{Mg}{II} $\lambda$2796 velocities using the data from Lockman--SpReSO, \citetalias{Prusinski2021} and \citetalias{Rubin2014} against the physical parameters of the galaxies, where the possible existence of a correlation was analysed in the same way as the EW (see Sect. \ref{sec:ew}). To investigate the correlation between outflow velocity and galaxy properties, we have restricted the sample to objects that satisfy $\delta v < 50$ km s$^{-1}$, excluding inflows from the study while accounting for measurement errors. The result was that the \ion{Mg}{II} $\lambda$2796 velocity is not correlated with $M_{*}$, SFR, or sSFR as the distributions obtained from the bootstrap simulations are similar to those expected in the uncorrelated case. The same analysis was carried out for the \ion{Fe}{II} $\lambda$2586 line using the \citetalias{Rubin2014} and Lockman data and is shown in Fig.\ \ref{fig:vel_analysis_rub_FeIIa}. Examining the distributions of the bootstrap simulations, we found that there is no significant correlation between the \ion{Fe}{II} $\lambda$2586 velocity and the $M_*$. However, there is a negative correlation between velocity and SFR with $\rho=-0.34$ at 3.6$\sigma$ of significance. The study on sSFR shows a weaker negative correlation ($\rho=-0.27$) at 2.9$\sigma$.

If we examine the mean velocity of all \ion{Fe}{II} lines weighted by the error, we can use the \citetalias{Prusinski2021} and Lockman--SpReSO data in conjunction. As can be seen in Fig.\ \ref{fig:vel_analysis_pru_FeII}, there is a strong correlation of the velocity with the $M_{*}$ ($\rho=-0.43$) and SFR ($\rho=-0.57$). The significance identified is lower due to the limited statistics available in this case and the narrow range of values that can be studied for the parameters. These results differ from that obtained in the framework of \citetalias{Prusinski2021}, where no correlations were found with a significance greater than 1.4$\sigma$. They argue that the lack of correlation is due to the short range of SFR in their objects, which, when combined with the intrinsic scatter of the relationship between SFR and velocity, makes it difficult to find a correlation. Additionally, the study may be affected by a systemic component of the ISM that scales with the SFR, as expected from the Schmidt-Kennicut law \citep{Kennicutt2012}. This component may have a greater impact on studies that use a single component fit of the spectral lines.

In contrast of our results, \cite{Rupke2005b} studying 78 starburst-dominated LIRGs and analysing the \ion{Na}{I} absorption, found that the outflow velocity and properties of these galaxies are independent. However, \cite{Martin2005}, analysing 18 ULIRGs, found that the \ion{Na}{I} velocity follows a relation $\Delta v\sim \mathrm{SFR}^{0.35}$. \cite{Weiner2009}, studying a co-added spectra of 1406 galaxies at $z\sim2$ from the DEEP2 redshift survey, also found that \ion{Mg}{II} absorption in outflows is stronger and reaches higher velocities for more massive galaxies; they obtained a similar relationship for the velocity $\Delta v\sim \mathrm{SFR}^{0.3}$. \cite{Heckman2015}, studying the UV absorption lines of 39 low-redshift starburst galaxies, found that there is a strong correlation of the velocity they measured with the SFR and SFR surface density ($\Sigma_{\rm SFR}$), but a weak correlation with $M_{*}$. This result is supported by \cite{Chisholm2015}, who analysed the \ion{Si}{II} absorption lines in 48 nearby SFGs. They found a correlation with a significance of 3--3.5$\sigma$ between the outflow velocity and galaxy properties such as SFR and $M_{*}$. \cite{Davis2023}
created a sample with a wide range in $M_*$ and SFR. They studied 46 late-stage galaxy mergers in conjunction with data from ten other papers about outflow winds and discovered a significant correlation between outflow velocities and SFR.

As can be seen, our results are inn good agreement with those found in other studies. The Lockman-SpReSO data is crucial for this study. Figures \ref{fig:vel_analysis_MgIIa}, \ref{fig:vel_analysis_rub_FeIIa}, and \ref{fig:vel_analysis_pru_FeII} demonstrate that the FIR-selected galaxies occupy regions in the charts where there were previously few objects, particularly in the higher SFR regions. This provides a more comprehensive sampling of the parameter space, allowing for a more detailed analysis of the general properties of galactic flows. In addition to higher SFRs, the velocities of the \ion{Fe}{II} line tend to be higher than those of \citetalias{Prusinski2021} and \citetalias{Rubin2014} sample (Figs. \ref{fig:vel_analysis_rub_FeIIa} and \ref{fig:vel_analysis_pru_FeII}), increasing the available range of velocities in the correlation study.

\subsection{Local covering factor and optical depth} \label{sec:covering}

The observed depth of an absorption line depends on the optical depth ($\tau$) of an absorbing cloud, and also on the fraction of the background source which the cloud covers, $C$. For a continuum-normalized spectrum, the residual intensity $I$ of the absorption feature is given by

$$I=(1-C)+C{\rm e}^{-\tau}.$$

If two absorption lines are observed closely in wavelength, for example two components of a multiplet of the same ion, for which the ratio of the optical depth is known ($\alpha$), we can solve for $C$ and $\tau$. The ratio of optical depths is the ratio of the oscillator strengths $f$ of the lines which can be found, for example, in \cite{Morton2003} and references therein. If $f_{\rm b}=\alpha f_{\rm r}$ then $\tau_{\rm b}=\alpha\tau_{\rm r}$, and the residual intensities in the red and blue components of the line are:

$$I_{\rm r}=(1-C)+C{\rm e}^{-\tau_{\rm b}/\alpha}$$

\noindent and 

$$I_{\rm b}=(1-C)+C{\rm e}^{-\tau_{\rm b}}.$$

These equations must be solved numerically. See a more detailed discussion in \cite{Benn2005}. 

In our case, we observe the pairs of absorption lines \MgII, \FeIIa, and \FeIIb. By measuring the residual intensities and knowing the oscillator strengths, we then estimate the covering factors ($C$), optical depths ($\tau_0$) and ionic column densities ($N$). Since our spectral resolution is not high enough, covering factors should be taken as lower limits. Results are shown in Table \ref{tab:taus} where we show only the results for the \ion{Fe}{II} absorption lines, since in most cases blending of the \ion{Mg}{II} doublet makes it impossible to obtain a physical solution or is unreliable. In some cases it is also very difficult even for \ion{Fe}{II} lines. 

From the above we have estimated ion column densities (see Table \ref{tab:taus}). Our data, limited by low spectral resolution and the lack of other ionized species, do not allow us to constrain ionization parameters or electron densities. We can, however, crudely estimate hydrogen column densities by assuming solar metallicity, no depletion of \ion{Fe}{}, and 100\% of ionization for the singly ionized iron \citep[Eq. 13 in][]{Martin2012}. For our outflows, \ion{H}{I} column densities are estimated to be in the range $\log\left(N\left(\ion{H}{I}\right)\right)\sim 19.7$--20.5 cm$^{-2}$, and in some cases can be up to $\log\left(N\left(\ion{H}{I}\right)\right)=21$. These large values seem unlikely, so we would expect higher than solar metallicities. Alternatively, for a control test, we can estimate \ion{H}{I} column densities using the relationship defined in \cite{Menard2009} between \ion{H}{I} column densities and rest frame \ion{Mg}{II} EW. We measured \ion{Mg}{II} EWs to be on average $\sim5\ \AA$, giving $\log\left(N\left(\ion{H}{I}\right)\right)\sim20.7\ \mathrm{cm}^{-2}$, consistent with our previous estimate.

Ionic densities and covering factors are similar to those in the literature \citep{Martin2012}. Also, average densities are $\log\left(N\right) \lesssim 16$ cm$^{-2}$ so that we could conclude that we observed similar galactic outflows as in previous studies; that is, typical mass flux in the low-ionization outflows can be of the order of $\sim 23\ M_\odot$ yr$^{-1}$ \citep[eq. 12 from][]{Martin2012}.

\begin{table*}
  \centering
  \caption{Covering factor, optical depth, and ionic column density obtained in the analysis of the \ion{Fe}{II} doublet lines.}
  \label{tab:taus}
  \begin{tabular}{c|ccc|ccc|ccc|ccc}
  \hline\hline
  \multicolumn{1}{c|}{\multirow{2}{*}{ID}} & \multicolumn{3}{c|}{\ion{Fe}{II} $\lambda$2374} & \multicolumn{3}{c|}{\ion{Fe}{II} $\lambda$2382} & \multicolumn{3}{c|}{\ion{Fe}{II} $\lambda$2586} & \multicolumn{3}{c}{\ion{Fe}{II} $\lambda$2600}  \\ \cline{2-13} 
  \multicolumn{1}{c|}{}            & $C$   & $\tau_{\rm 0}$   & $\log N $ (cm$^{-2}$)   & $C$   & $\tau_{\rm 0}$   & $\log N $ (cm$^{-2}$)   & $C$  & $\tau_{\rm 0}$ & $\log N $ (cm$^{-2}$)       & $C$  & $\tau_{\rm 0}$ & $\log N $ (cm$^{-2}$)        \\
  \hline
  123207                   & -   & -    &  -   & -    & -   & -   & 0.33 & 1.2 & 15.7  & 0.33 & 4.2  & 15.4        \\
  96864                    & -   & -    &  -   & -    & -   & -   & 0.3 & $\infty$ & >16  & 0.3 & $\infty$ & >15.7       \\
  101926                   & 0.31  & 2.5   & 16.0  & 0.31   & 2.7  & 15.7  & 0.3 & $\infty$ & >16  & 0.3 & $\infty$ & >15.8       \\
  120080                   & -   &  -   &  -   & -    & -   & -   & 0.33 & $\infty$ & >16  & 0.33 & $\infty$ & >16.2       \\
  118338                   & -   &  -   &  -   & -    & -   & -   & 0.20 & $\infty$ & >16  & 0.20 & $\infty$ & >15.6       \\
  95738                    & -   &  -   &  -   & -    & -   & -   & 0.28 & 4.8 & 16.2  & 0.28 & 16.6 & 16.0        \\
  109219                   & -   &  -   &  -   & -    & -   & -   & 0.19 & 1.8 & 15.9  & 0.19 & 6.3  & 15.7        \\
  94458                    & -   &  -   &  -   & -    & -   & -   & 0.24 & 1.2 & 15.4  & 0.24 & 4.1  & 15.6        \\
  92467                    & -   &  -   &  -   & -    & -   & -   & 0.30 & 1.3 & 15.4  & 0.30 & 4.4  & 15.5        \\
  120257                   & 0.54  & 0.25  & 15.3  & 0.54   & 2.6  & 15.2  & 0.43 & 0.7 & 15.7  & 0.43 & 2.3  & 15.3        \\
  95958                    & 0.17  & 3    & 16.4  & 0.17   & 31   & 16.1  & 0.35 & 0.7 & 15.3  & 0.35 & 2.5  & 15.2        \\
  116662                   & 0.20  & 1.6   & 16.0  & 0.20   & 17   & 15.8  & 0.32 & 1.1 & 15.5  & 0.32 & 3.5  & 15.5        \\
  133957                   & -   & -   & -   & -    & -   & -   & 0.38 & 1.3 & 15.9  & 0.38 & 4.5  & 15.5        \\
  186820                   & -   & -   & -   & -    & -   & -   & 0.19 & 1.5 & 15.8  & 0.19 & 5.3  & 15.3        \\
  97778                    & 0.15  & 1.8   & 16.0  & 0.15   & 19   & 15.8  & 0.25 & 2.5 & 15.7  & 0.25 & 8.6  & 15.8        \\
  77155                    & 0.34  & 1.1   & 16.0  & 0.34   & 12   & 15.8  & 0.43 & 2.9 & 15.9  & 0.43 & 10  & 15.8        \\
  102473                   & 0.26  & 3.2   & 16.3  & 0.26   & 34   & 16.1  & 0.34 & 0.7 & 15.8  & 0.34 & 2.2  & 15.5        \\
  120237$^{a}$                & 0.17  & 4.4   & 16.5  & 0.17   & 47   & 16.3  & 0.30 & 1.0 & 15.7  & 0.30 & 3.4  & 15.5        \\
  206641$^{a}$                & 0.08  & 0.2   & 15.2  & 0.08   & 2.2  & 15.1  & 0.08 & 0.8 & 15.2  & 0.08 & 2.9  & 15.6        \\
  78911                    & 0.41  & 0.5   & 15.6  & 0.41   & 5.6  & 15.5  & 0.53 & 1.8 & 15.9  & 0.53 & 6.2  & 15.7        \\
  206679$^{a}$                & 0.41  & 0.9   & 16.0  & 0.41   & 10   & 15.8  & 0.60 & 1.3 & 15.6  & 0.60 & 4.5  & 15.4        \\
  \hline
  \end{tabular}\\[0.2cm]
  \raggedright\small{$^{a}$ Objects classified as AGN by González-Otero et al. (in prep). }\\
\end{table*}

\section{Summary and conclusions} \label{sec:conclusions}

\begin{table}
  \centering
  \caption{Summary of the Spearman rank correlation coefficient and the significance found in the study of the \ion{Mg}{II} $\lambda$2796 line.}
  \label{tab:spearman_coefs_MgII}
  \begin{tabular}{ccccccccccccccc}
  \hline \hline
  \multicolumn{1}{c|}{} & \multicolumn{6}{c}{\ion{Mg}{II} $\lambda$2796}  \\ \cline{2-7}
  \multicolumn{1}{c|}{$_{\rho_s}$} & \multicolumn{2}{c|}{\citetalias{Prusinski2021}} & \multicolumn{2}{c|}{\citetalias{Rubin2014}} & \multicolumn{2}{c}{This work} \\ \cline{2-7}
  \multicolumn{1}{c|}{} & \multicolumn{1}{c|}{EW} & \multicolumn{1}{c|}{$\Delta v$} & \multicolumn{1}{c|}{EW} & \multicolumn{1}{c|}{$\Delta v$} & \multicolumn{1}{c|}{EW} & $\Delta v$  \\ 
  \hline
  $M_{*}$           & - & - & $0.44_{3.2\sigma}$ & -          & $0.43_{4.5\sigma}$ & $0.06_{0.7\sigma}$  \\ [0.5ex]
  $\mathrm{SFR}$    & $0.67_{3.2\sigma}$ & - & $0.48_{3.5\sigma}$ & -          & $0.42_{4.4\sigma}$ & $-0.11_{1.1\sigma}$   \\ [0.5ex]
  $\mathrm{sSFR}$   & - & - & -                  & -          & $0.11_{1.1\sigma}$ & $-0.16_{1.6\sigma}$  \\ [0.5ex]
  \hline
   \end{tabular}
\end{table}

\begin{table}
  \centering
  \caption{Summary of the Spearman rank correlation coefficient and the significance found in the study of the \ion{Fe}{II} $\lambda$2586 line.}
  \label{tab:spearman_coefs_FeII}
  \begin{tabular}{cccccccccccccccccccc}
  \hline \hline
  \multicolumn{1}{c|}{} & \multicolumn{4}{c}{\ion{Fe}{II} $\lambda$2586}  \\ \cline{2-5}
  \multicolumn{1}{c|}{$_{\rho_s}$} & \multicolumn{2}{c|}{\citetalias{Rubin2014}} & \multicolumn{2}{c}{This work} \\ \cline{2-5}
  \multicolumn{1}{c|}{} & \multicolumn{1}{c|}{EW} & \multicolumn{1}{c|}{$\Delta v$} & \multicolumn{1}{c|}{EW} & $\Delta v$  \\ 
  \hline
  $M_{*}$     & $0.07_{0.4\sigma}$ & - & $0.43_{3.3\sigma}$ & $0.02_{0.2\sigma}$ \\ [0.5ex]
  $\mathrm{SFR}$    & $0.46_{2.4\sigma}$ & - & $0.69_{5.3\sigma}$ & $-0.34_{3.6\sigma}$ \\ [0.5ex]
  $\mathrm{sSFR}$   & -         & - & $0.33_{2.5\sigma}$ & $-0.27_{2.9\sigma}$ \\ [0.5ex]
  \hline
  \end{tabular}
\end{table}

\begin{table}
  \centering
  \caption{Summary of the Spearman rank correlation coefficient and the significance found in the study of the weighted average of the \ion{Fe}{II} lines. }
  \label{tab:spearman_coefs_FeII_avg}
  \begin{tabular}{cccccccccccccccccccc}
  \hline \hline
  \multicolumn{1}{c|}{} & \multicolumn{4}{c}{\ion{Fe}{II}$_{\rm avg}$}  \\ \cline{2-5}
  \multicolumn{1}{c|}{$_{\rho_s}$} & \multicolumn{2}{c|}{\citetalias{Prusinski2021}} & \multicolumn{2}{c}{This work} \\ \cline{2-5}
  \multicolumn{1}{c|}{} & \multicolumn{1}{c|}{EW$_{\rm avg}$} & \multicolumn{1}{c|}{$\Delta v_{\rm avg}$} & \multicolumn{1}{c|}{EW$_{\rm avg}$} & \multicolumn{1}{c}{$\Delta v_{\rm avg}$}  \\ 
  \hline
  $M_{*}$     & $_{\leq1.4\sigma}$ & $_{\leq1.4\sigma}$ & $0.31_{1.9\sigma}$ & $-0.43_{2.3\sigma}$ \\ [0.5ex]
  $\mathrm{SFR}$    & $0.65_{2.9\sigma}$ & $_{\leq1.4\sigma}$ & $0.62_{3.9\sigma}$ & $-0.57_{3.1\sigma}$ \\ [0.5ex]
  $\mathrm{sSFR}$   & $_{\leq1.4\sigma}$ & $_{\leq1.4\sigma}$ & $0.52_{3.2\sigma}$ & $-0.25_{1.3\sigma}$ \\ [0.5ex]
  \hline
  \end{tabular}
\end{table}

In this paper, we have studied galactic flows in objects from the Lockman--SpReSO project \citep{Gonzalez2022}. The objects were selected for having \ion{Fe}{II} lines in absorption, and \ion{Mg}{II} lines in absorption and emission. We found 21 objects with redshift in the range $0.5\lesssim z \lesssim1.45$, of which three were classified as AGN (González-Otero in prep.). The objects in the sample span an $M_{*}$ range of $9.89 < \log( M_{*} /M_\odot) < 11.50$ and are (U)LIRGs ($10.84 < \log(L_{\rm TIR}/L_\odot) < 12.53$) with relatively high $\log(\mathrm{SFR})$ between 1.01 and 2.70. We have determined the systemic velocities of the objects and measured the EW and velocities of the \FeIIa, \FeIIb, \FeIIc, \MgII, and \MgI\ spectral lines.

The EWs and line velocities were used to explore possible correlations with the $M_{*}$, SFR and sSFR values of the galaxies. In order to perform this study with better statistics we made a joint analysis of our sample of 18 objects with an additional 22 and 105 galaxies from \citetalias{Prusinski2021} and \citetalias{Rubin2014} respectively. Bootstrap simulations were performed on the Spearman rank test to check for the existence of correlations between the properties. The inclusion of Lockman--SpReSO objects, selected for their FIR emission, adds great value to the sample since, to the best of our knowledge, there are very few such studies based on LIRGs \citep[for example][]{Banerji2011}. This helps us to validate whether the results obtained with other types of objects are also valid for the FIR-selected objects.

Using the three samples as a whole, the EW of \ion{Mg}{II} $\lambda$2796 was found to correlate strongly, $\rho_s=0.43$ ($\rho_s=0.42$) and significantly at 4.5$\sigma$ (4.4$\sigma$) with $M_{*}$ (SFR). This result is in good agreement with \citetalias{Rubin2014} but only for SFR with \citetalias{Prusinski2021}, who found no correlation of EW with $M_{*}$. The Lockman--SpReSO sample has enabled a more in-depth examination of the correlations between EW and galaxy properties. The EWs of these objects occupy regions of the parameter space where object density is very low, with higher values than in the \citetalias{Rubin2014} and \citetalias{Prusinski2021} samples. This enables us to expand the boundaries of the relationships. The \ion{Mg}{II} $\lambda$2796 line velocity has no correlation ($\rho_s\sim-0.1$) with the properties of the galaxies in the sample. \cite{Davis2023} found a positive correlation between velocity and SFR for the \ion{Mg}{II} line. The discrepancy observed could be attributed to the effect of emission line infilling, which dilutes the lines and introduces errors. Table \ref{tab:spearman_coefs_MgII} contains a summary of the Spearman coefficient obtained in the analysis of \ion{Mg}{II} $\lambda$2796 line.

For the analysis of the \ion{Fe}{II} lines, separate studies were carried out for the \citetalias{Prusinski2021} and \citetalias{Rubin2014} samples. In the sample with Lockman--SpReSO and \citetalias{Rubin2014} data, a positive correlation $\rho_s=0.43$ ($\rho_s=0.69$) and a significant one at 3.3$\sigma$ (5.3$\sigma$) were found with the $M_{*}$ (SFR) and the EW of \ion{Fe}{II} $\lambda$2586. With the sSFR the correlation found is not very strong ($\rho_s=0.33$) and of marginal significance (2.5$\sigma$). The velocity has no significant correlation with $M_*$, but a strong correlation ($\rho=-0.34$) with high significance ($3.6\sigma$) was found with SFR. This result implies that the velocity of \ion{Fe}{II} spectral lines remains decoupled from $M_{*}$ but not from SFR; in other words, the energy suffered by the material depends more on SFR than on $M_*$. A summary of the correlations can be seen in Table \ref{tab:spearman_coefs_FeII}. To study the Lockman--SpReSO and \citetalias{Prusinski2021} sample as a whole, we performed weighted averages of the EW and velocities of the \ion{Fe}{II} lines measured in the Lockman--SpReSO objects. We found that EW$_{\rm avg}$ is strongly correlated with the SFR ($\rho_s=0.62$) at 3.9$\sigma$ and more marginally correlated with the $M_{*}$ ($\rho_s=0.31$) and sSFR ($\rho_s=0.52$) at 1.9$\sigma$ and 3.2$\sigma$ respectively. The average velocity also show strong correlation with $M_*$ ($\rho=-0.43$) and SFR ($\rho=-0.57$). This is consistent with the previously found result that the velocity is dependent of the properties of the SFR. In general, thanks to the inclusion of the Lockman--SpReSO objects selected for their FIR emission, which have been little studied with respect to galactic flows in distant FIR objects, we have confirmed the results of previous studies and found discrepancies in others. A summary of the correlations obtained for the weighted average of \ion{Fe}{II} lines can be seen Table \ref{tab:spearman_coefs_FeII_avg}.

Although the spectral resolution and dispersion used in the Lockman--SpReSO observations ($R\sim500$, $\sim4$ $\AA\,\mathrm{pix}^{-1}$) are not commonly used for outflow analysis, where higher resolutions are usually required, we were able to detect the existence of nine (ten) galactic outflows, five (one) inflows and four (seven) absorption-only objects based on the velocities obtained for the \ion{Fe}{II} $\lambda$2586 (\ion{Mg}{II} $\lambda$2796) line. In addition, of the three objects classified as AGN, object 206679 was found to be a clear candidate for \textit{loitering outflows}, a new class of FeLoBALQs characterized by low flow velocities ($v\lesssim2000$ kms$^{-1}$) and high column density winds located at $\log\left(R_\mathrm{d}\right) \lesssim 1$ pc. 

Finally, it is important to highlight that this study was not initially planned, but rather emerged from the serendipitous discovery of these objects in the Lockman--SpReSO data. Despite the unplanned nature of the study, the findings have yielded valuable insights into the characteristics and behaviour of these objects. Higher-resolution observations will help us to better constrain and study the velocities of the objects with increased accuracy, as well as provide more information about the FeLoBALQ candidate. But this study undoubtedly sheds more light on the study of galactic flows by adding objects of a different nature to what has been studied so far, as they are objects selected for their FIR emission.

\begin{acknowledgements}
We thank the anonymous referee for their useful report. 
This work was supported by the Evolution of Galaxies project, of references: 
PRE2018-086047,
AYA2017-88007-C3-1-P, 
AYA2017-88007-C3-2-P, 
AYA2018-RTI-096188-BI00, 
PID2019-107408GB-C41, 
PID2019-106027GB-C41, 
PID2021-122544NB-C41, 
and 
MDM-2017-0737 (Unidad de Excelencia María de Maeztu, CAB), within the \textit{Programa estatal de fomento de la investigación científica y técnica de excelencia del Plan Estatal de Investigación Científica y Técnica y de Innovación (2013-2016)} of the Spanish Ministry of Science and Innovation/State Agency of Research MCIN/AEI/ 10.13039/501100011033 and by `ERDF A way of making Europe'. 
this work is based on observations made with the Gran Telescopio Canarias (GTC) at Roque de los Muchachos Observatory on the island of La Palma, with the Willian Herschel Telescope (WHT) at Roque de los Muchachos Observatory on the island of La Palma and on observations at Kitt Peak National Observatory, NSF's National Optical-Infrared Astronomy Research Laboratory (NOIRLab Prop. ID: 2018A-0056; PI: Gonz\'alez-Serrano, J.I.), which is operated by the Association of Universities for Research in Astronomy (AURA) under a cooperative agreement with the National Science Foundation. 
This research has made use of the NASA/IPAC Extragalactic Database (NED), which is funded by the National Aeronautics and Space Administration and operated by the California Institute of Technology.
EB and ICG acknowledge support from DGAPA-UNAM grant IN119123.
J.N acknowledges the support of the National Science Centre, Poland through the SONATA BIS grant 2018/30/E/ST9/00208
Y.K. acknowledges support from DGAPA-PAPIIT grant IN102023.
The authors thank Terry Mahoney (at the IAC's Scientific Editorial Service) for his substantial improvements of the manuscript.
   
\end{acknowledgements}

%  \bibliographystyle{aa} % style aa.bst
%  \bibliography{Yourfile} % your references Yourfile.bib

\bibliographystyle{aa}
\bibliography{biblio} 

\onecolumn
\begin{appendix} %First appendix
\section{Cutouts and SED fits of the Lockman--SpReSO objects} \label{sec:appendix}

\begin{figure*}[!ht]
  \begin{center}
    \includegraphics[width=0.40\textwidth]{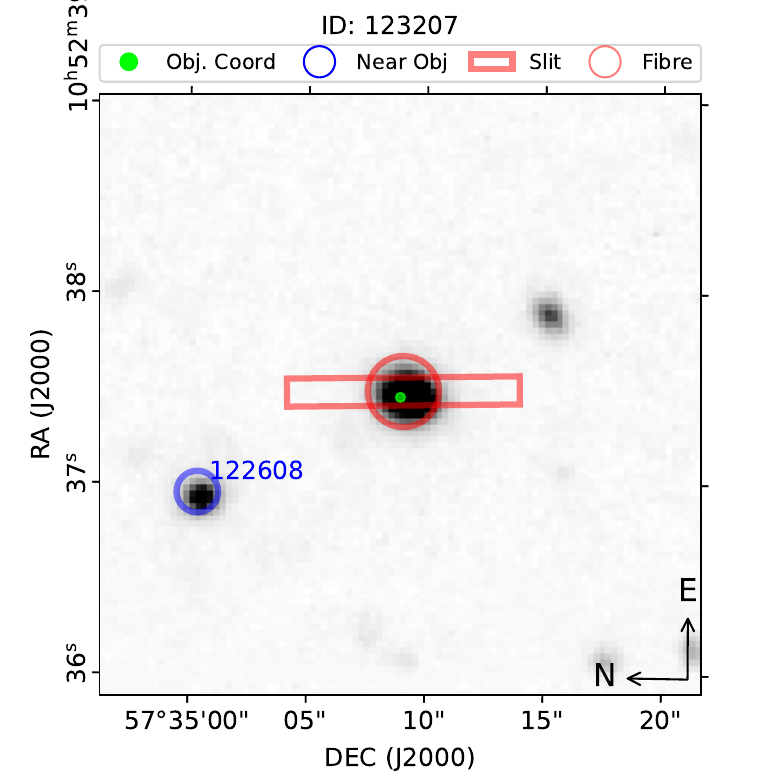}
    \includegraphics[width=0.53\textwidth]{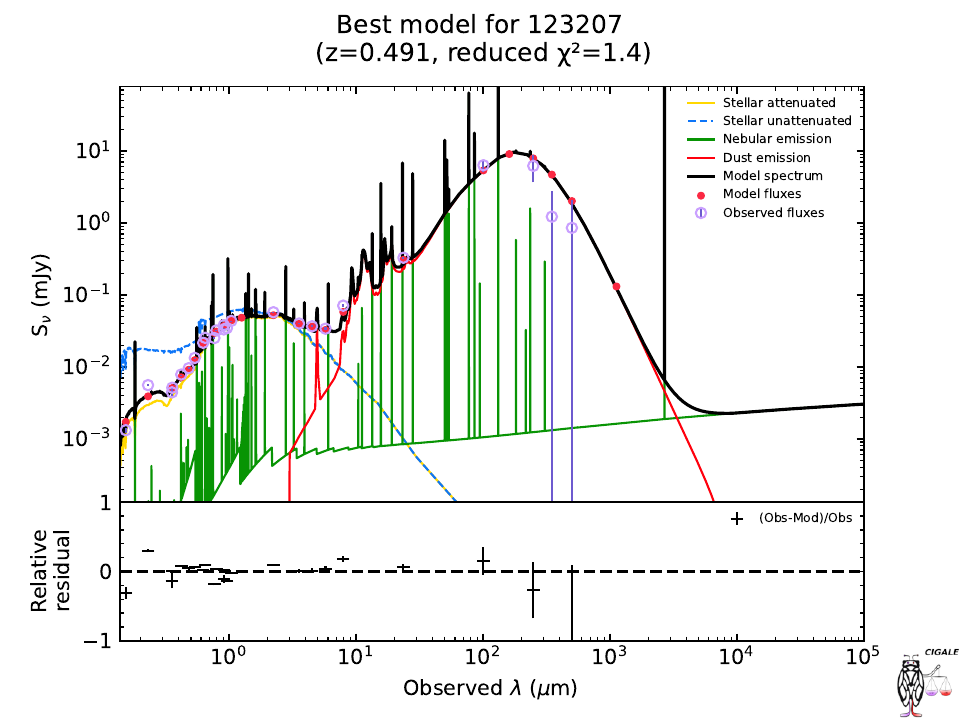}
    \includegraphics[width=0.30\textwidth]{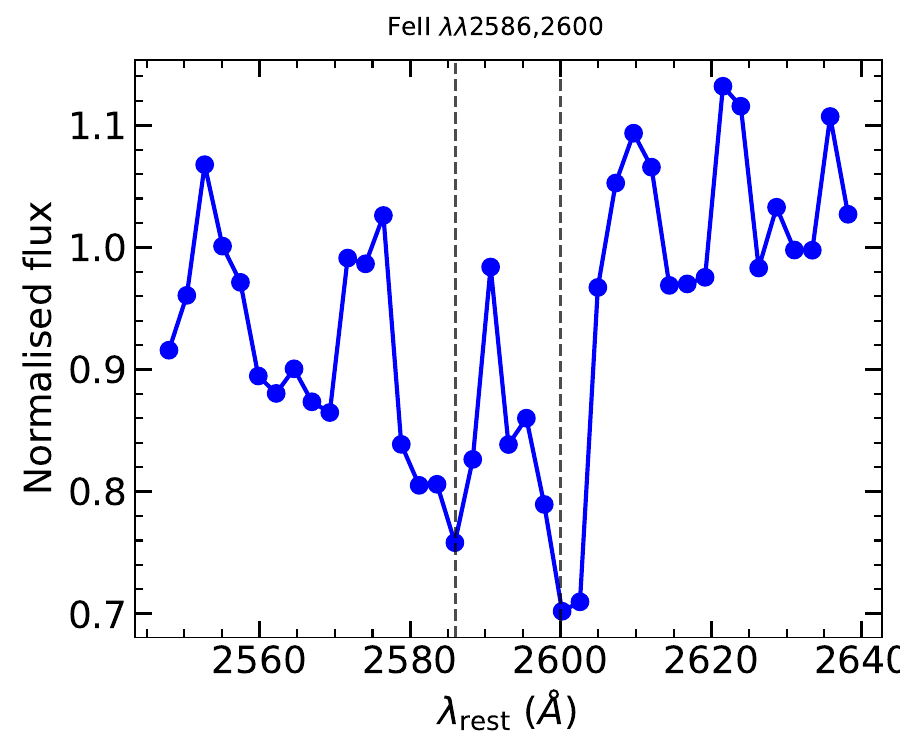}
    \includegraphics[width=0.30\textwidth]{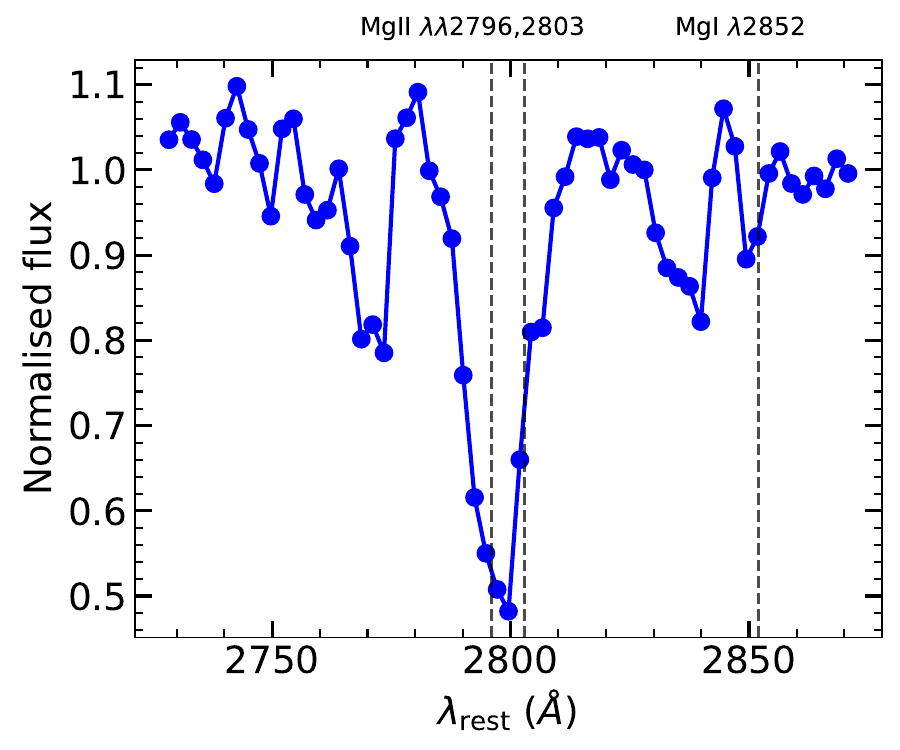}
  \end{center}
  \caption{Cutouts (top left), SED fits (top right) and spectra slices (bottom) for the object ID 123207 studied in this paper. The cutouts are made on the GTC image of the Lockman-SpReSO survey field (see \citealt{Gonzalez2022} for details). The green dot represents the optical coordinates of the object. The blue circle marks nearby objects in the Lockman-SpReSO catalogue. The red rectangle represents the position and size of the slit used to observe the object. There are two types of slit: a small slit of 3 arcsec and a large slit of 10 arcsec (see \citealt{Gonzalez2022} for more details). The red circle represents the position of the fibre. In the SED fits the best model is plotted as a solid black line, the photometric information of the object is plotted as empty violet circles, and the red filled circles are the fluxes obtained by the best model. The individual contributions of the models used are also plotted where the yellow line represents the attenuated stellar component, the blue dashed line is the unattenuated stellar component, the green line illustrates the nebular emission, and the red line is the dust emission. The relative residuals of the flux for the best model are plotted at the bottom. The spectral slices show the absorption lines studied in this paper. The spectra have been normalised to the continuum. The grey vertical dashed lines represent the rest-frame wavelength of the lines.}
  \label{fig:cut_out_SED_all}
\end{figure*}
\begin{figure*}[!ht]
  \begin{center}
    \includegraphics[width=0.40\textwidth]{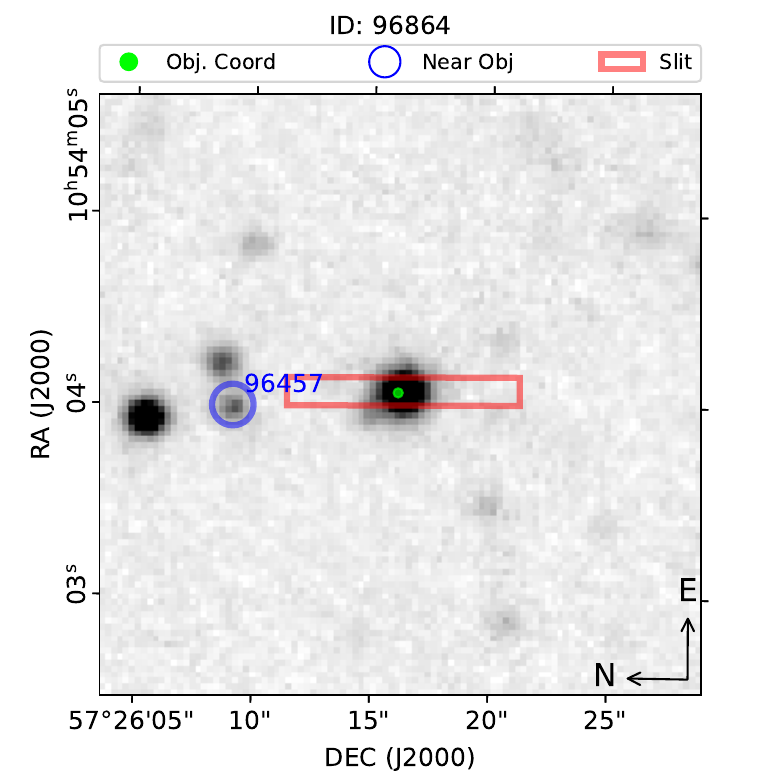}
    \includegraphics[width=0.53\textwidth]{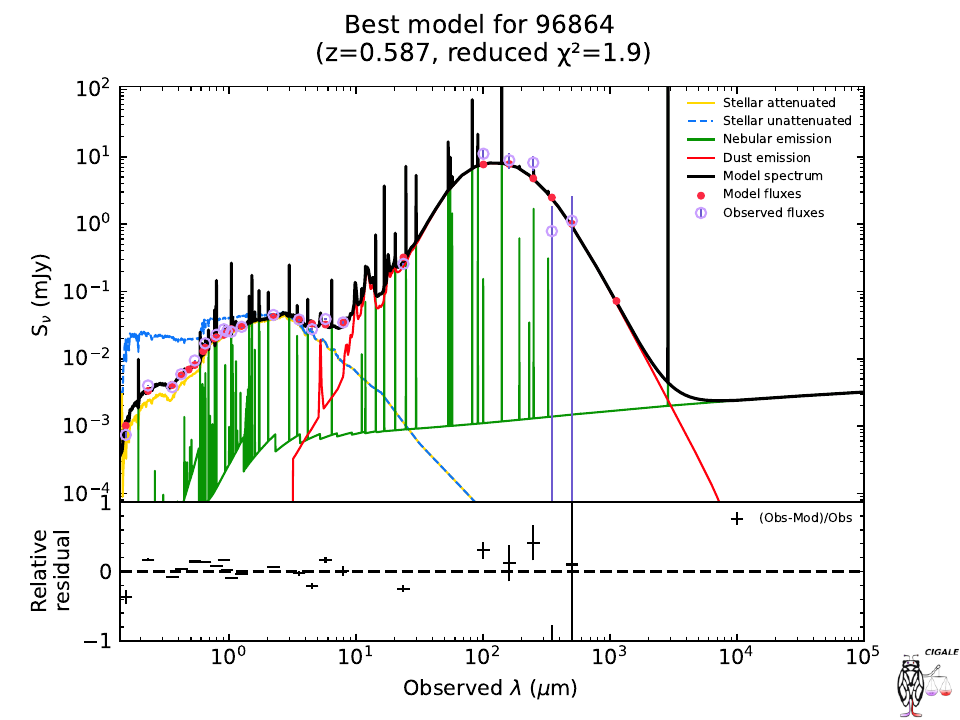}
    \includegraphics[width=0.20\textwidth]{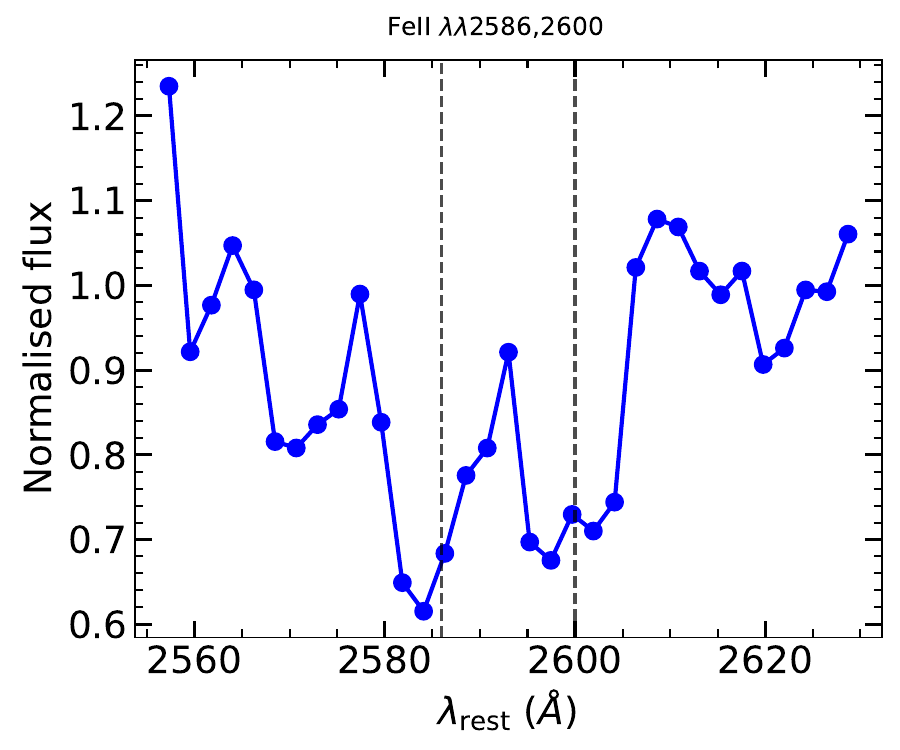}
    \includegraphics[width=0.20\textwidth]{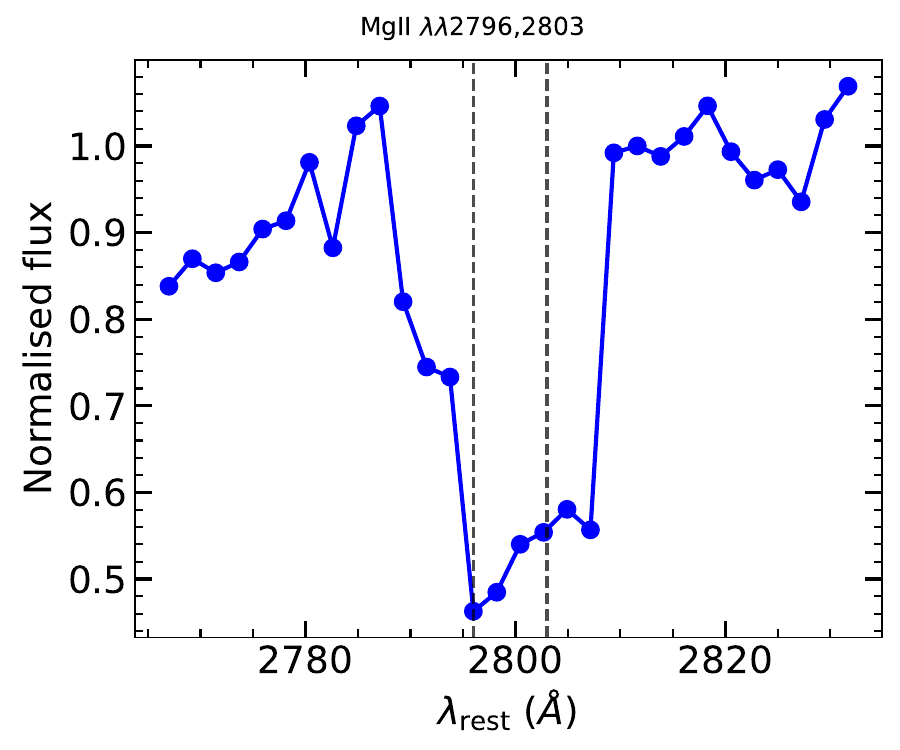}
  \end{center}
  \caption{Same as Fig. \ref{fig:cut_out_SED_all} but for object ID 96864.}
  \label{fig:cut_out_SED_all1}
\end{figure*}
\begin{figure*}[!ht]
  \begin{center}
    \includegraphics[width=0.40\textwidth]{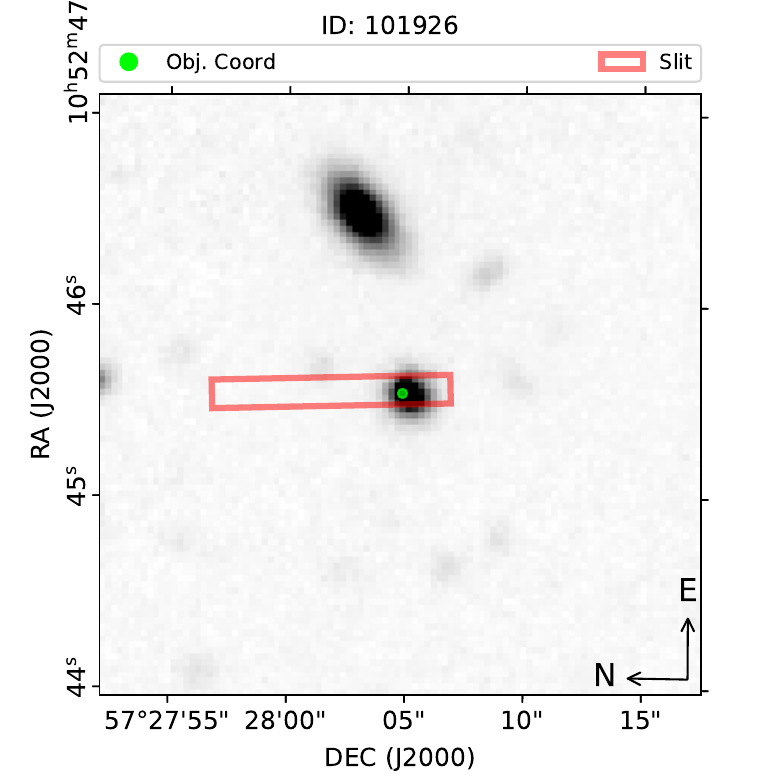}
    \includegraphics[width=0.53\textwidth]{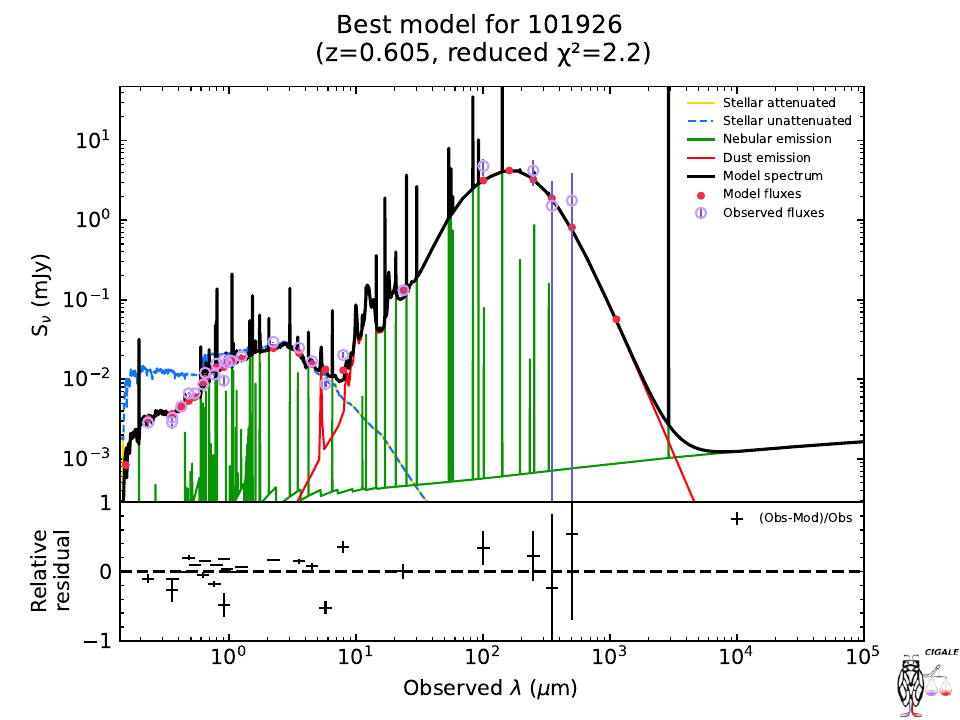}
    \includegraphics[width=0.20\textwidth]{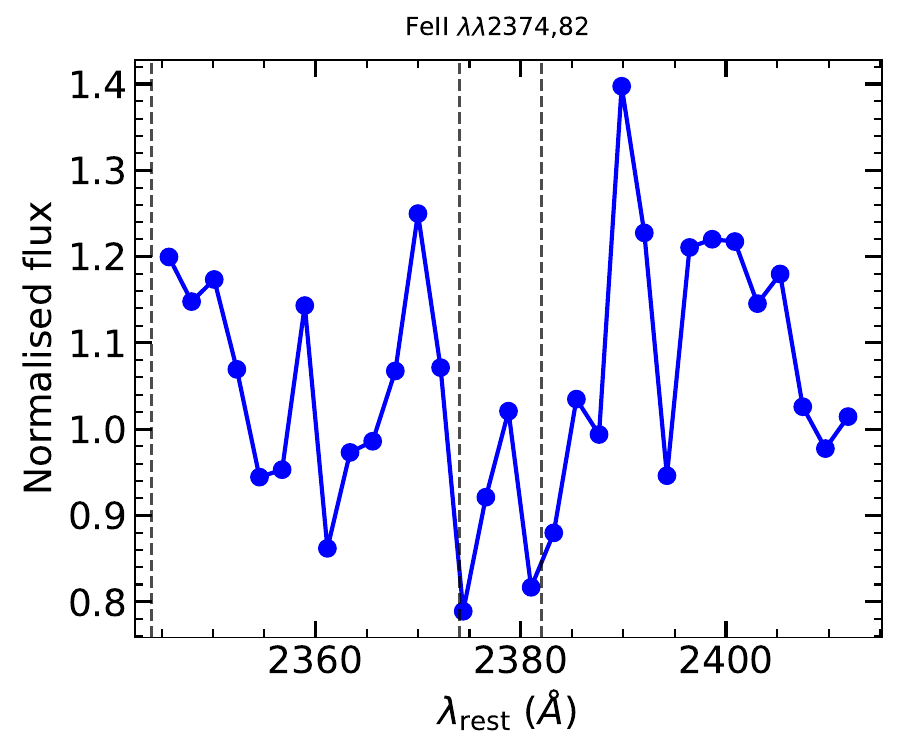}
    \includegraphics[width=0.20\textwidth]{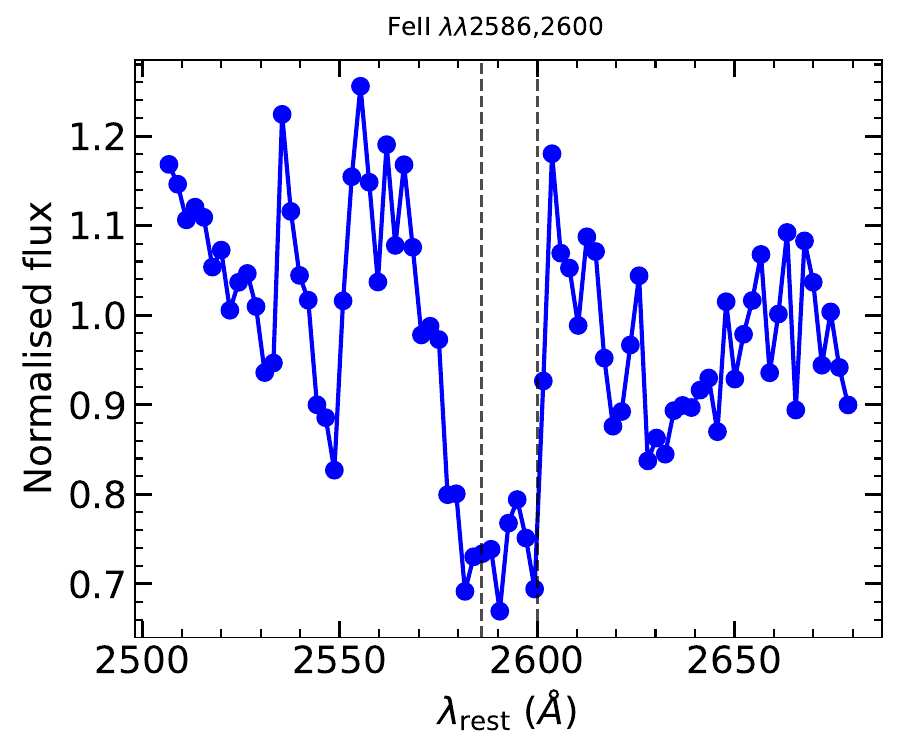}
    \includegraphics[width=0.20\textwidth]{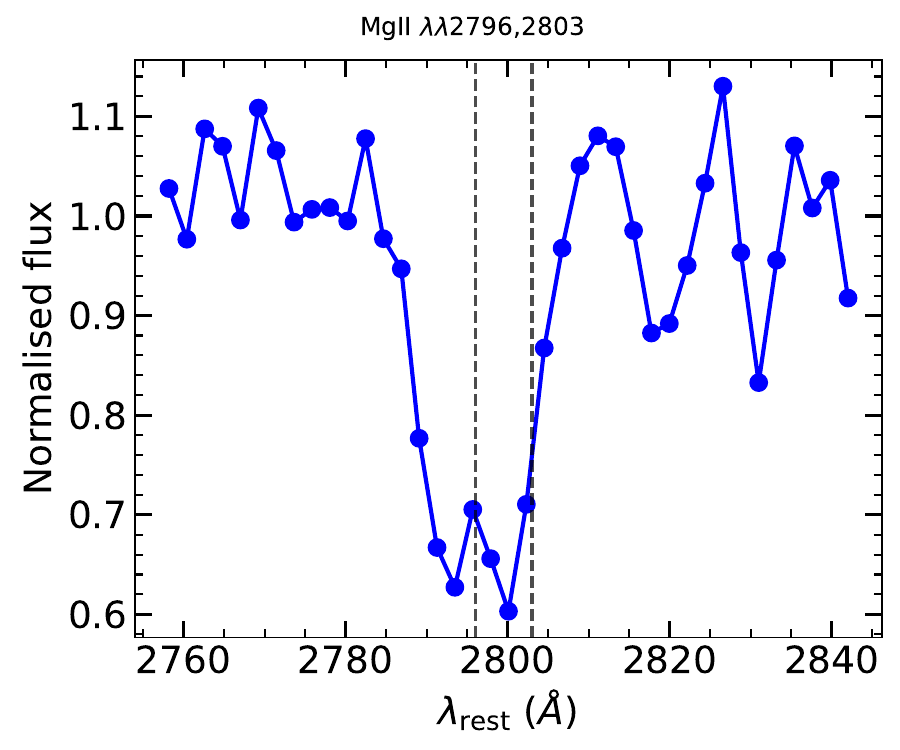}
  \end{center}
  \caption{Same as Fig. \ref{fig:cut_out_SED_all} but for object ID 101926.}
  \label{fig:cut_out_SED_all2}
\end{figure*}
\begin{figure*}[!ht]
  \begin{center}
    \includegraphics[width=0.40\textwidth]{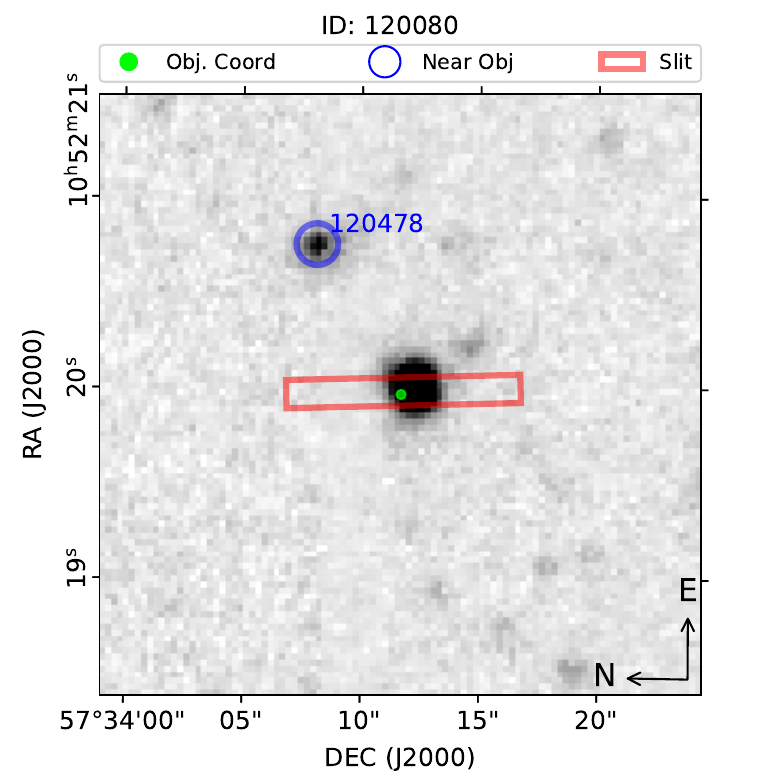}
    \includegraphics[width=0.53\textwidth]{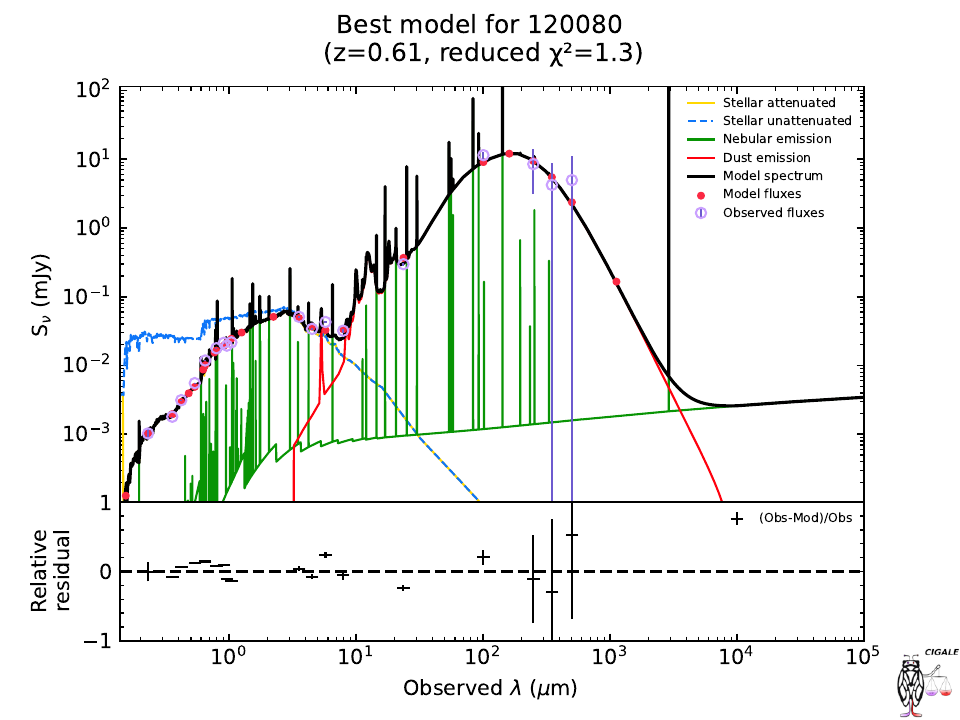}
    \includegraphics[width=0.20\textwidth]{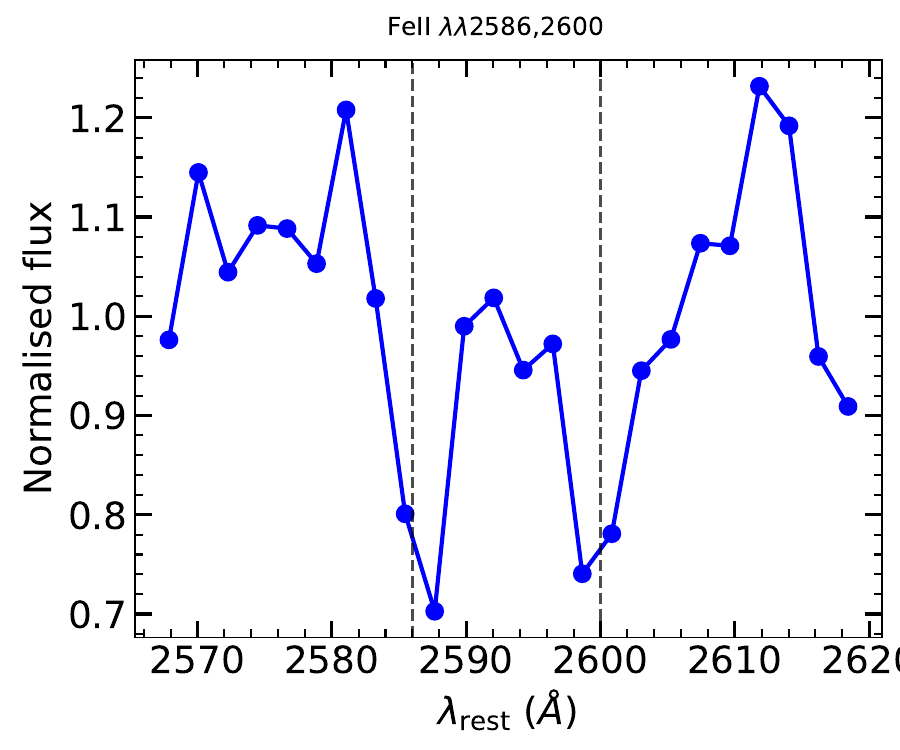}
    \includegraphics[width=0.20\textwidth]{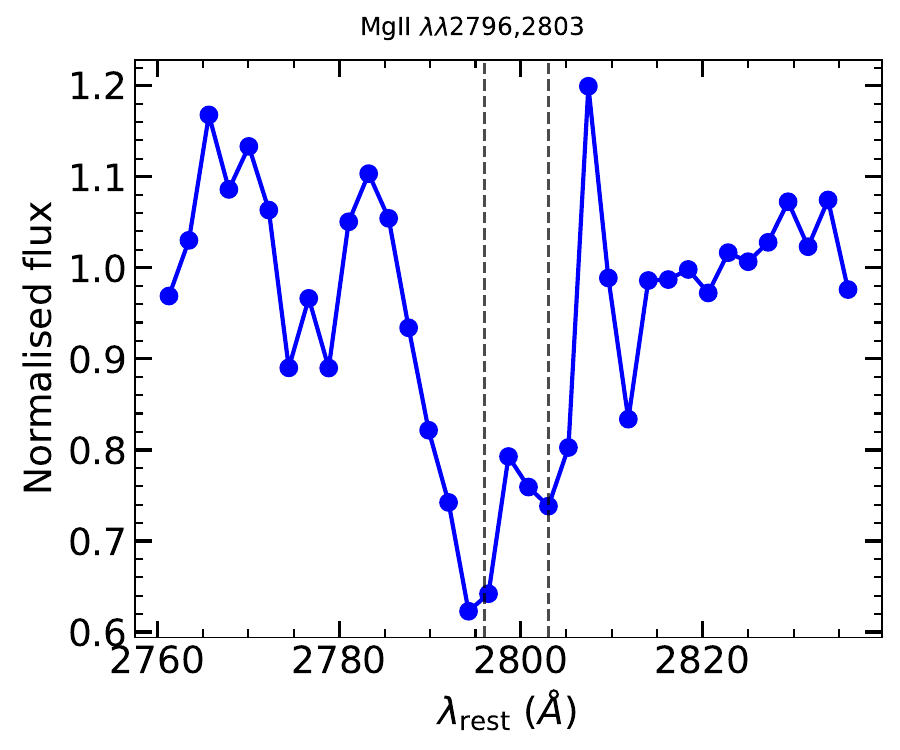}
  \end{center}
  \caption{Same as Fig. \ref{fig:cut_out_SED_all} but for object ID 120080.}
  \label{fig:cut_out_SED_all3}
\end{figure*}
\begin{figure*}[!ht]
  \begin{center}
    \includegraphics[width=0.40\textwidth]{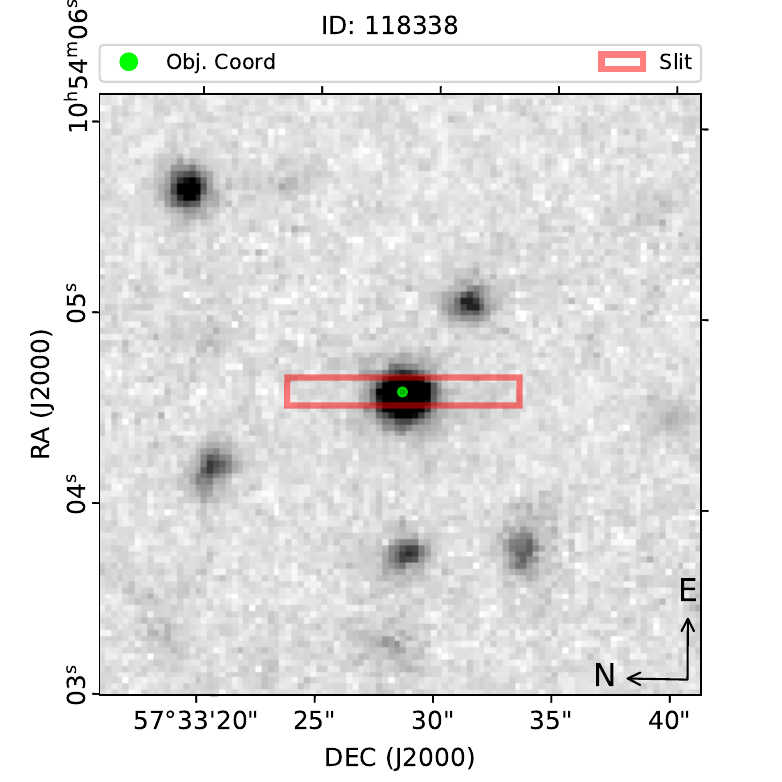}
    \includegraphics[width=0.53\textwidth]{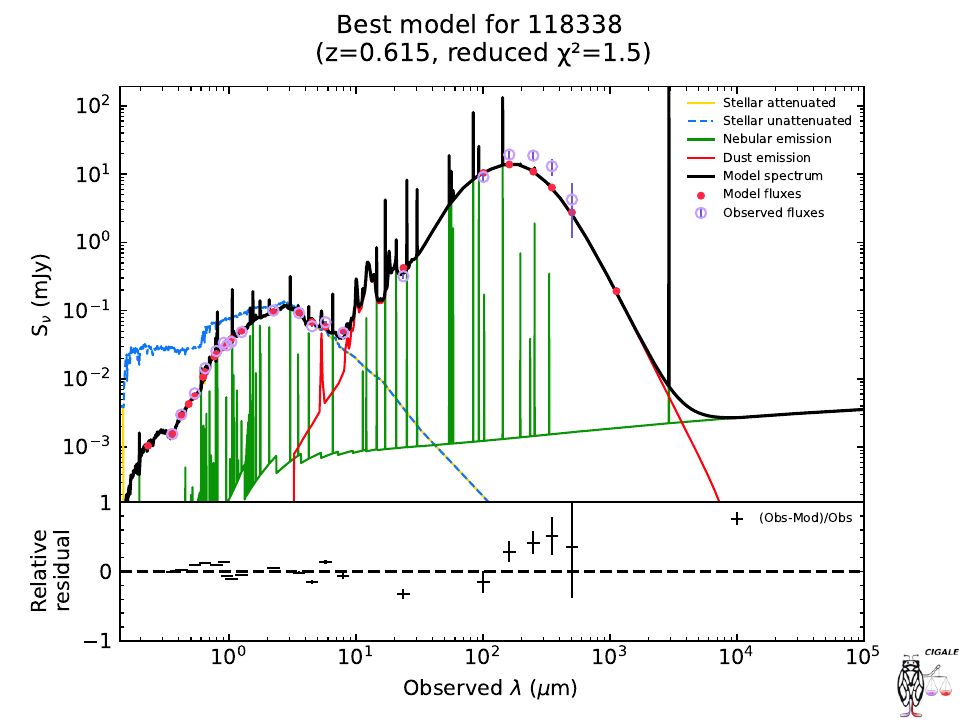}
    \includegraphics[width=0.20\textwidth]{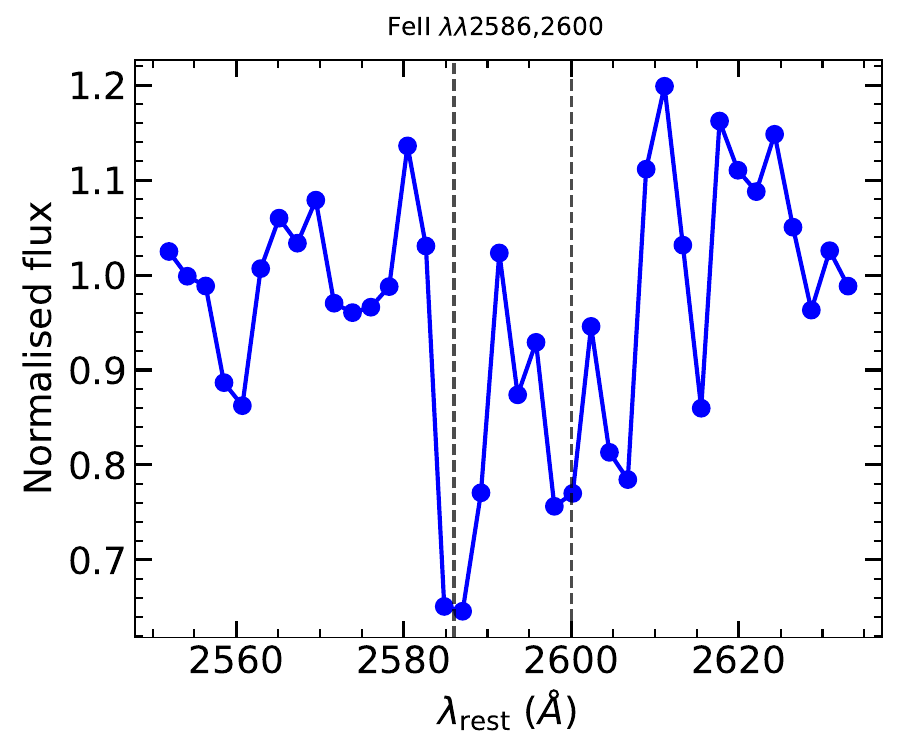}
    \includegraphics[width=0.20\textwidth]{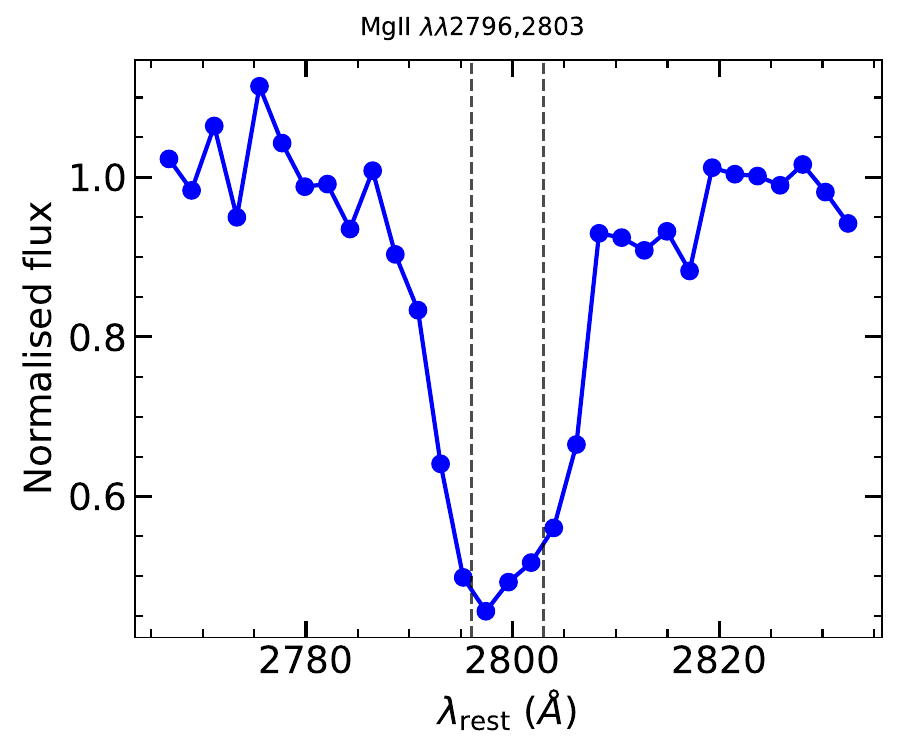}
  \end{center}
  \caption{Same as Fig. \ref{fig:cut_out_SED_all} but for object ID 118338.}
  \label{fig:cut_out_SED_all4}
\end{figure*}
\begin{figure*}[!ht]
  \begin{center}
    \includegraphics[width=0.40\textwidth]{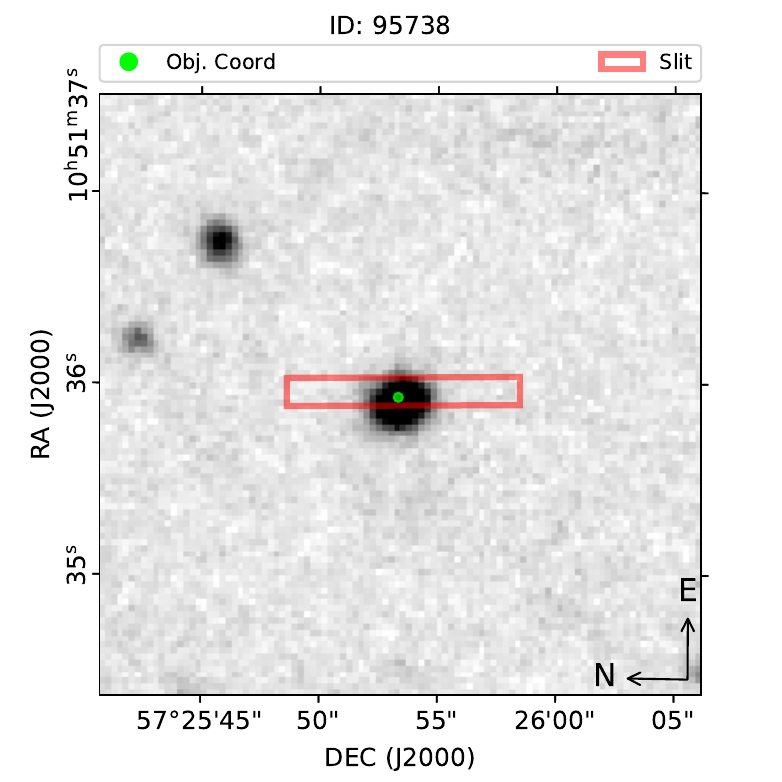}
    \includegraphics[width=0.53\textwidth]{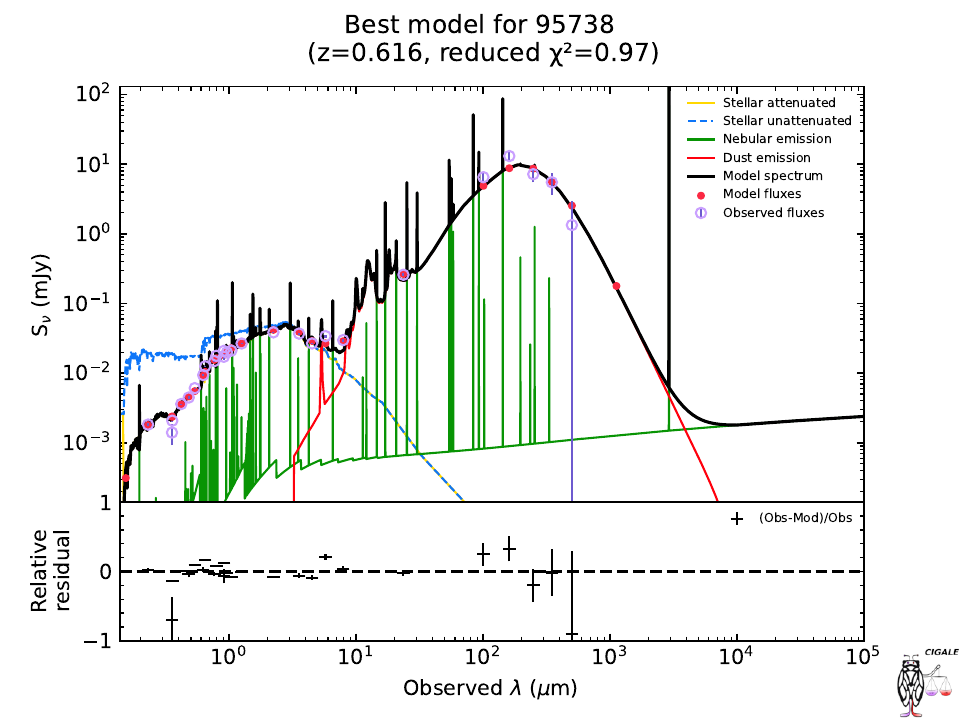}
    \includegraphics[width=0.20\textwidth]{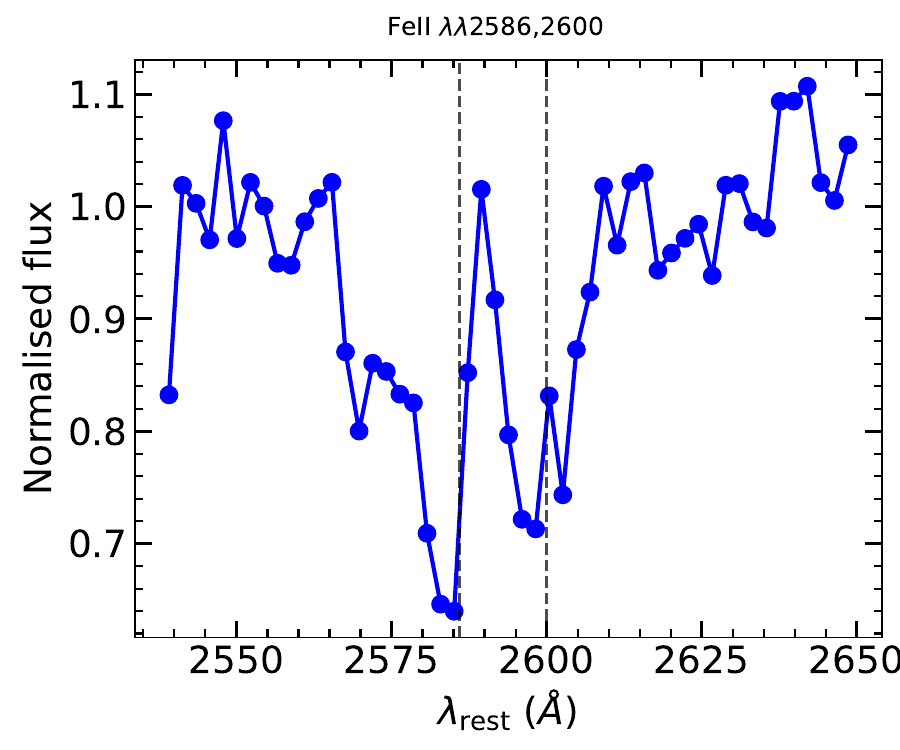}
    \includegraphics[width=0.20\textwidth]{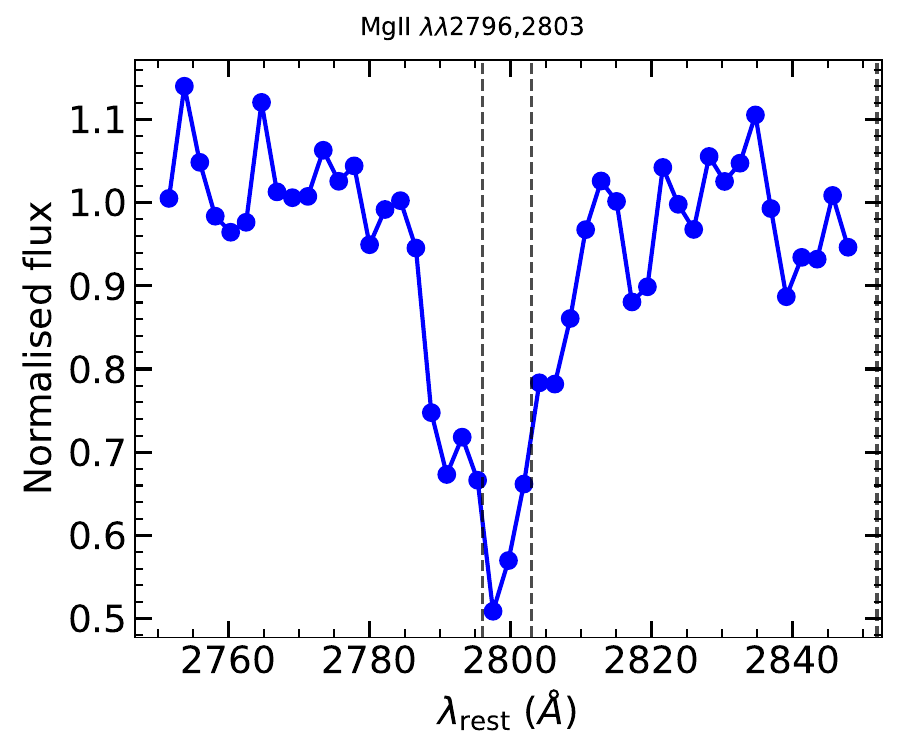}
  \end{center}
  \caption{Same as Fig. \ref{fig:cut_out_SED_all} but for object ID 95738.}
  \label{fig:cut_out_SED_all5}
\end{figure*}
\begin{figure*}[!ht]
  \begin{center}
    \includegraphics[width=0.40\textwidth]{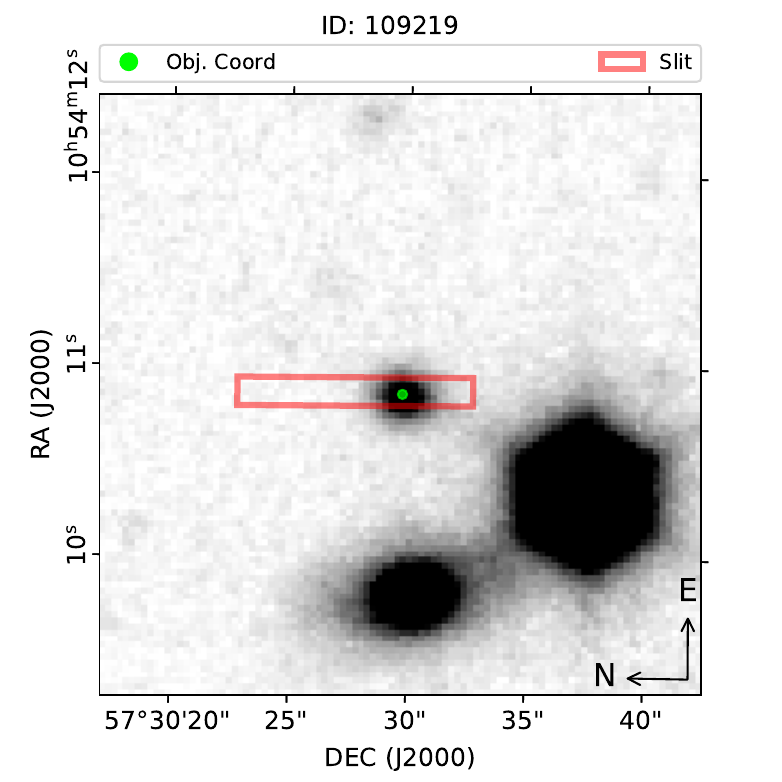}
    \includegraphics[width=0.53\textwidth]{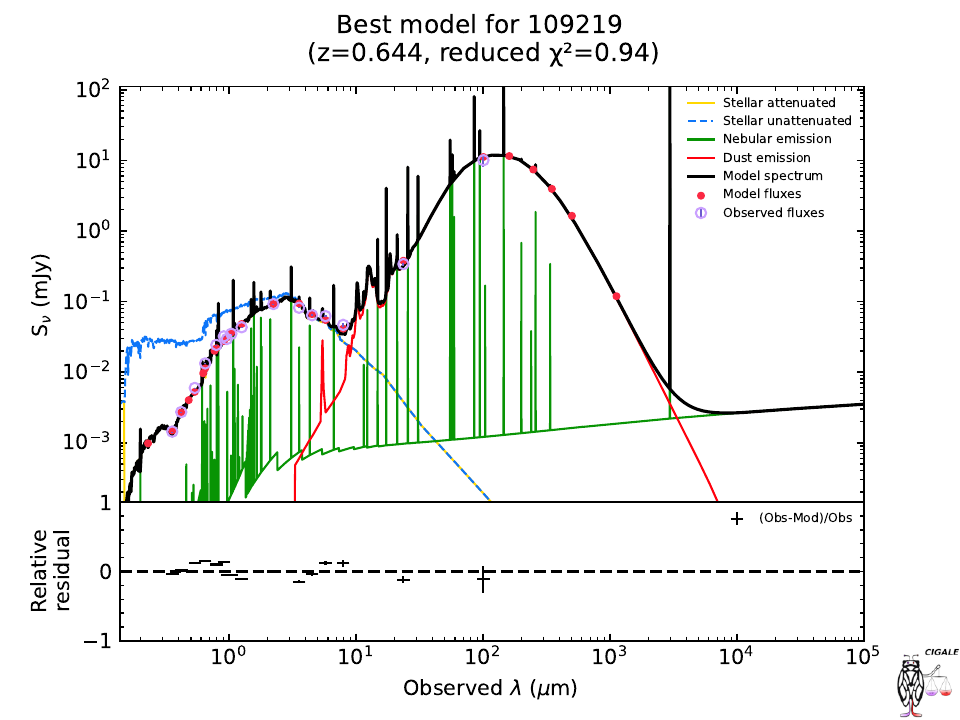}
    \includegraphics[width=0.20\textwidth]{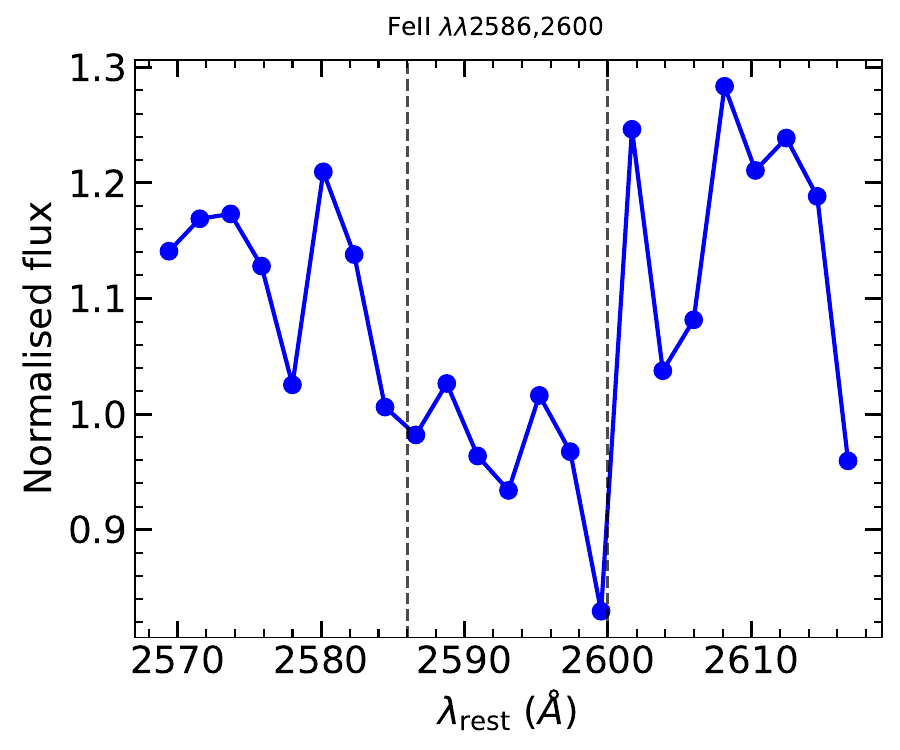}
    \includegraphics[width=0.20\textwidth]{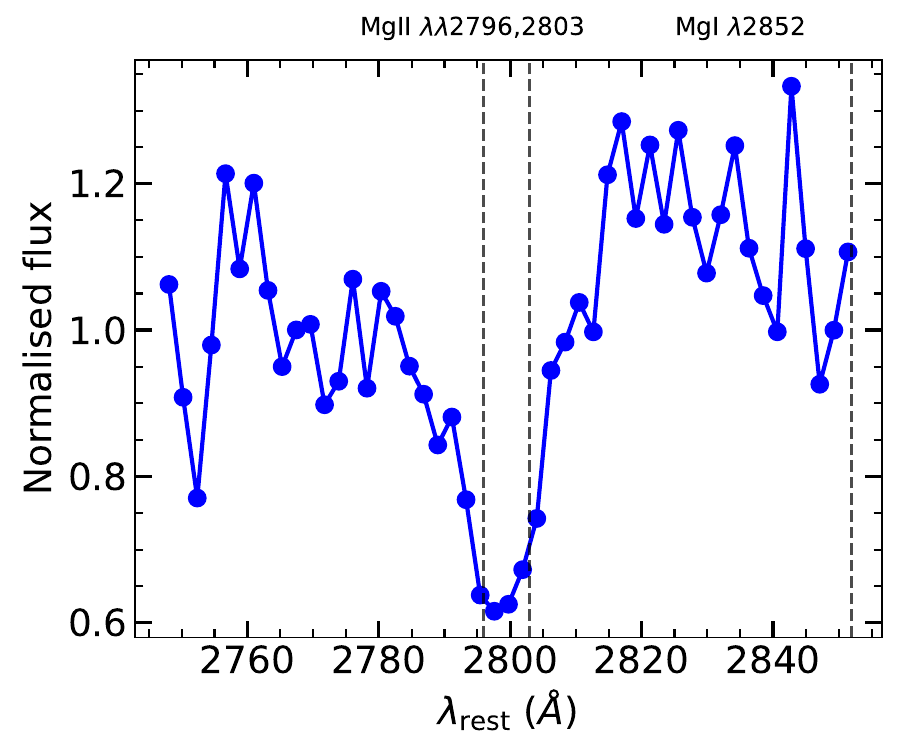}
  \end{center}
  \caption{Same as Fig. \ref{fig:cut_out_SED_all} but for object ID 109219.}
  \label{fig:cut_out_SED_all6}
\end{figure*}
\begin{figure*}[!ht]
  \begin{center}
    \includegraphics[width=0.40\textwidth]{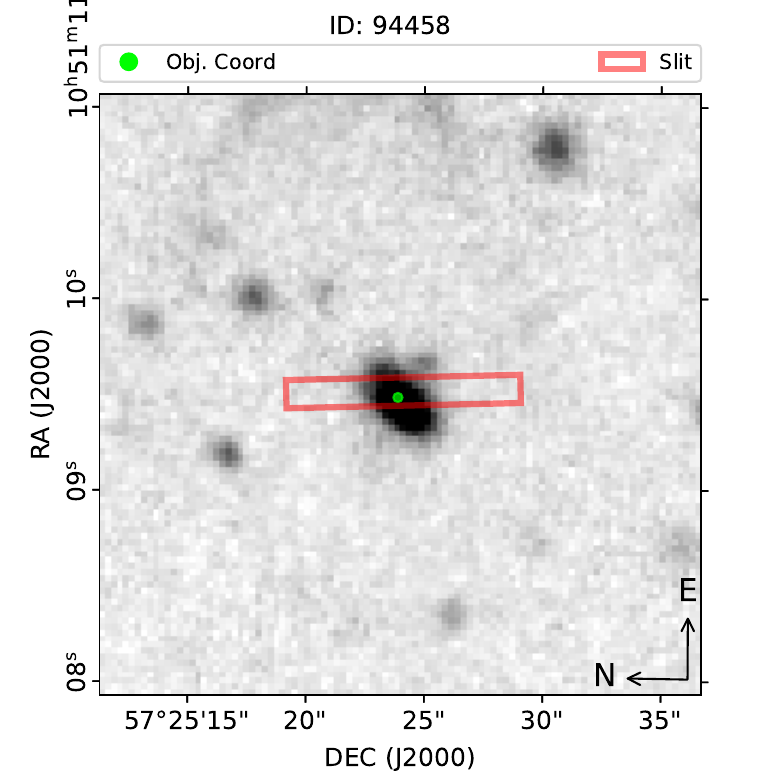}
    \includegraphics[width=0.53\textwidth]{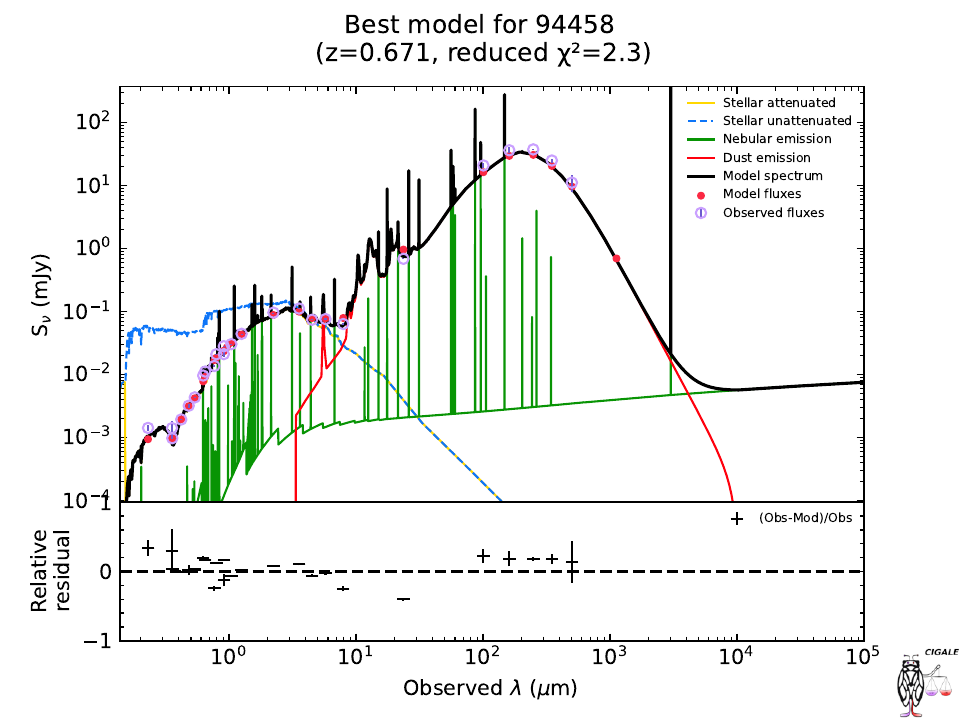}
    \includegraphics[width=0.20\textwidth]{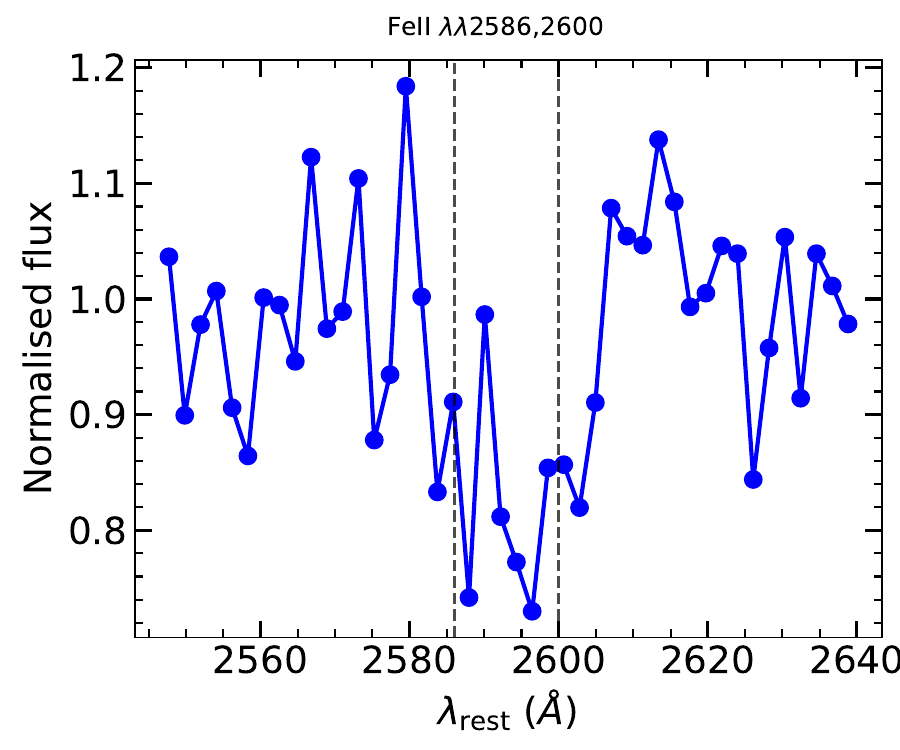}
    \includegraphics[width=0.20\textwidth]{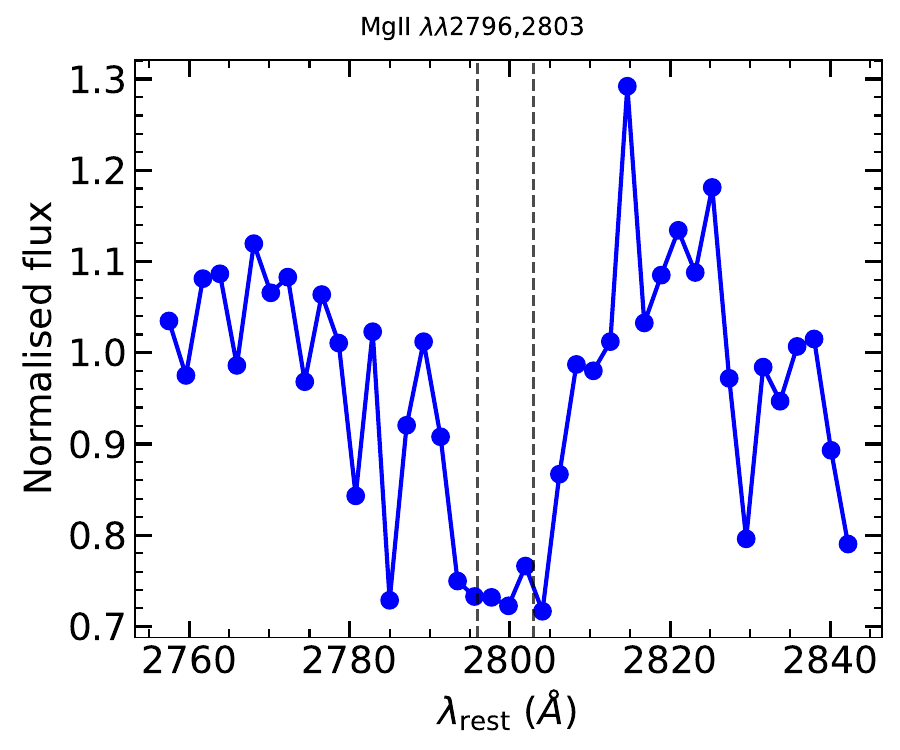}
  \end{center}
  \caption{Same as Fig. \ref{fig:cut_out_SED_all} but for object ID 94458.}
  \label{fig:cut_out_SED_all7}
\end{figure*}
\begin{figure*}[!ht]
  \begin{center}
    \includegraphics[width=0.40\textwidth]{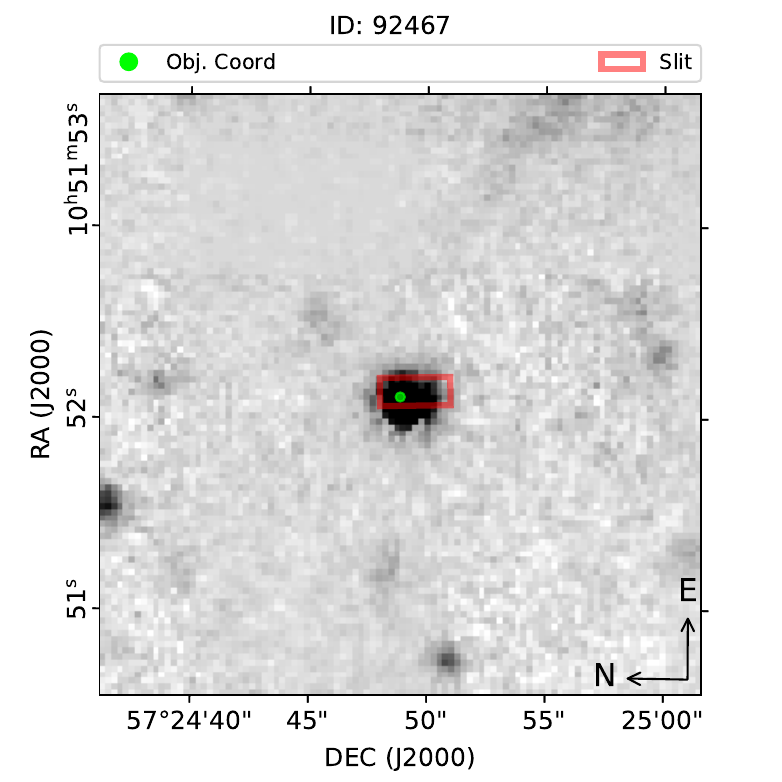}
    \includegraphics[width=0.53\textwidth]{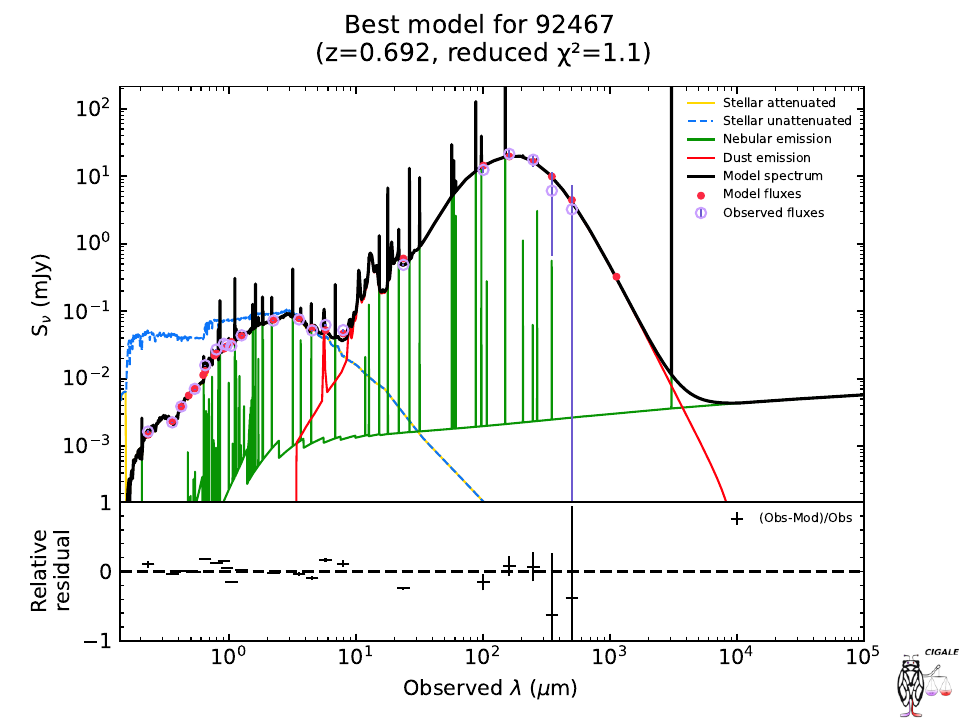}
    \includegraphics[width=0.20\textwidth]{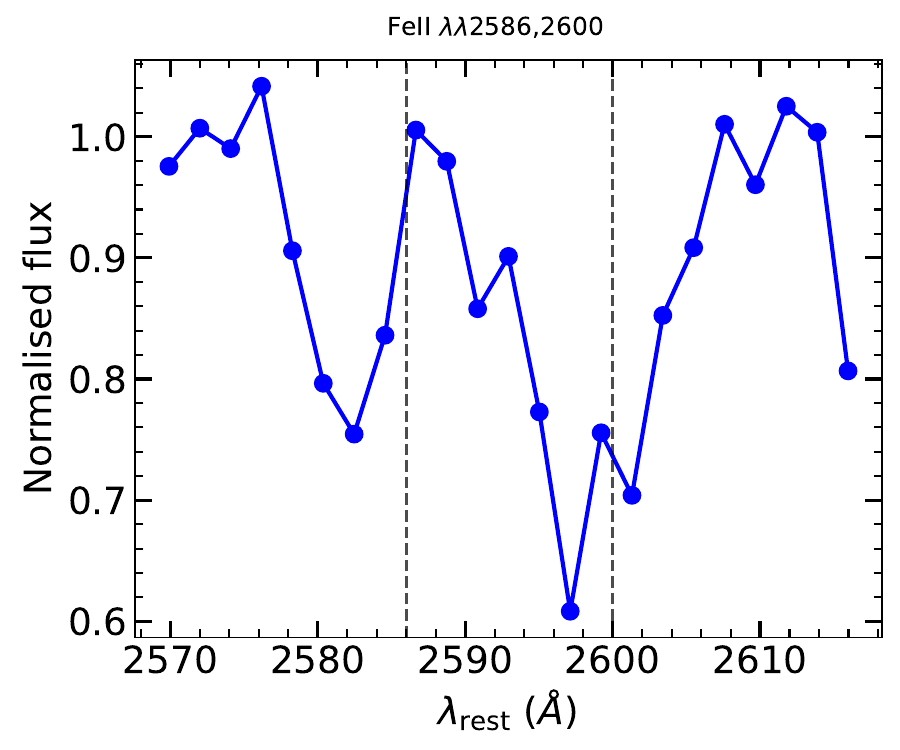}
    \includegraphics[width=0.20\textwidth]{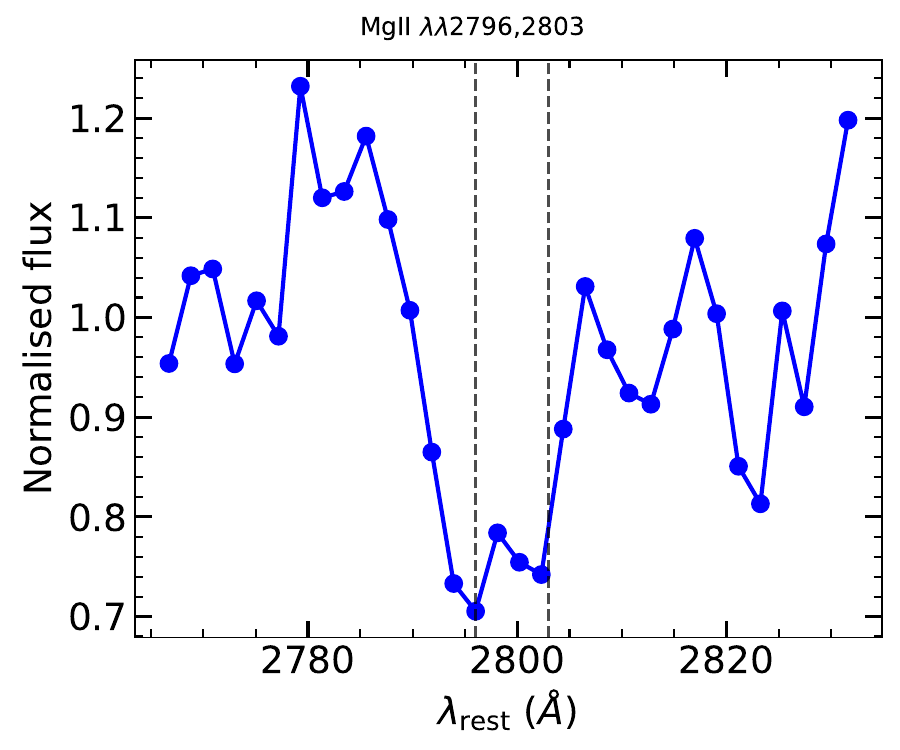}
  \end{center}
  \caption{Same as Fig. \ref{fig:cut_out_SED_all} but for object ID 92467.}
  \label{fig:cut_out_SED_all8}
\end{figure*}
\begin{figure*}[!ht]
  \begin{center}    
    \includegraphics[width=0.40\textwidth]{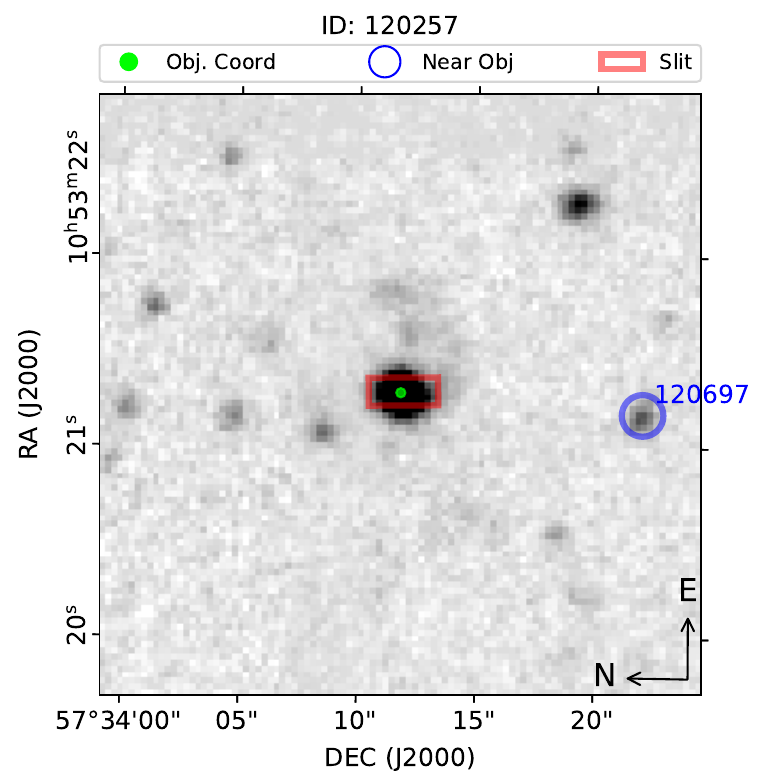}
    \includegraphics[width=0.53\textwidth]{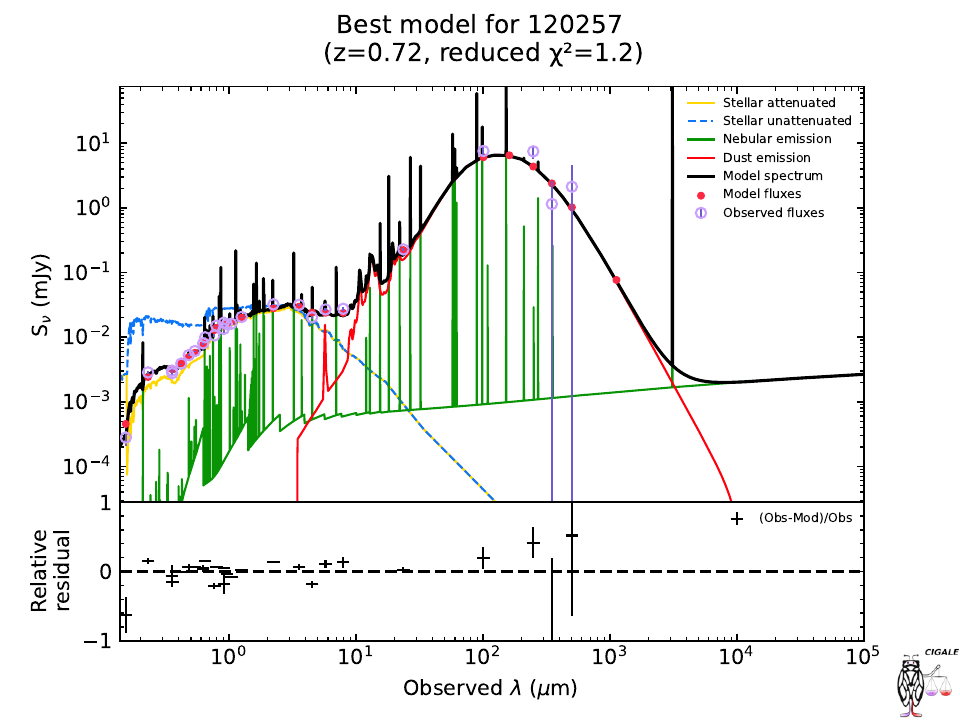}
    \includegraphics[width=0.20\textwidth]{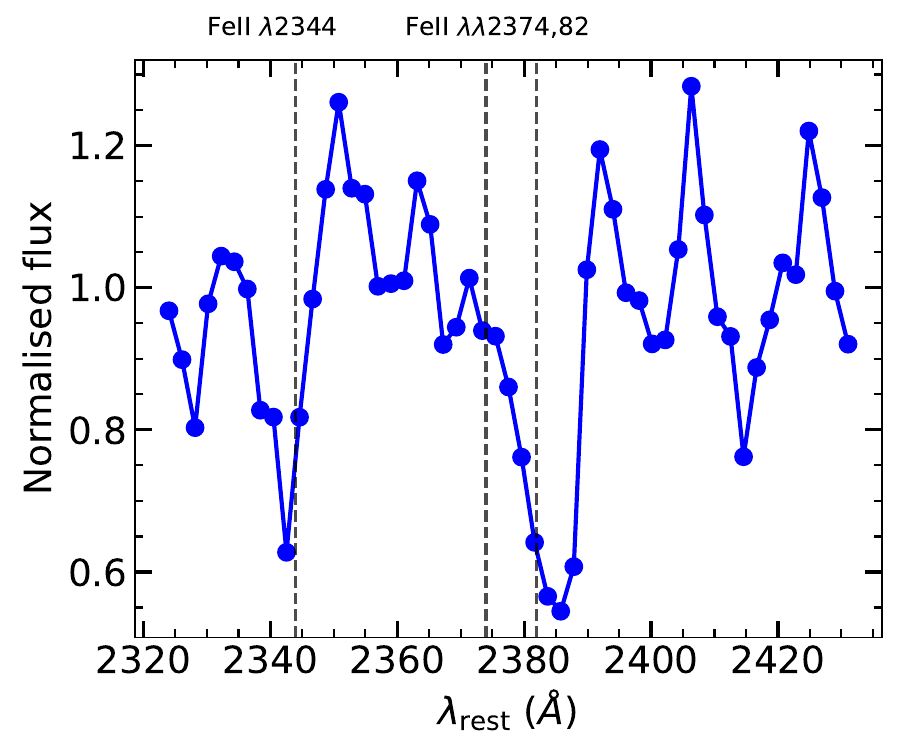}
    \includegraphics[width=0.20\textwidth]{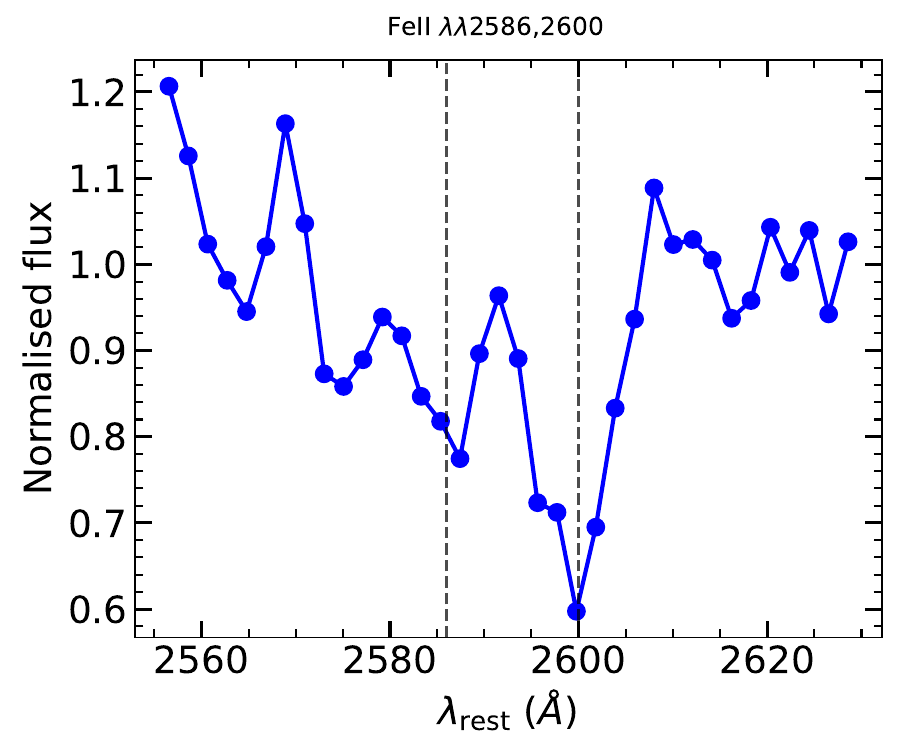}
    \includegraphics[width=0.20\textwidth]{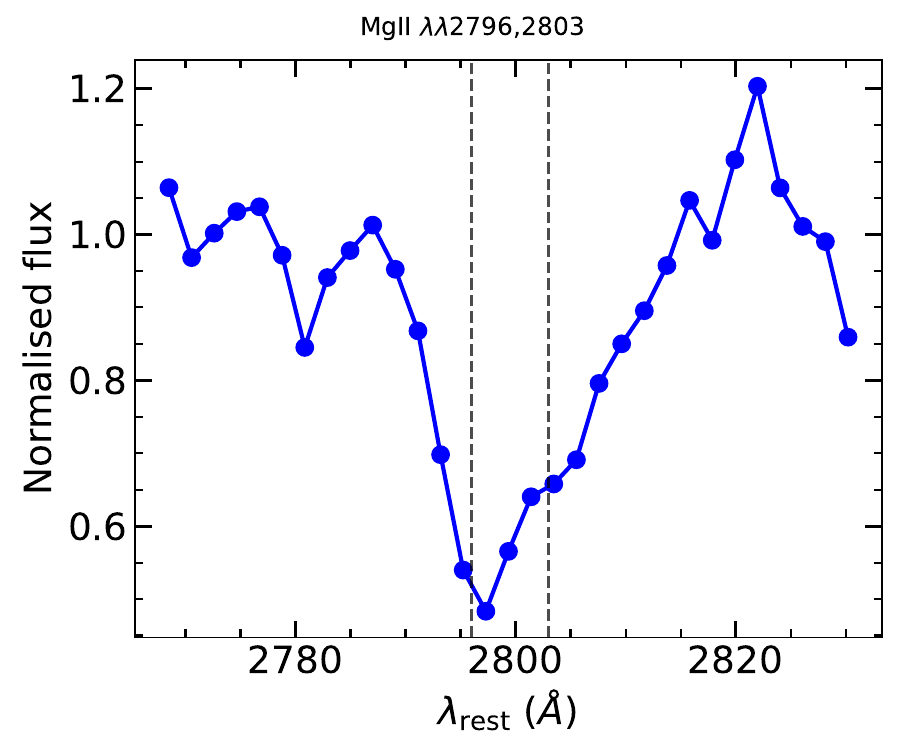}
  \end{center}
  \caption{Same as Fig. \ref{fig:cut_out_SED_all} but for object ID 120257.}
  \label{fig:cut_out_SED_all9}
\end{figure*}
\begin{figure*}[!ht]
  \begin{center}
    \includegraphics[width=0.40\textwidth]{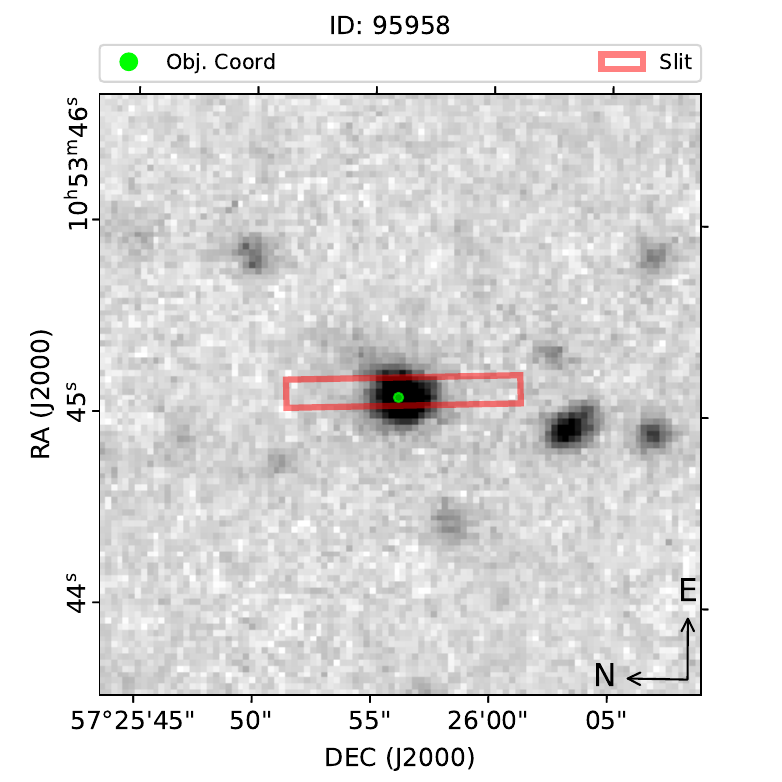}
    \includegraphics[width=0.53\textwidth]{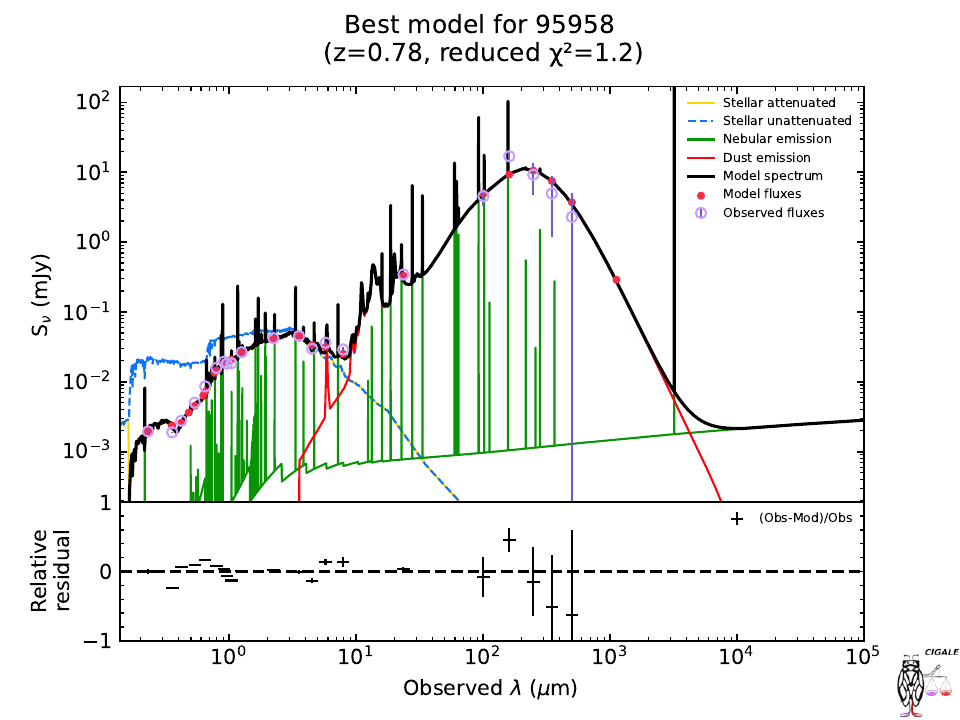}
    \includegraphics[width=0.20\textwidth]{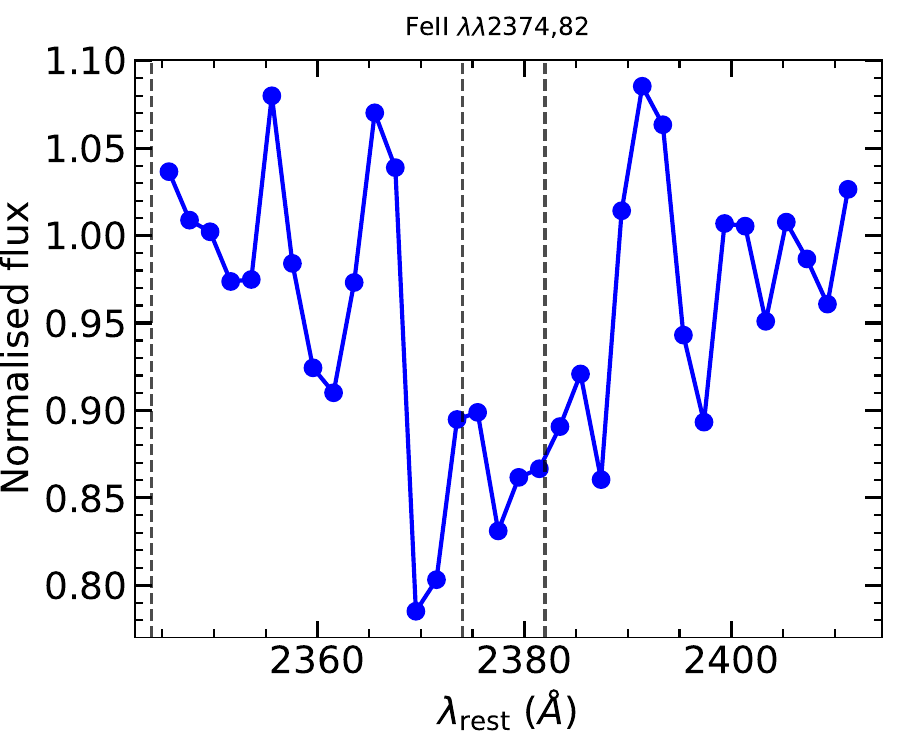}
    \includegraphics[width=0.20\textwidth]{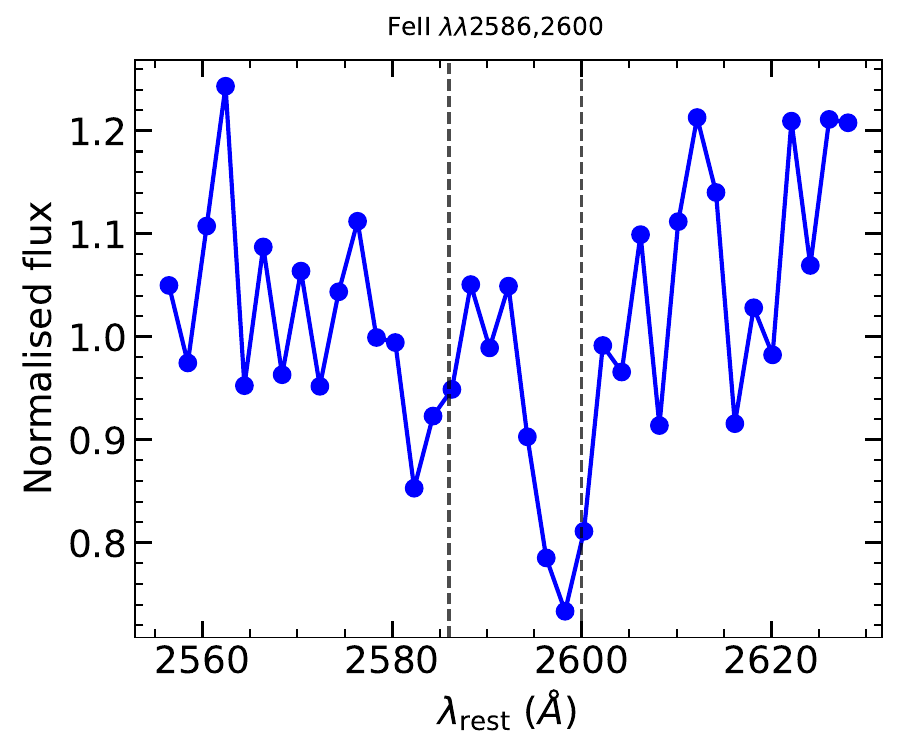}
    \includegraphics[width=0.20\textwidth]{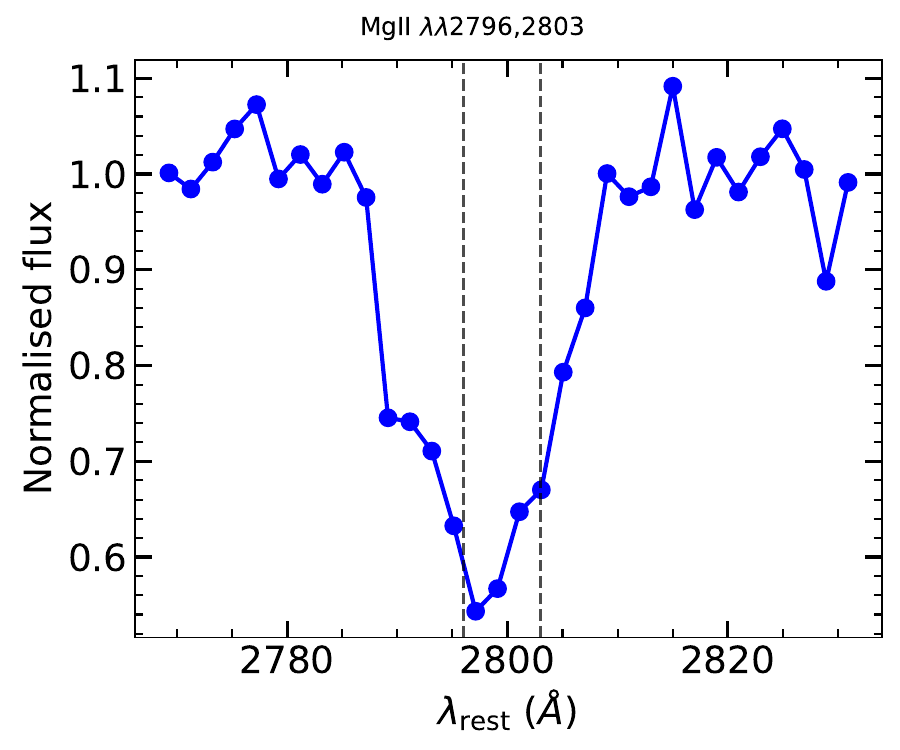}
  \end{center}
  \caption{Same as Fig. \ref{fig:cut_out_SED_all} but for object ID 95958.}
  \label{fig:cut_out_SED_all10}
\end{figure*}
\begin{figure*}[!ht]
  \begin{center}
    \includegraphics[width=0.40\textwidth]{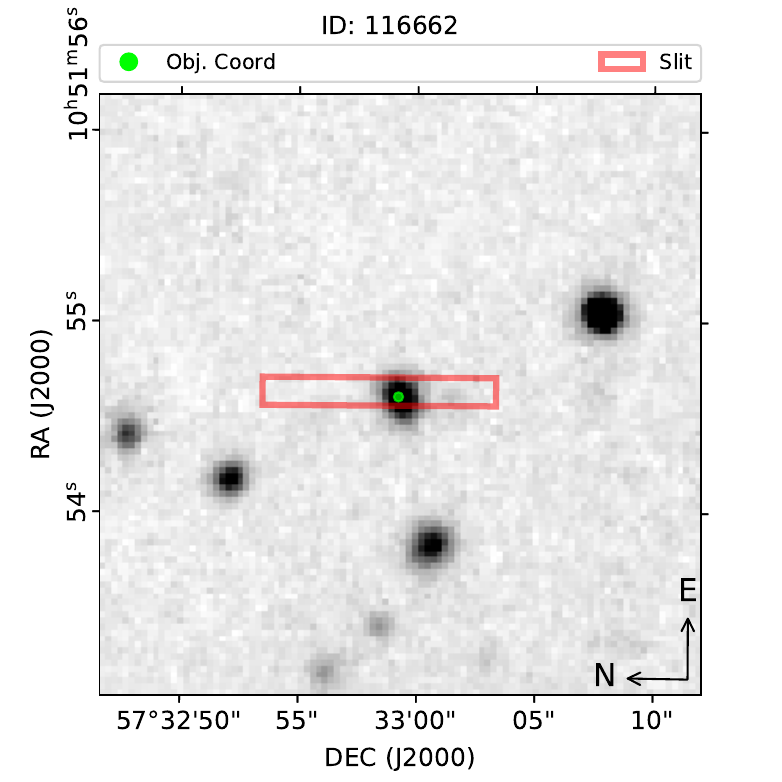}
    \includegraphics[width=0.53\textwidth]{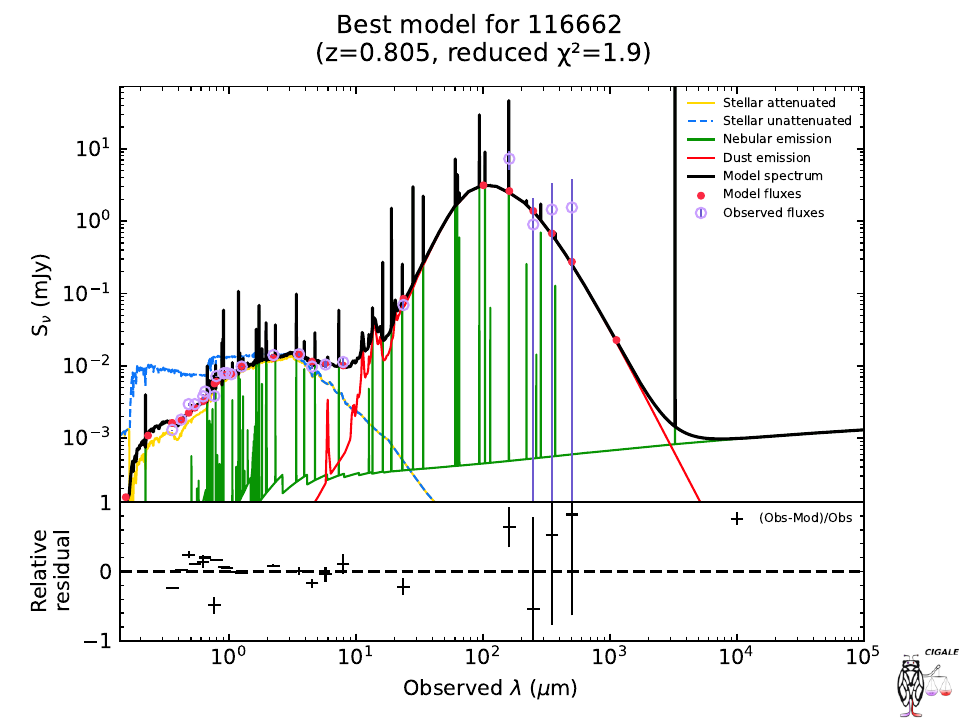}
    \includegraphics[width=0.20\textwidth]{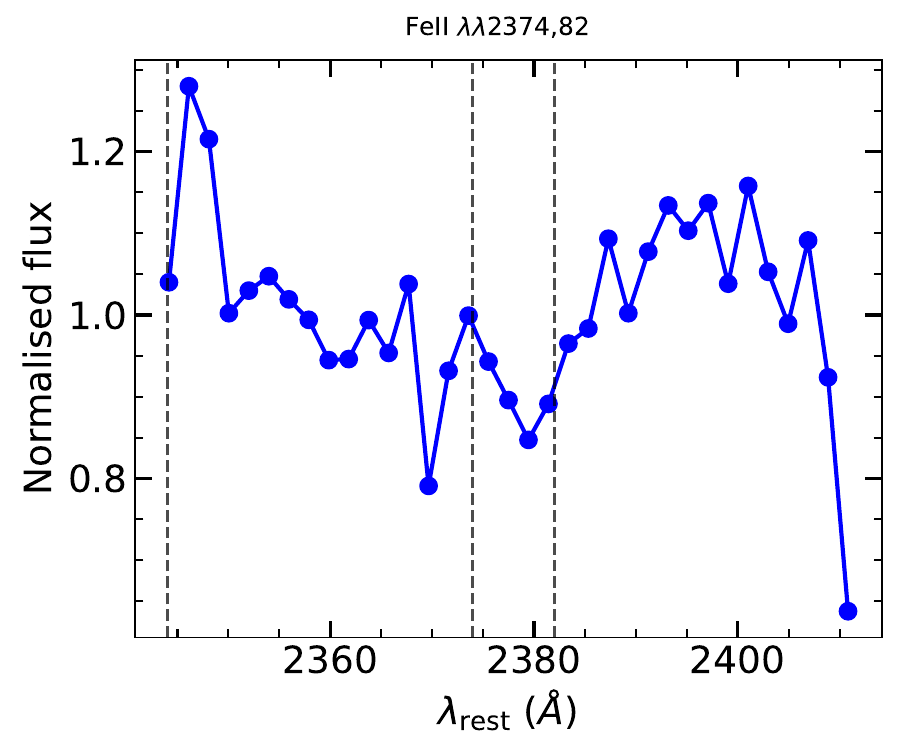}
    \includegraphics[width=0.20\textwidth]{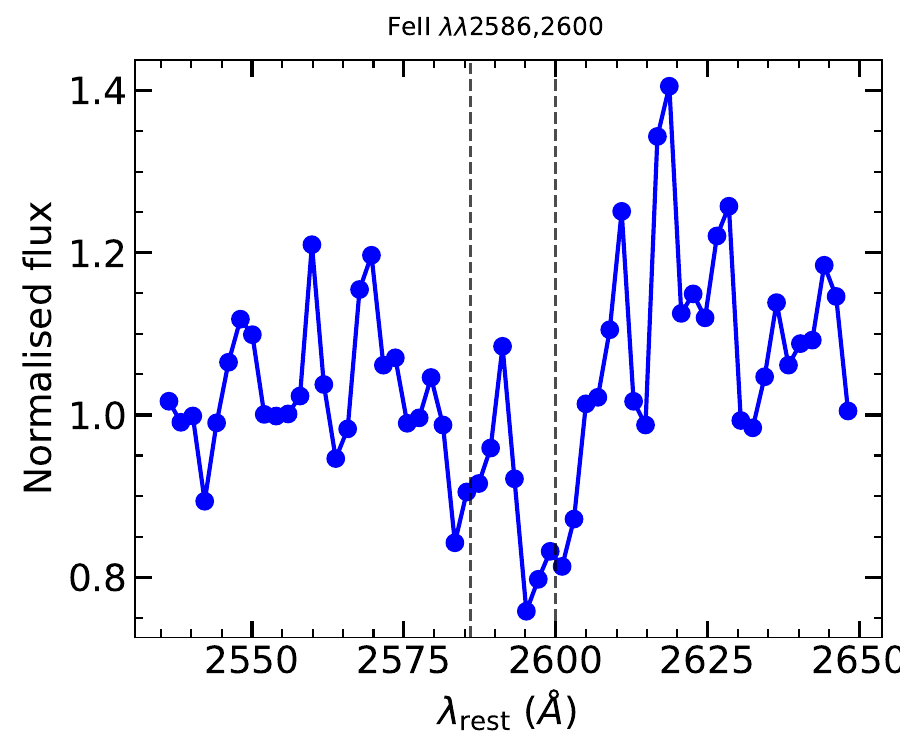}
    \includegraphics[width=0.20\textwidth]{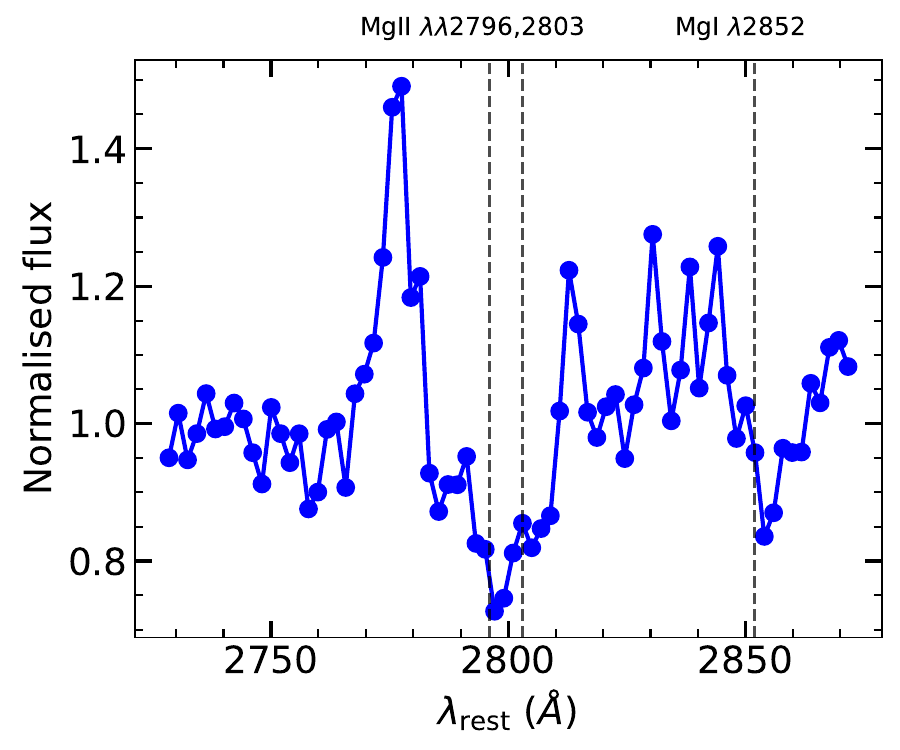}
  \end{center}
  \caption{Same as Fig. \ref{fig:cut_out_SED_all} but for object ID 116662.}
  \label{fig:cut_out_SED_all11}
\end{figure*}
\begin{figure*}[!ht]
  \begin{center}
    \includegraphics[width=0.40\textwidth]{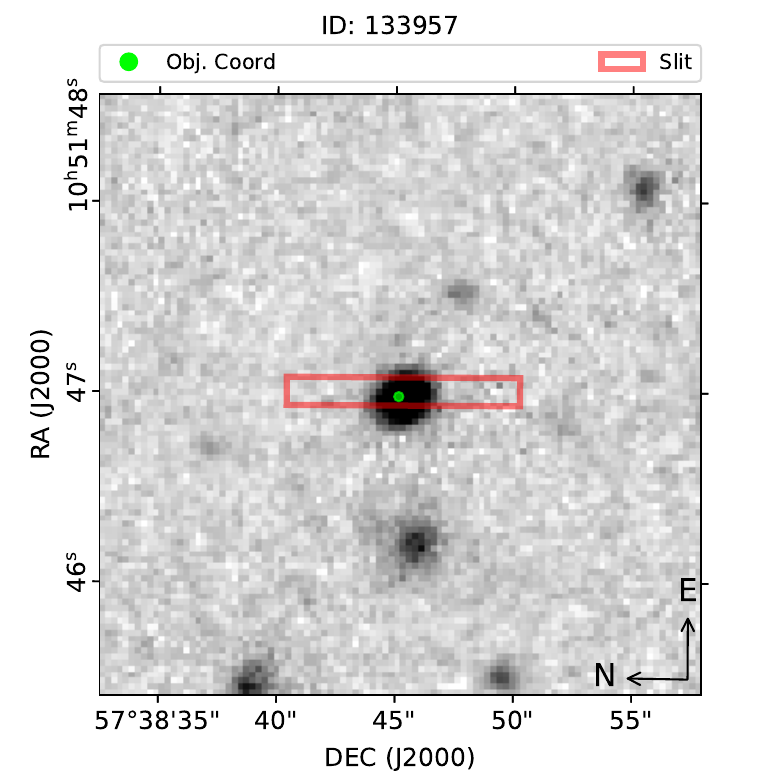}
    \includegraphics[width=0.53\textwidth]{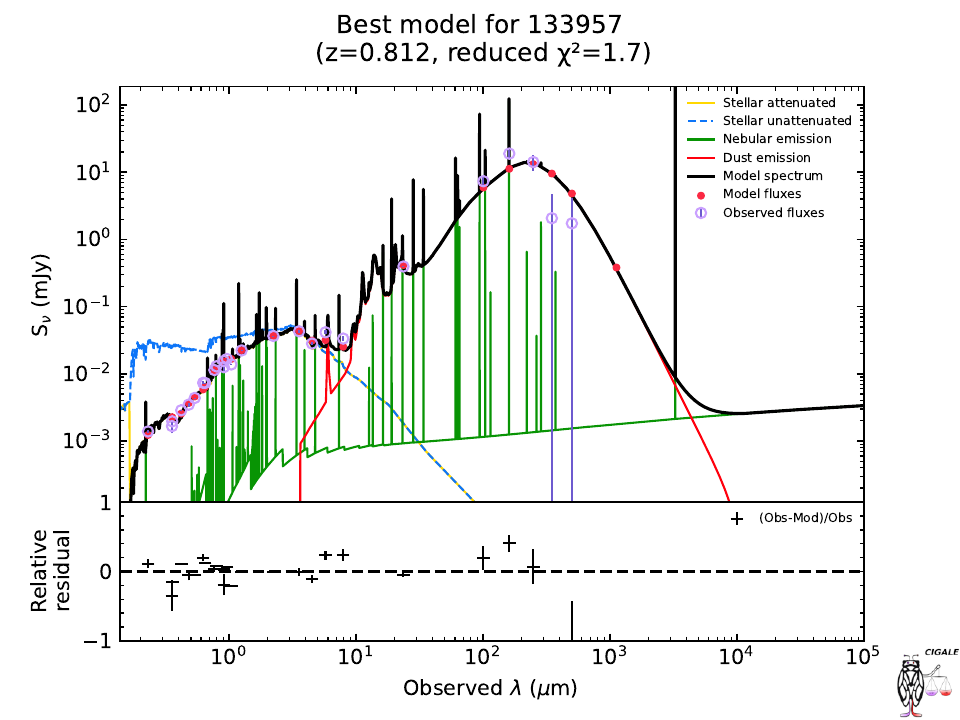}
    \includegraphics[width=0.20\textwidth]{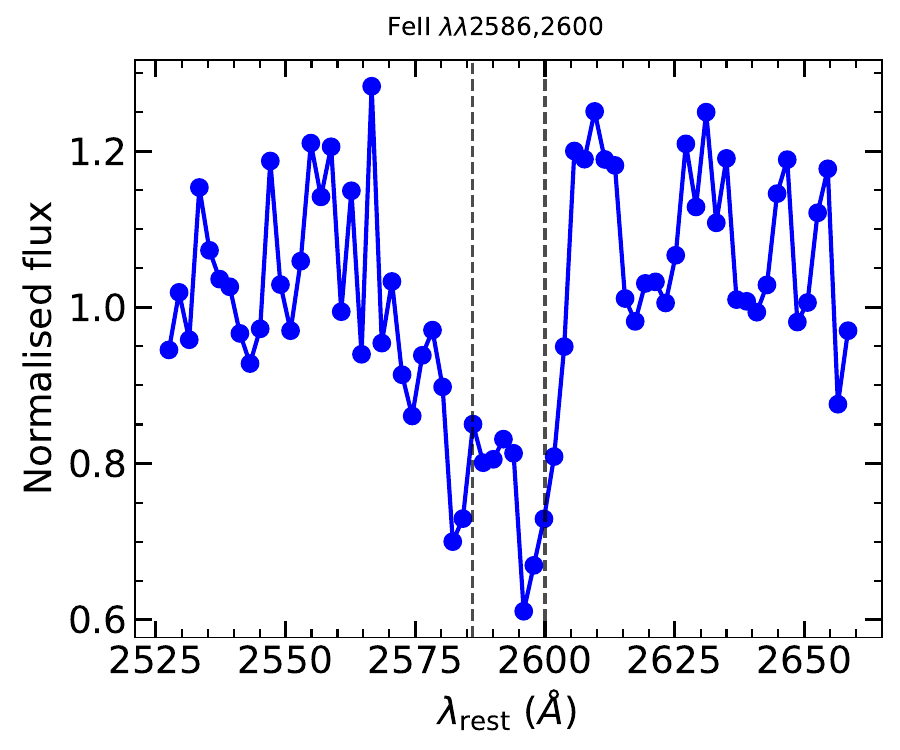}
    \includegraphics[width=0.20\textwidth]{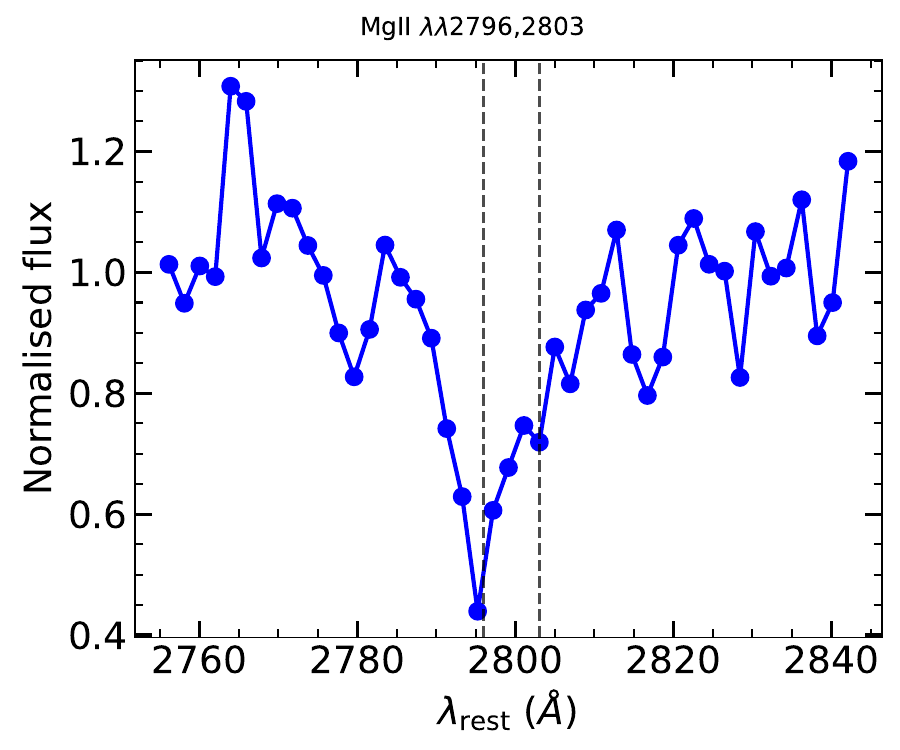}
  \end{center}
  \caption{Same as Fig. \ref{fig:cut_out_SED_all} but for object ID 133957.}
  \label{fig:cut_out_SED_all12}
\end{figure*}
\begin{figure*}[!ht]
  \begin{center}
    \includegraphics[width=0.40\textwidth]{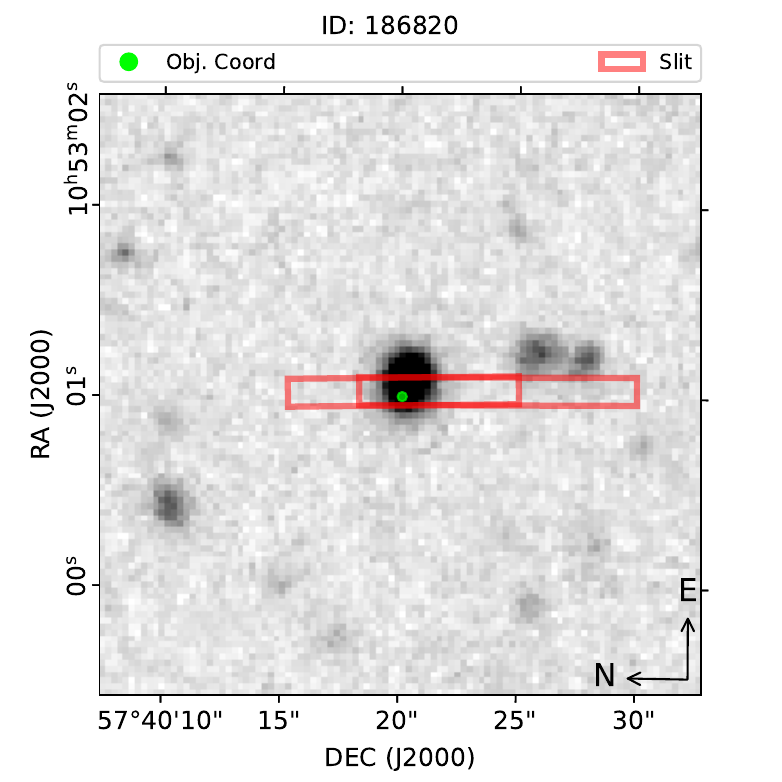}
    \includegraphics[width=0.53\textwidth]{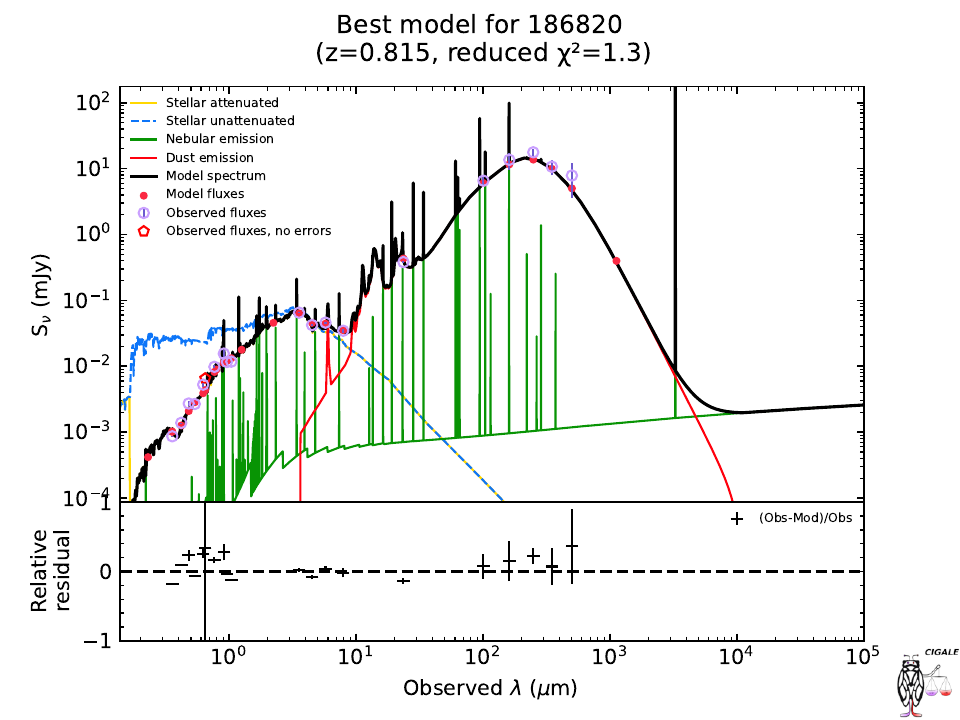}
    \includegraphics[width=0.20\textwidth]{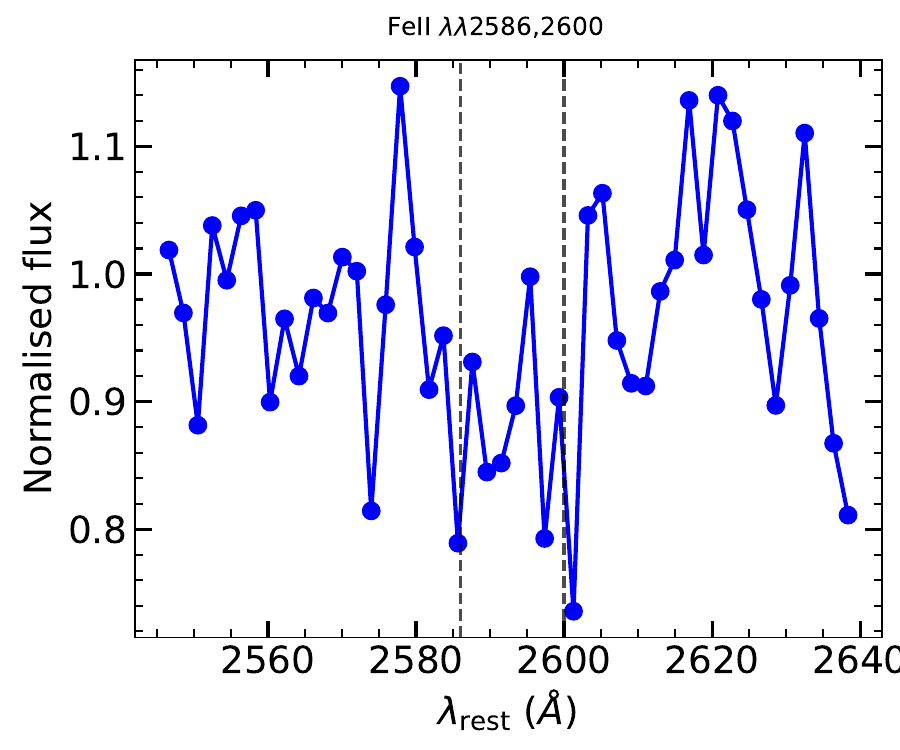}
    \includegraphics[width=0.20\textwidth]{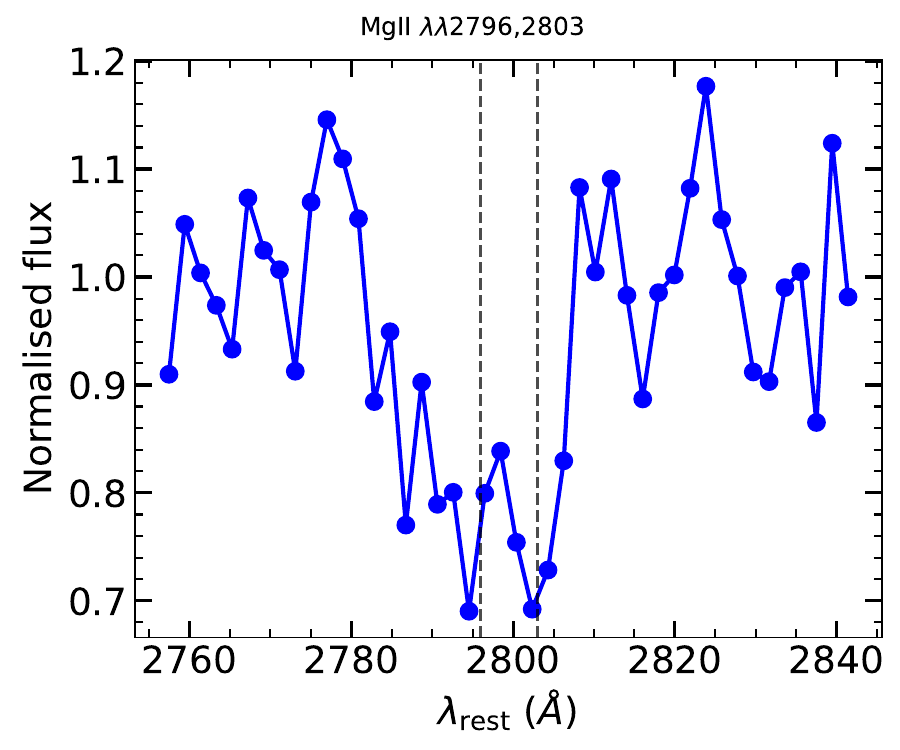}
  \end{center}
  \caption{Same as Fig. \ref{fig:cut_out_SED_all} but for object ID 186820.}
  \label{fig:cut_out_SED_all13}
\end{figure*}
\begin{figure*}[!ht]
  \begin{center}
    \includegraphics[width=0.40\textwidth]{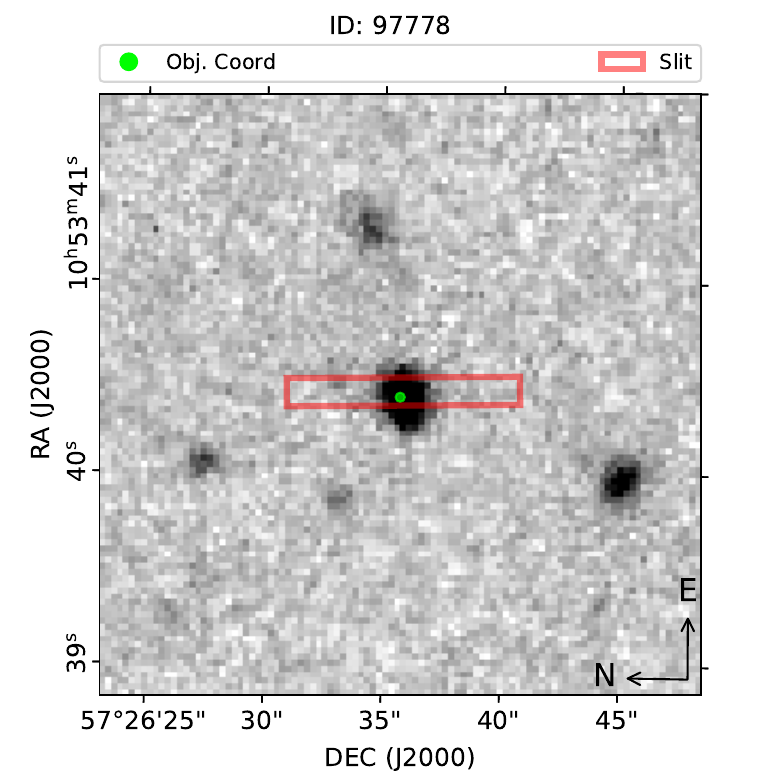}
    \includegraphics[width=0.53\textwidth]{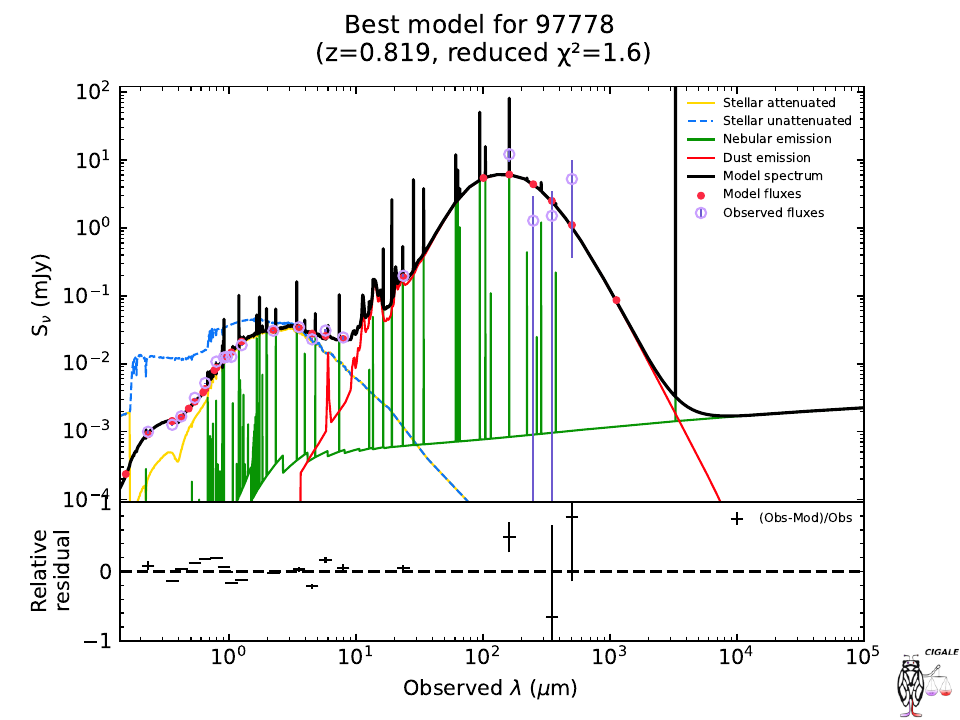}
    \includegraphics[width=0.20\textwidth]{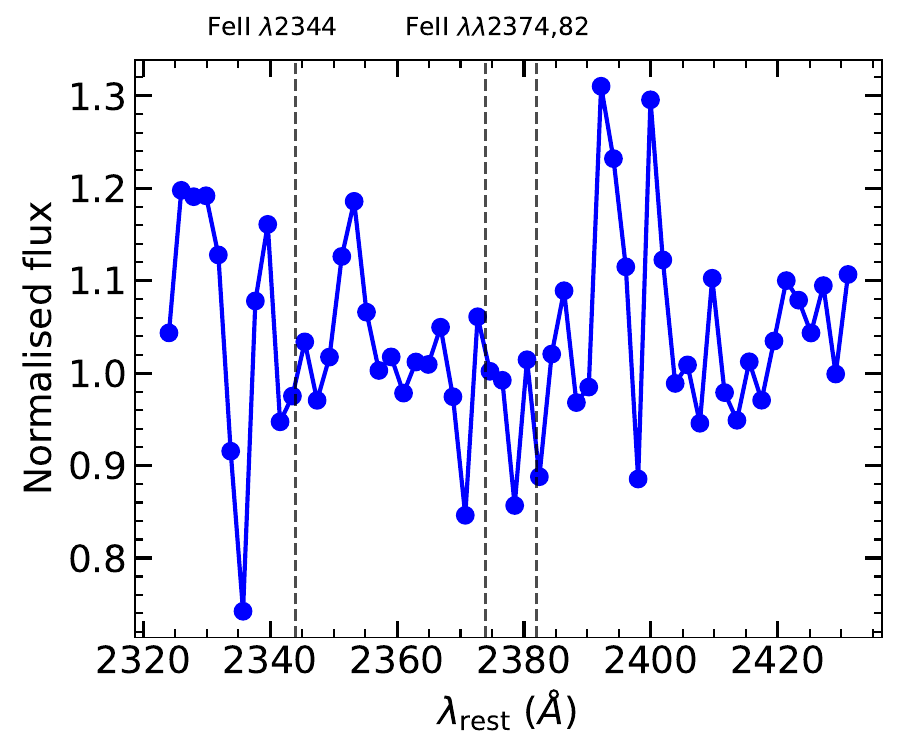}
    \includegraphics[width=0.20\textwidth]{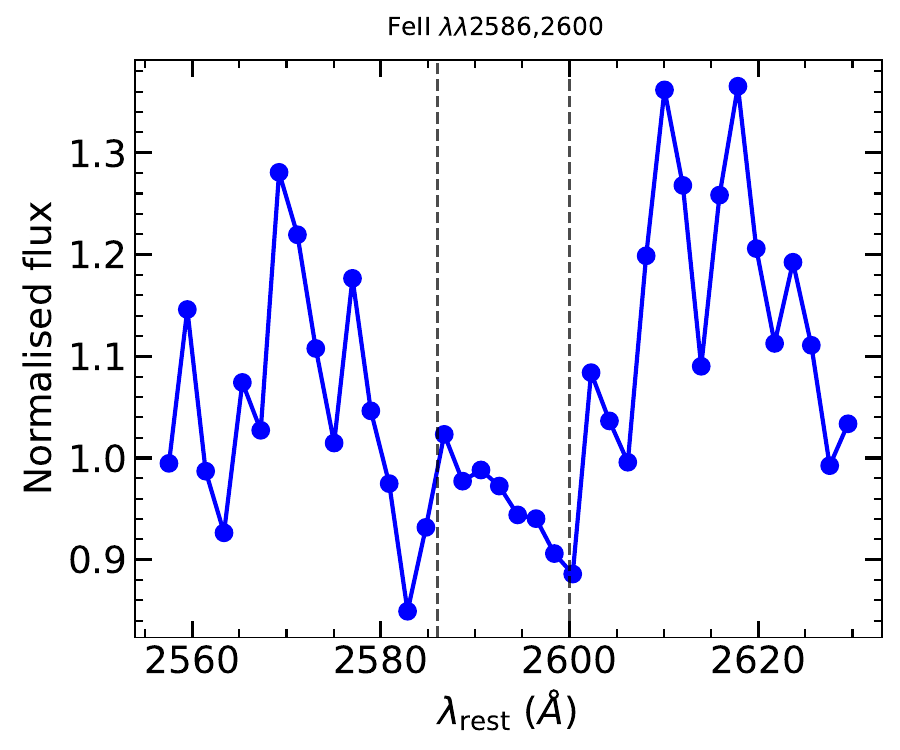}
    \includegraphics[width=0.20\textwidth]{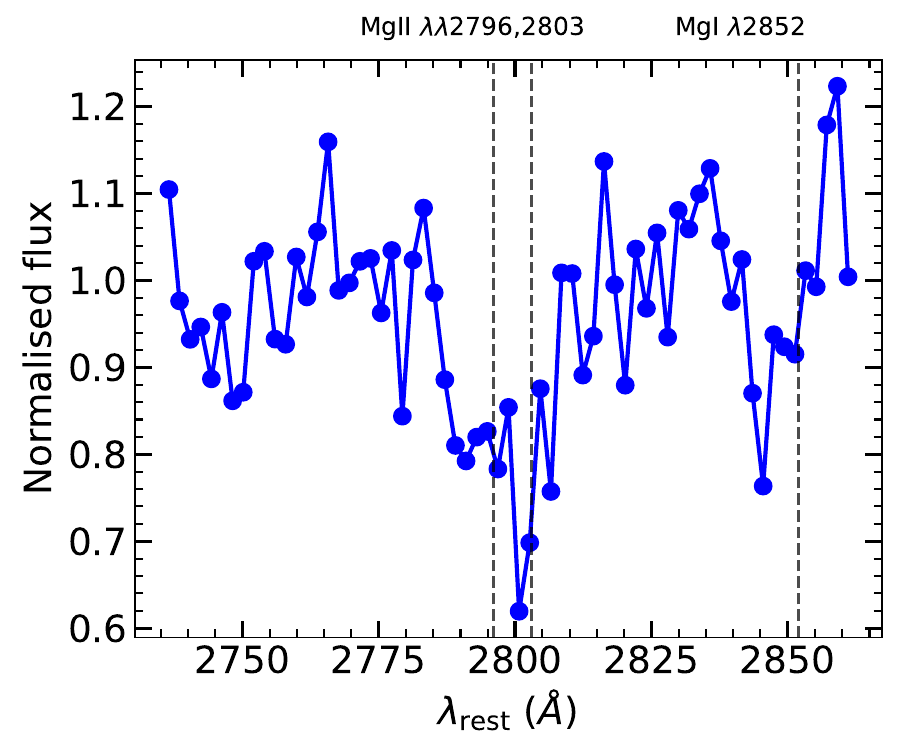}
  \end{center}
  \caption{Same as Fig. \ref{fig:cut_out_SED_all} but for object ID 97778.}
  \label{fig:cut_out_SED_all14}
\end{figure*}
\begin{figure*}[!ht]
  \begin{center}    
    \includegraphics[width=0.40\textwidth]{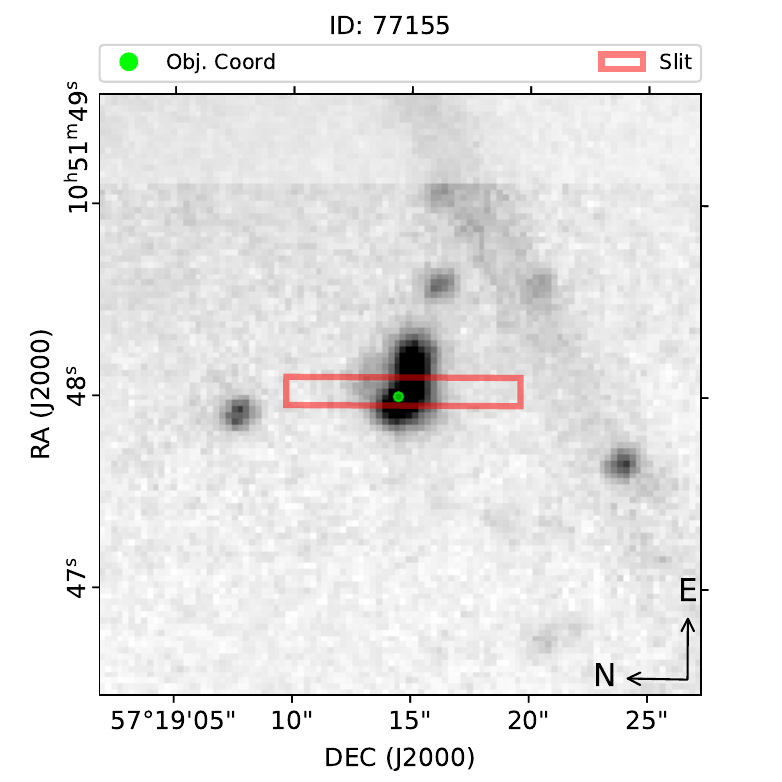}
    \includegraphics[width=0.53\textwidth]{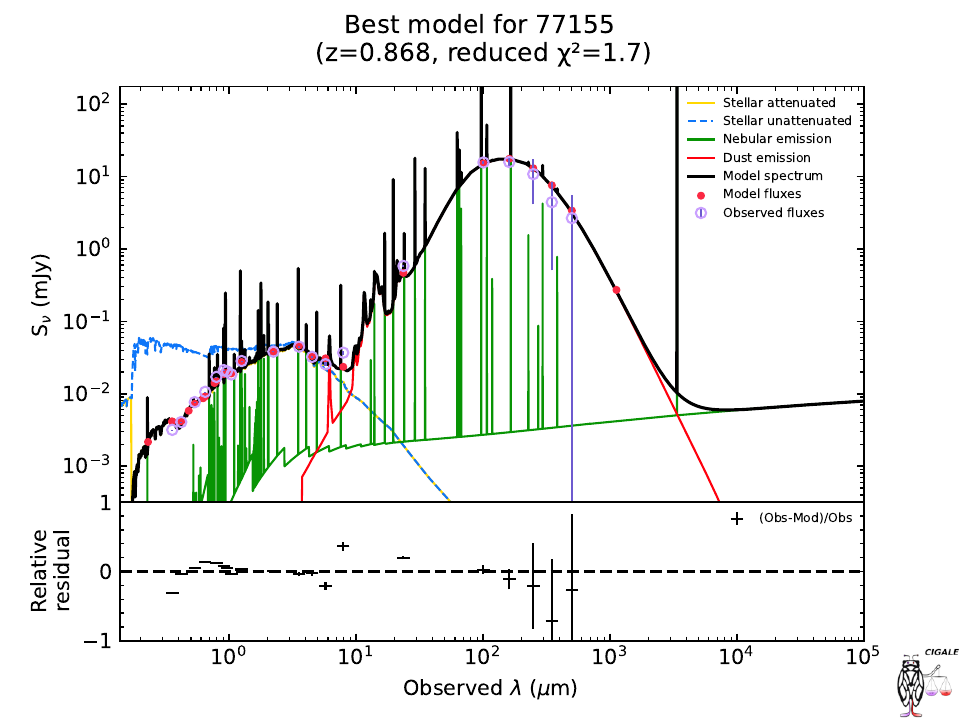}
    \includegraphics[width=0.20\textwidth]{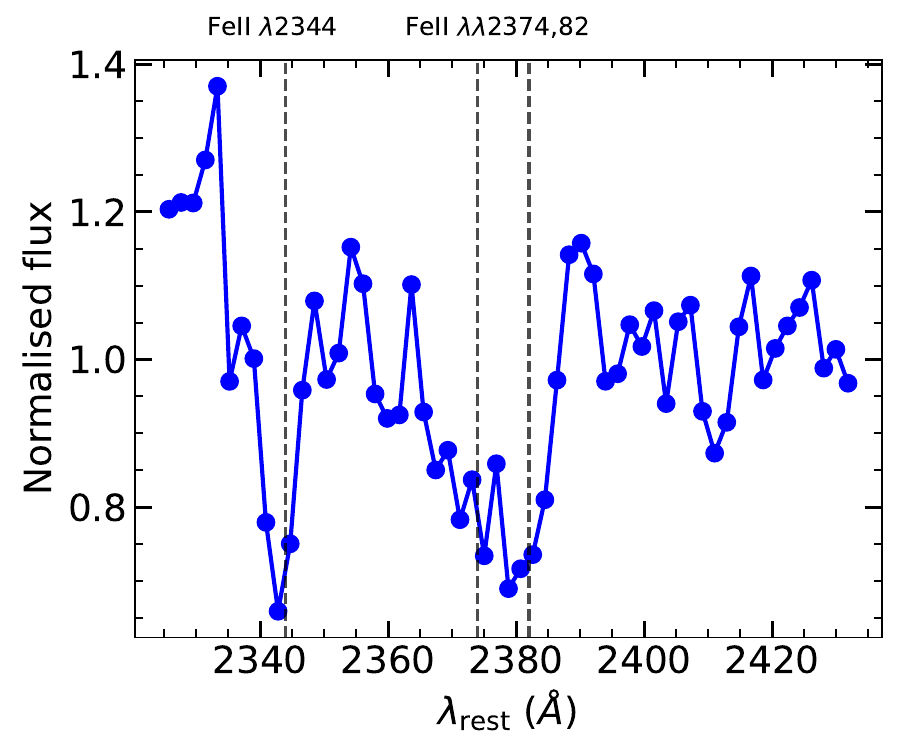}
    \includegraphics[width=0.20\textwidth]{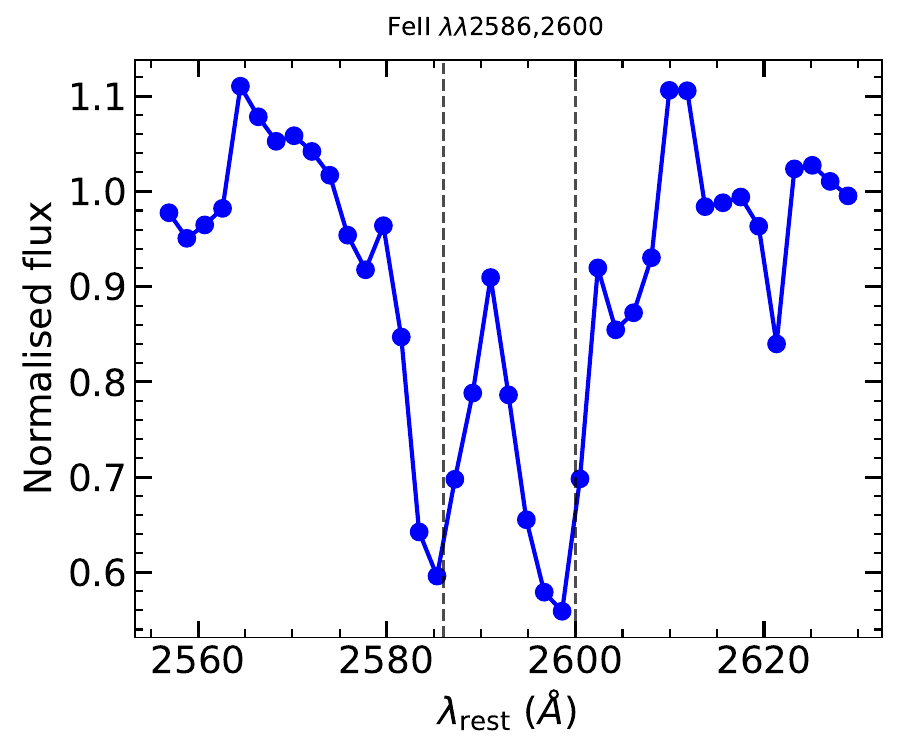}
    \includegraphics[width=0.20\textwidth]{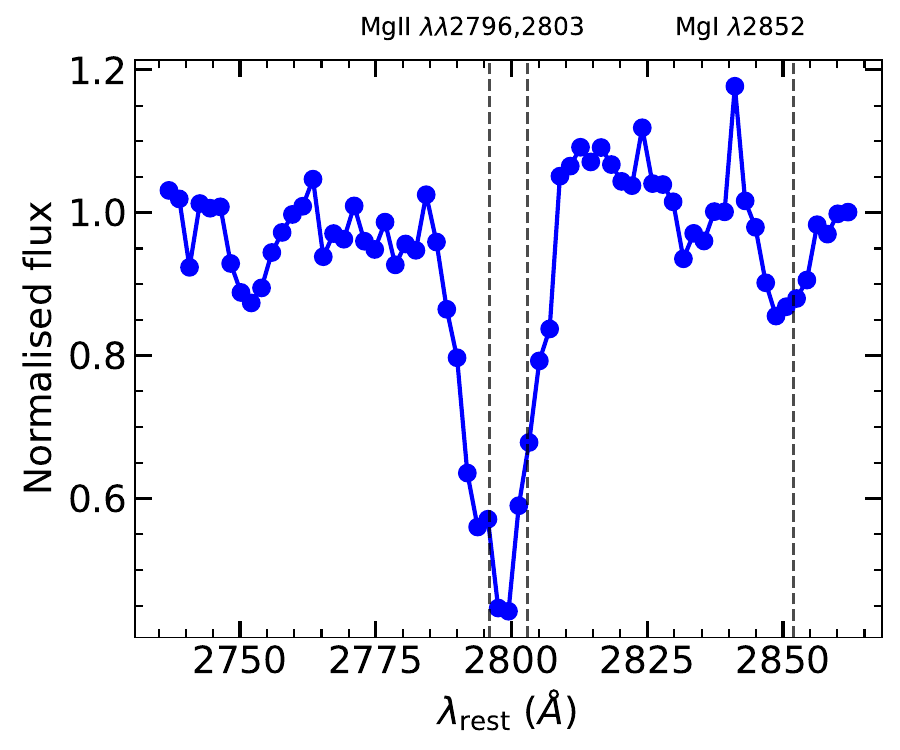}
  \end{center}
  \caption{Same as Fig. \ref{fig:cut_out_SED_all} but for object ID 77155.}
  \label{fig:cut_out_SED_all15}
\end{figure*}
\begin{figure*}[!ht]
  \begin{center}
    \includegraphics[width=0.40\textwidth]{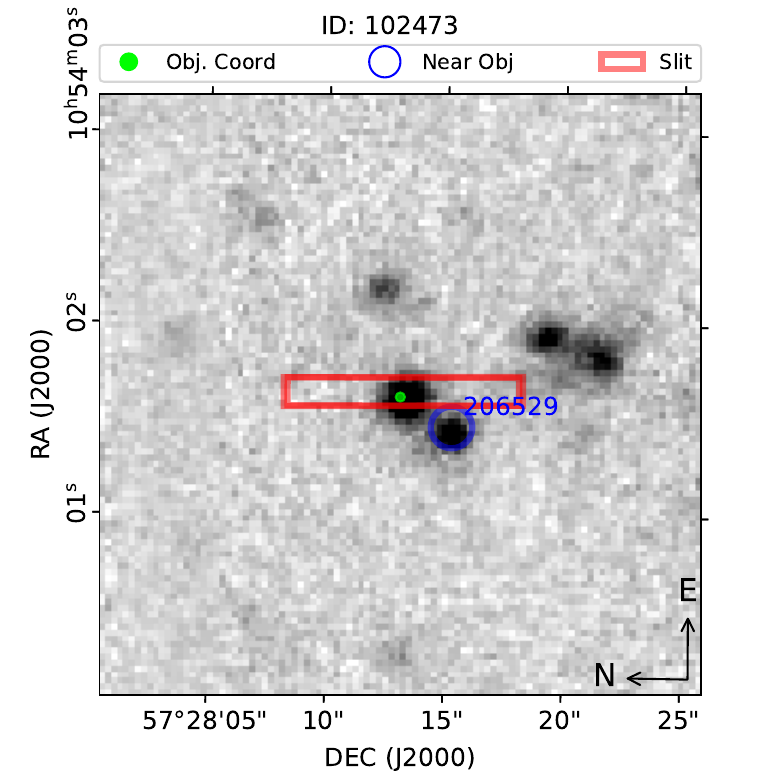}
    \includegraphics[width=0.53\textwidth]{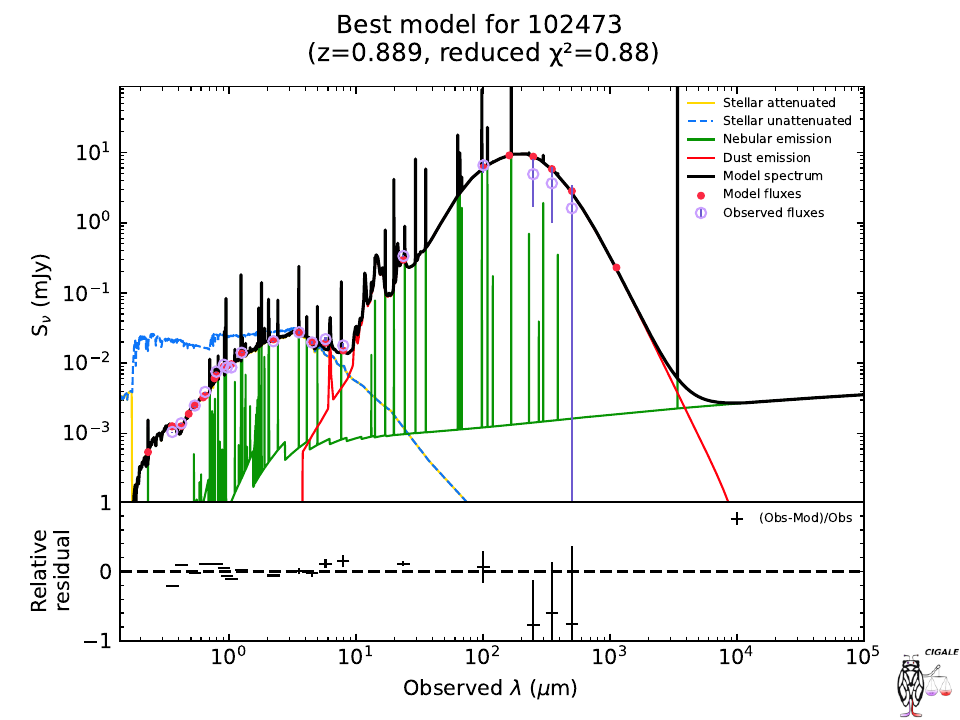}
    \includegraphics[width=0.20\textwidth]{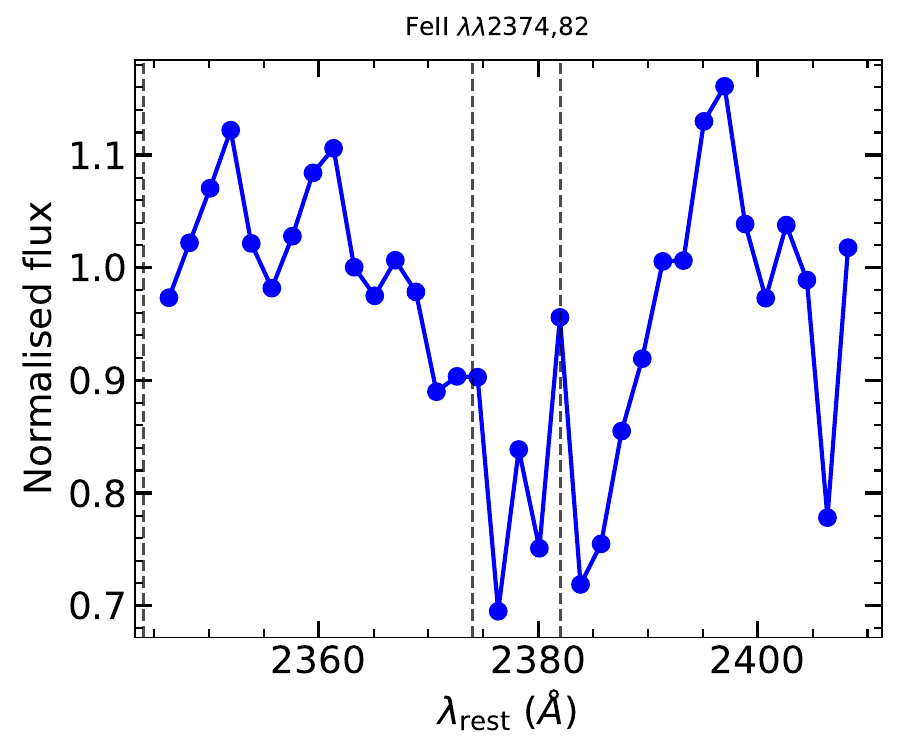}
    \includegraphics[width=0.20\textwidth]{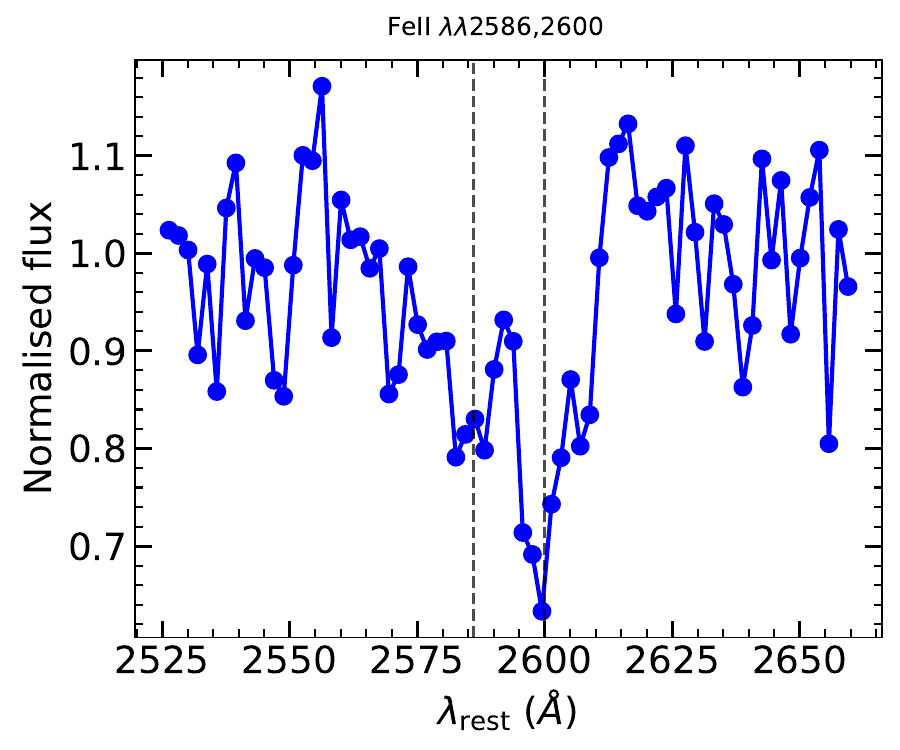}
    \includegraphics[width=0.20\textwidth]{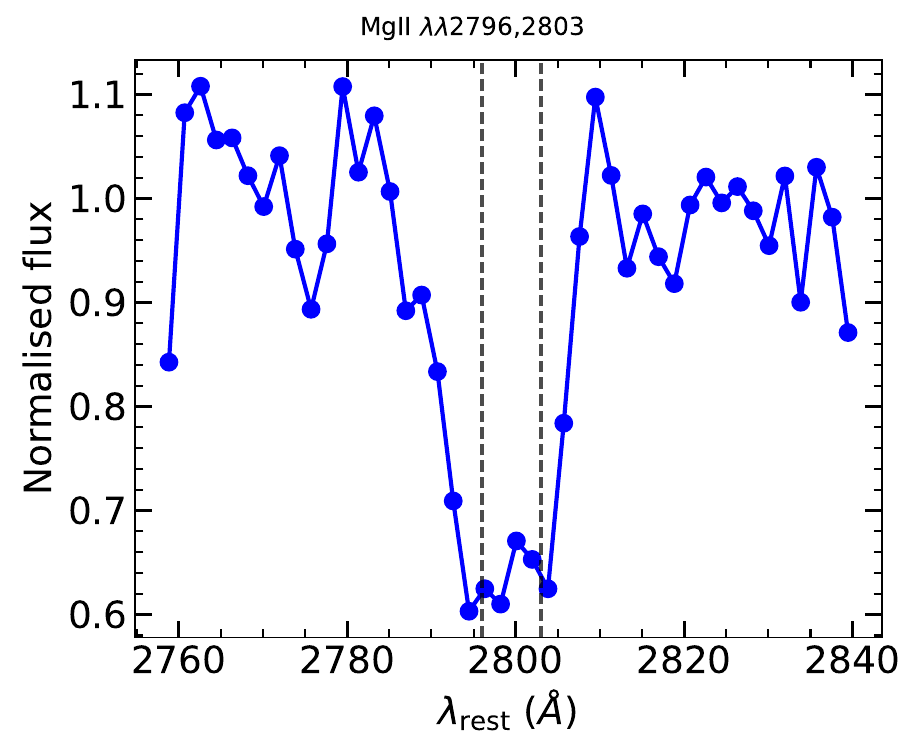}
  \end{center}
  \caption{Same as Fig. \ref{fig:cut_out_SED_all} but for object ID 102473.}
  \label{fig:cut_out_SED_all16}
\end{figure*}
\begin{figure*}[!ht]
  \begin{center}
    \includegraphics[width=0.40\textwidth]{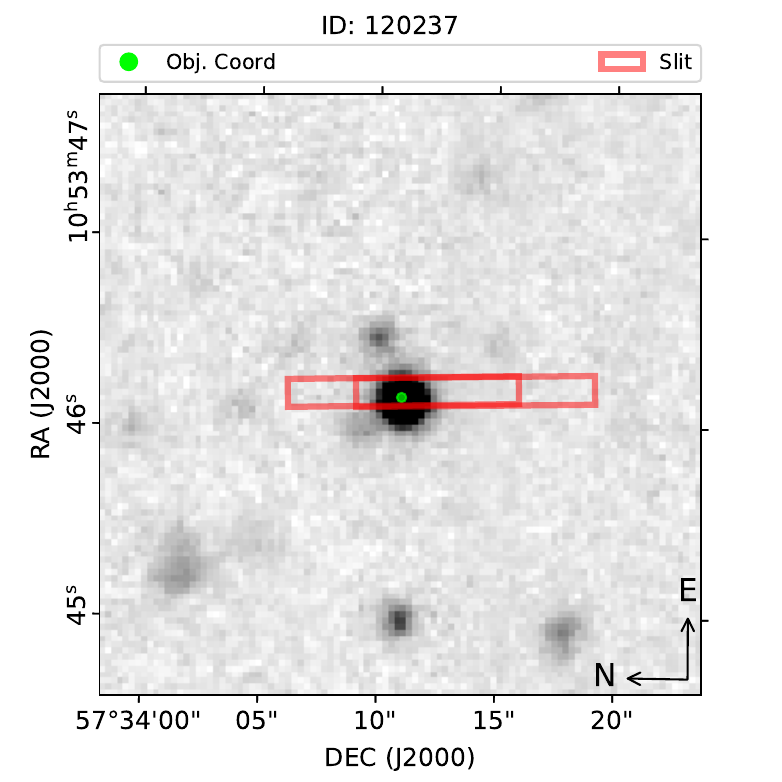}
    \includegraphics[width=0.53\textwidth]{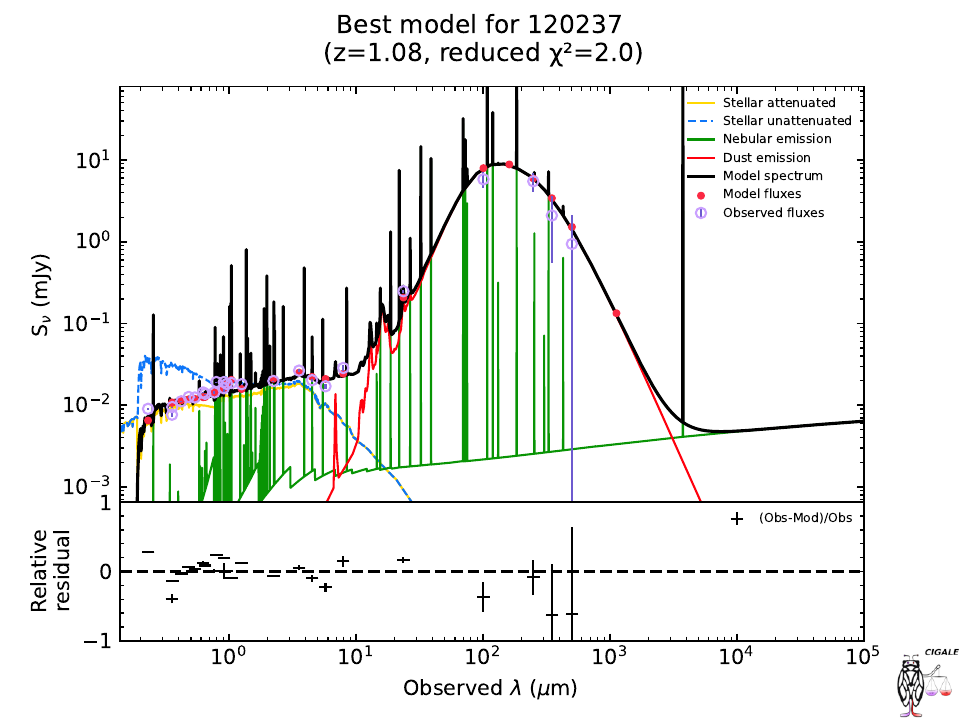}
    \includegraphics[width=0.20\textwidth]{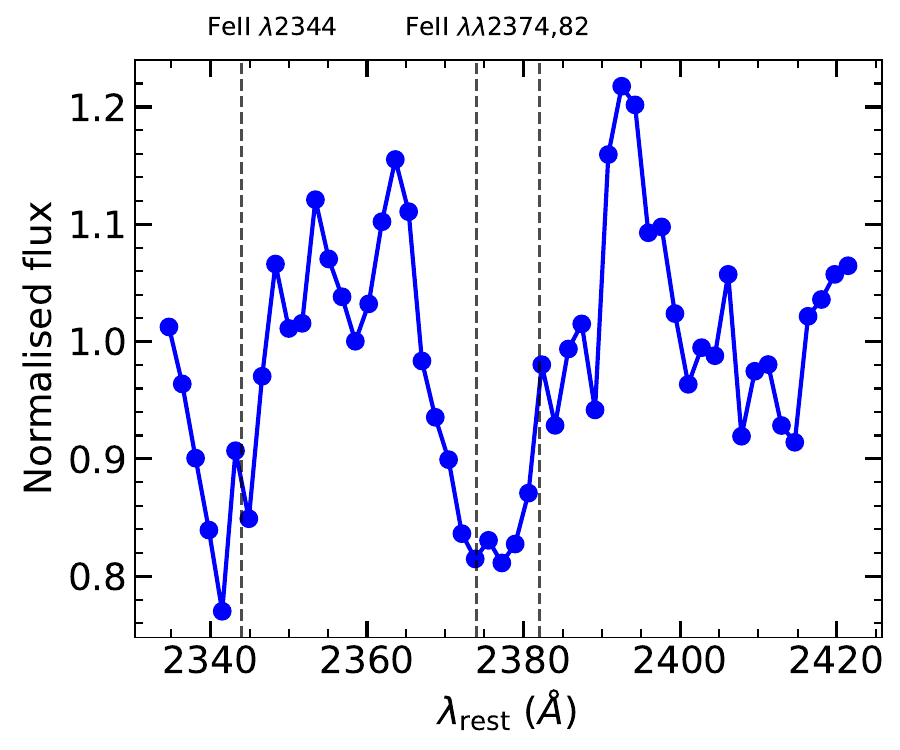}
    \includegraphics[width=0.20\textwidth]{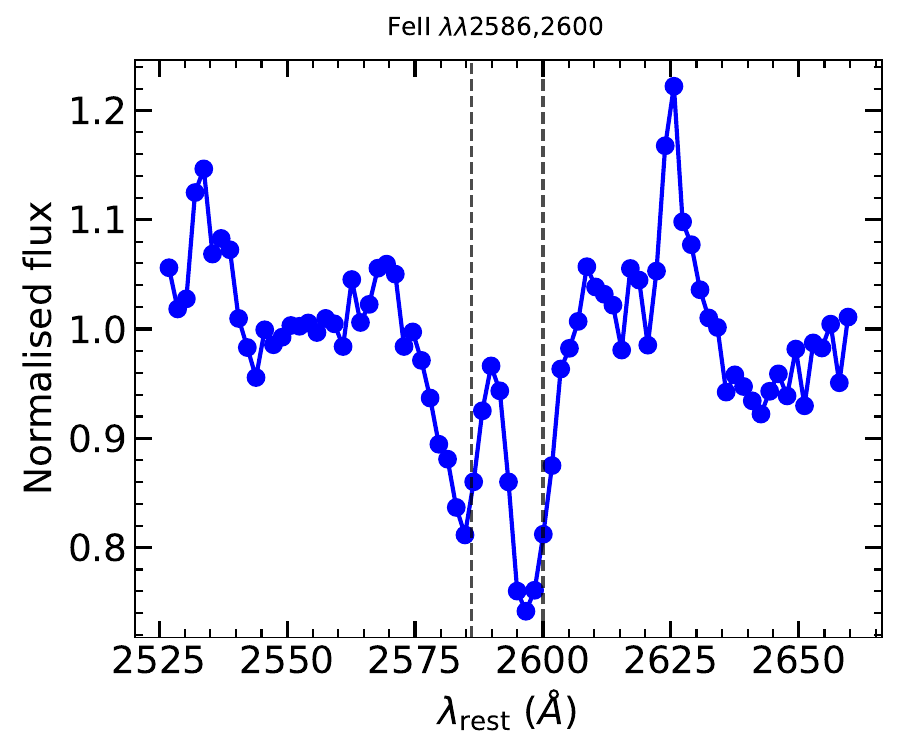}
  \end{center}
  \caption{Same as Fig. \ref{fig:cut_out_SED_all} but for object ID 120237.}
  \label{fig:cut_out_SED_all17}
\end{figure*}
\begin{figure*}[!ht]
  \begin{center}
    \includegraphics[width=0.40\textwidth]{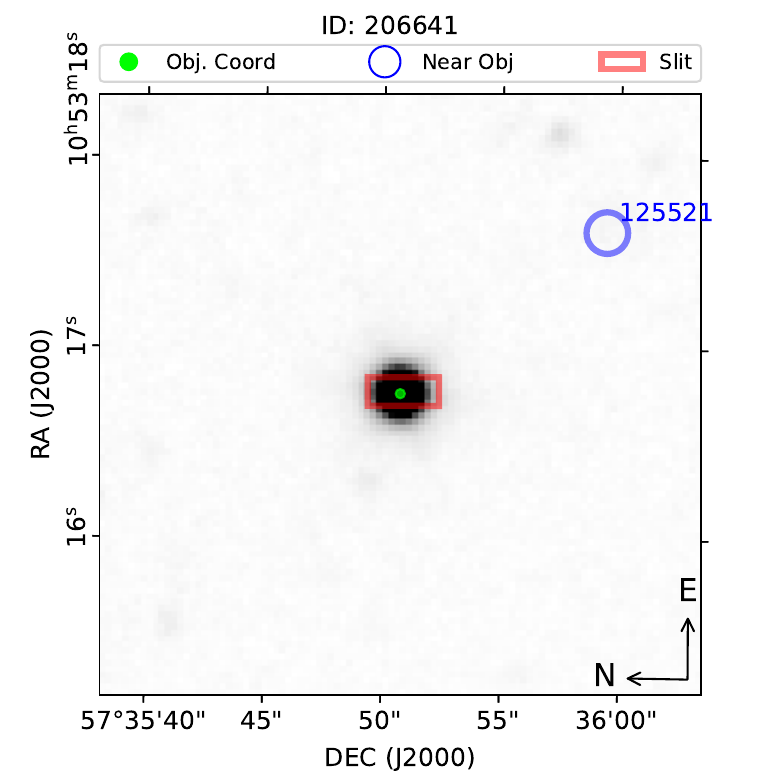}
    \includegraphics[width=0.53\textwidth]{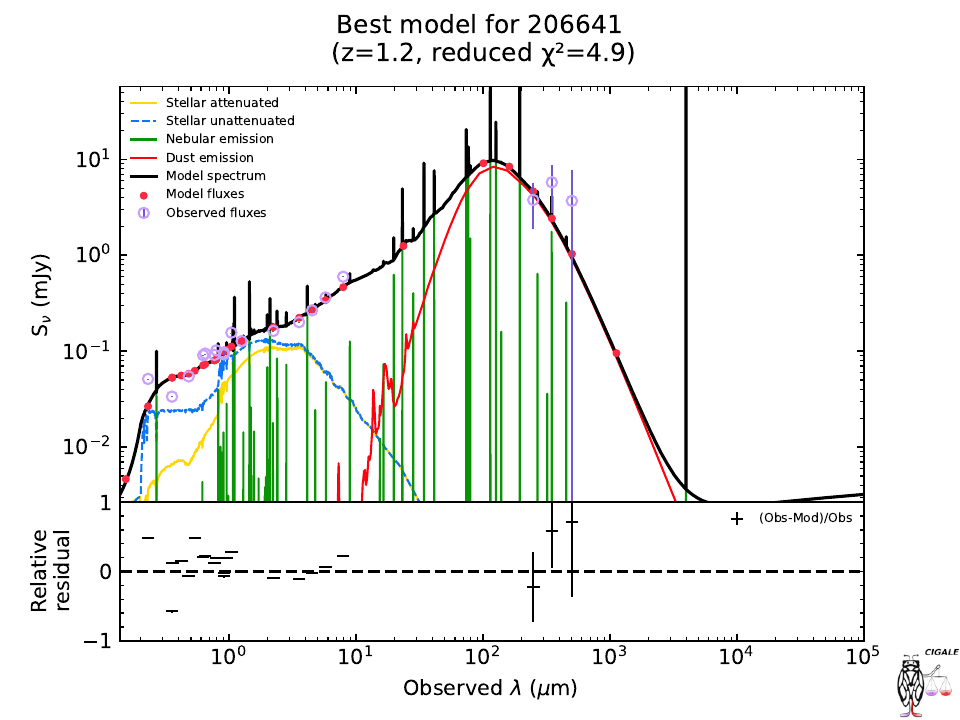}
    \includegraphics[width=0.20\textwidth]{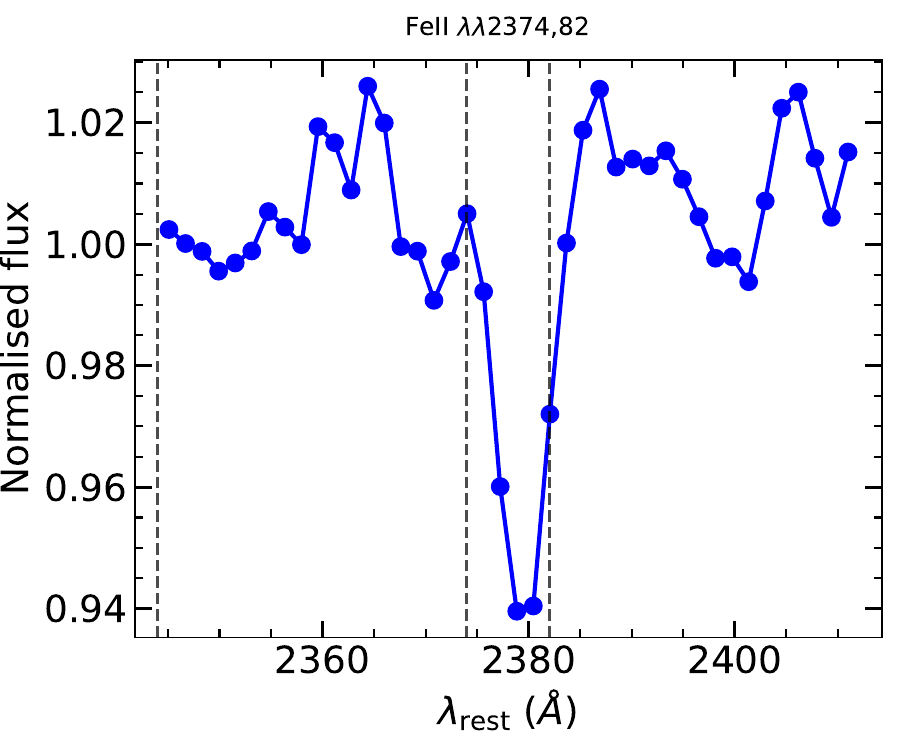}
    \includegraphics[width=0.20\textwidth]{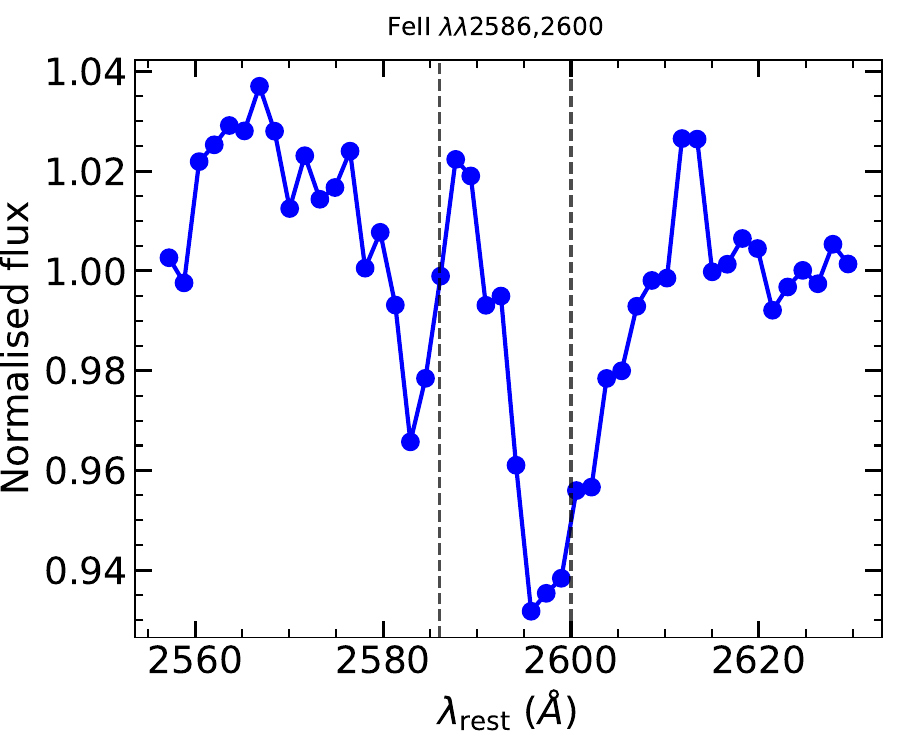}
    \includegraphics[width=0.20\textwidth]{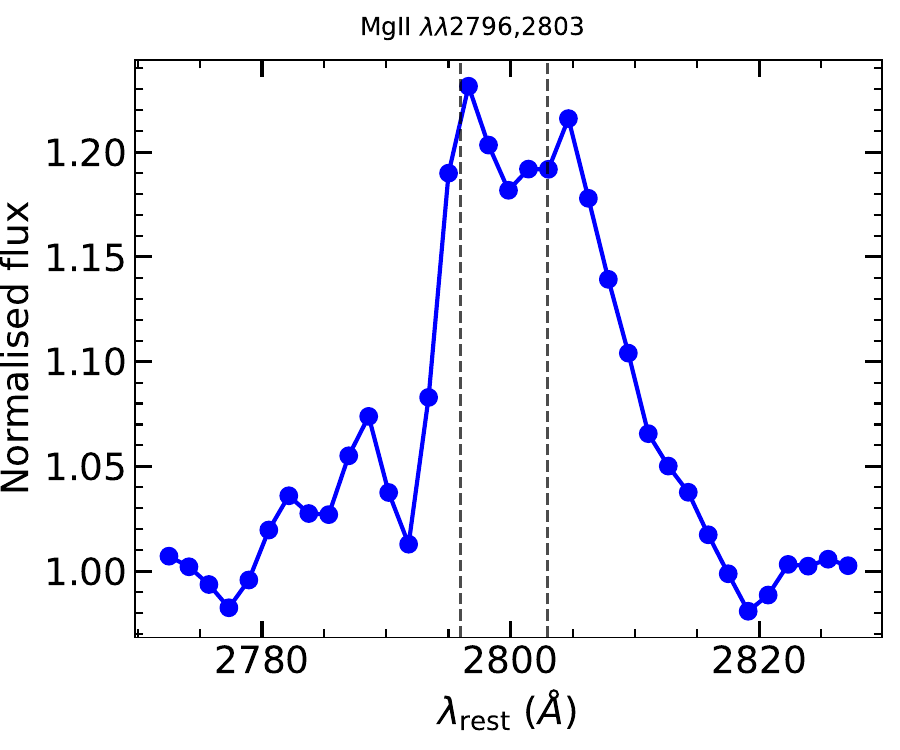}
  \end{center}
  \caption{Same as Fig. \ref{fig:cut_out_SED_all} but for object ID 206641.}
  \label{fig:cut_out_SED_all18}
\end{figure*}
\begin{figure*}[!ht]
  \begin{center}
    \includegraphics[width=0.40\textwidth]{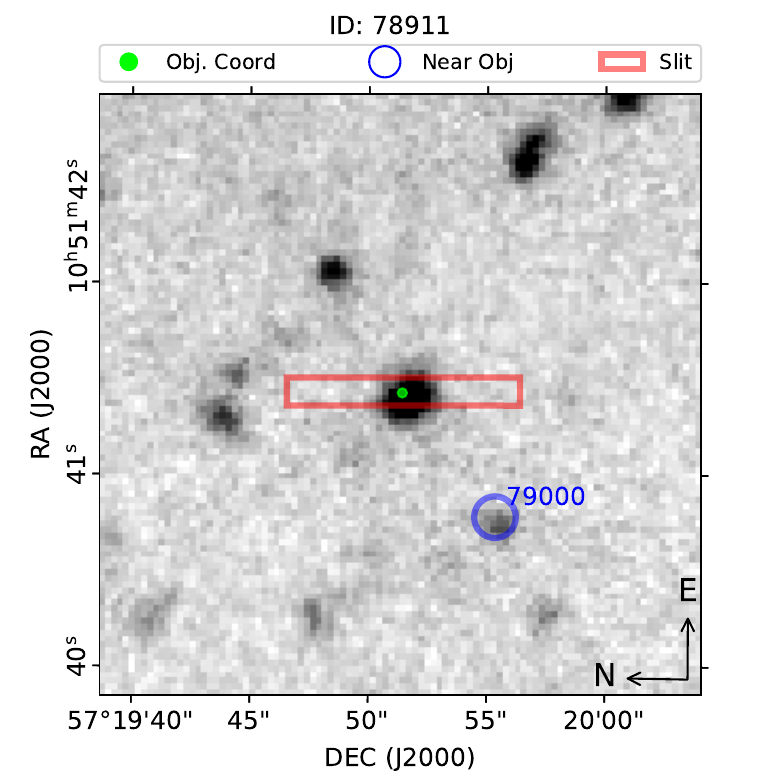}
    \includegraphics[width=0.53\textwidth]{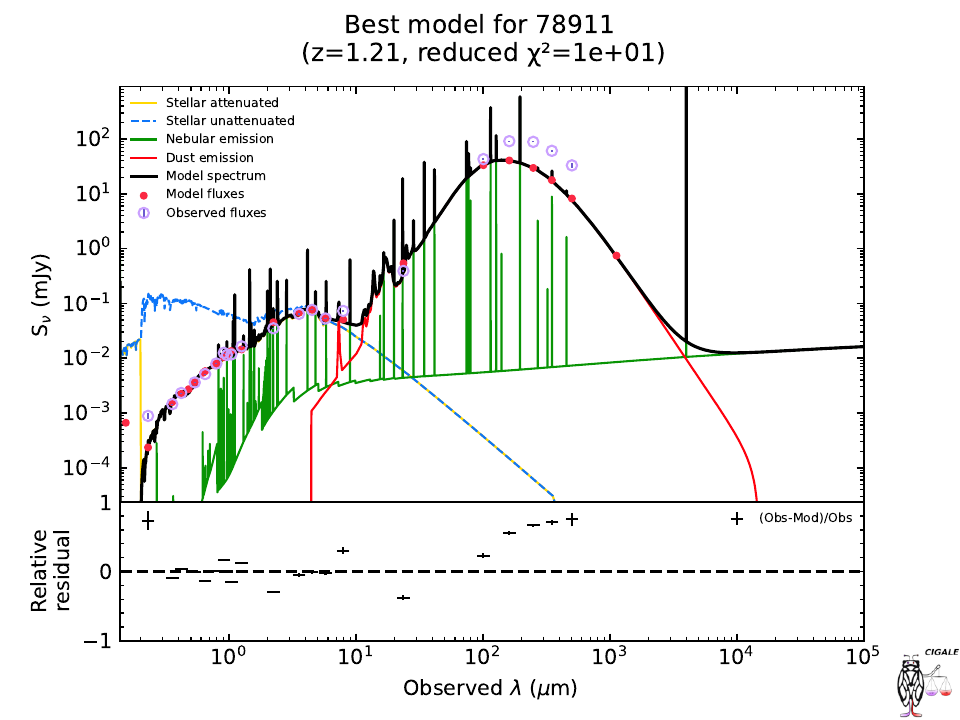}
    \includegraphics[width=0.20\textwidth]{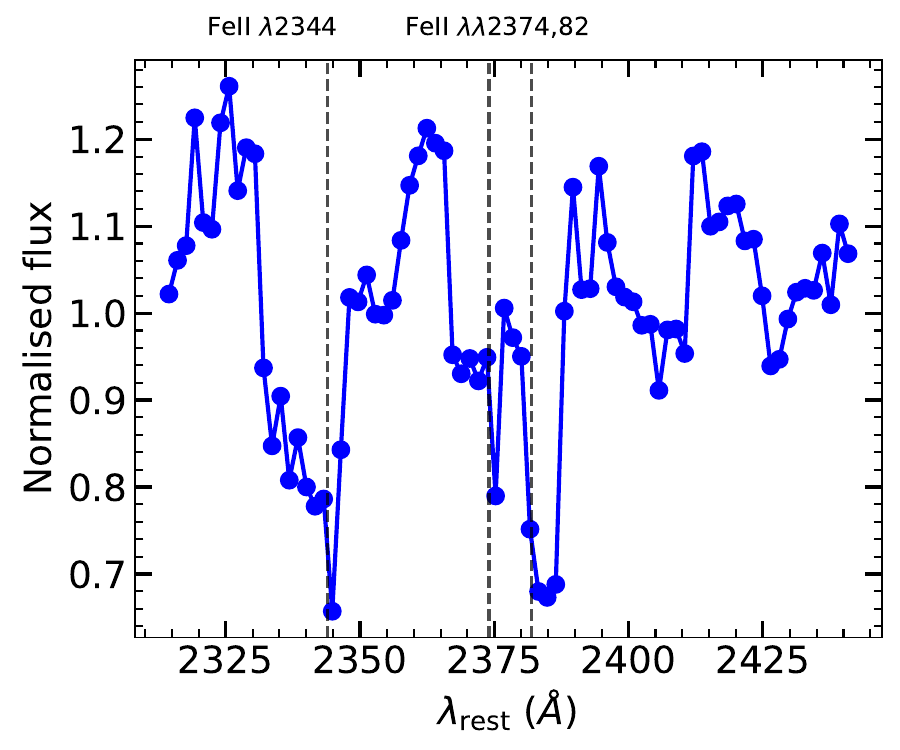}
    \includegraphics[width=0.20\textwidth]{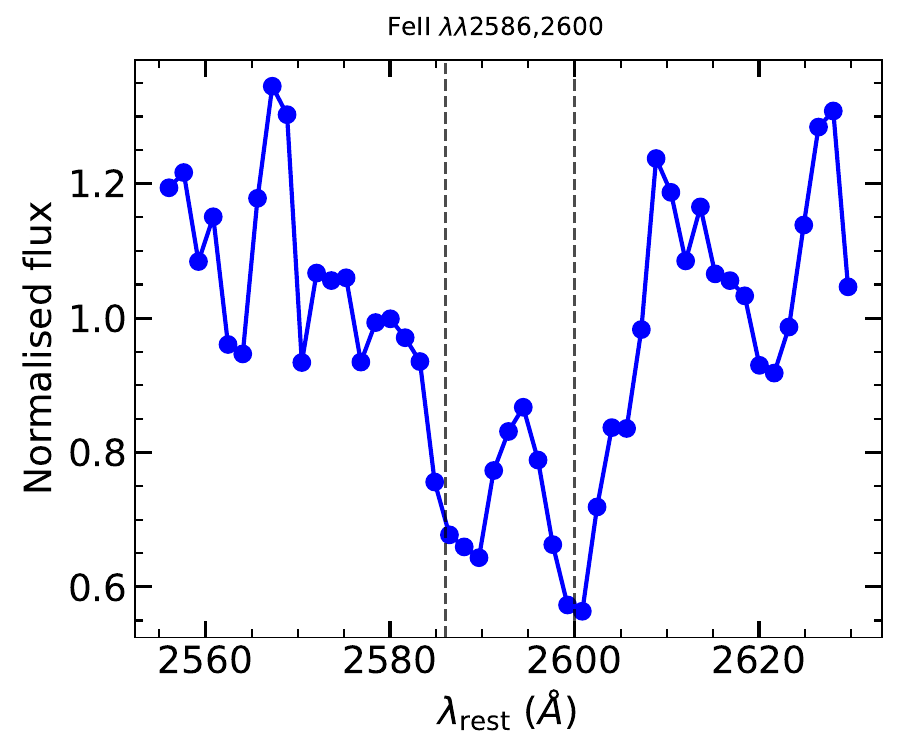}
    \includegraphics[width=0.20\textwidth]{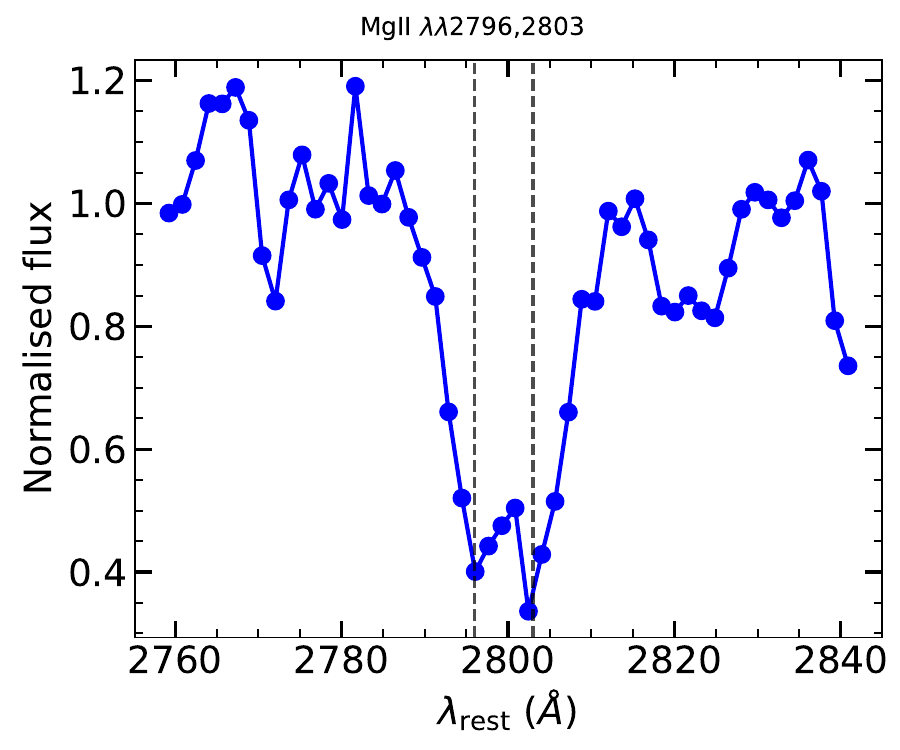}
  \end{center}
  \caption{Same as Fig. \ref{fig:cut_out_SED_all} but for object ID 78911.}
  \label{fig:cut_out_SED_all19}
\end{figure*}
\begin{figure*}[!ht]
  \begin{center}
    \includegraphics[width=0.40\textwidth]{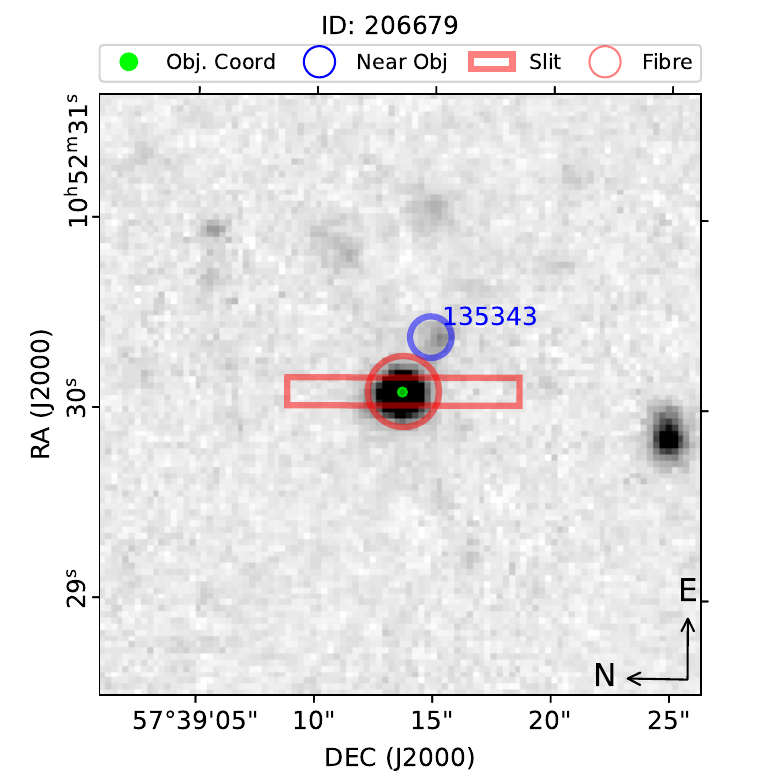}
    \includegraphics[width=0.53\textwidth]{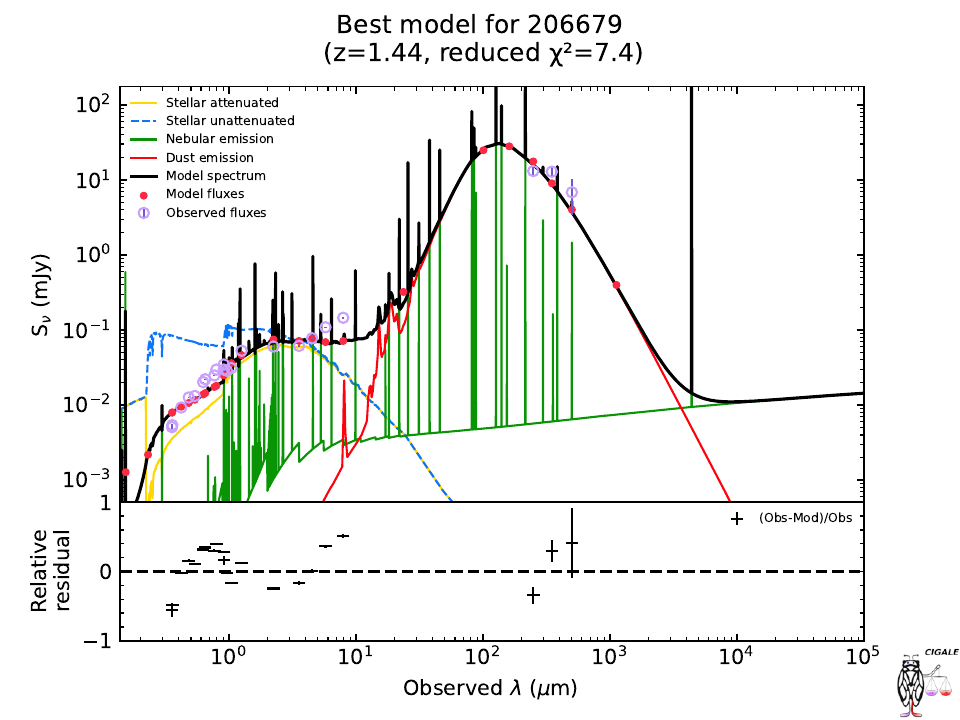}
    \includegraphics[width=0.20\textwidth]{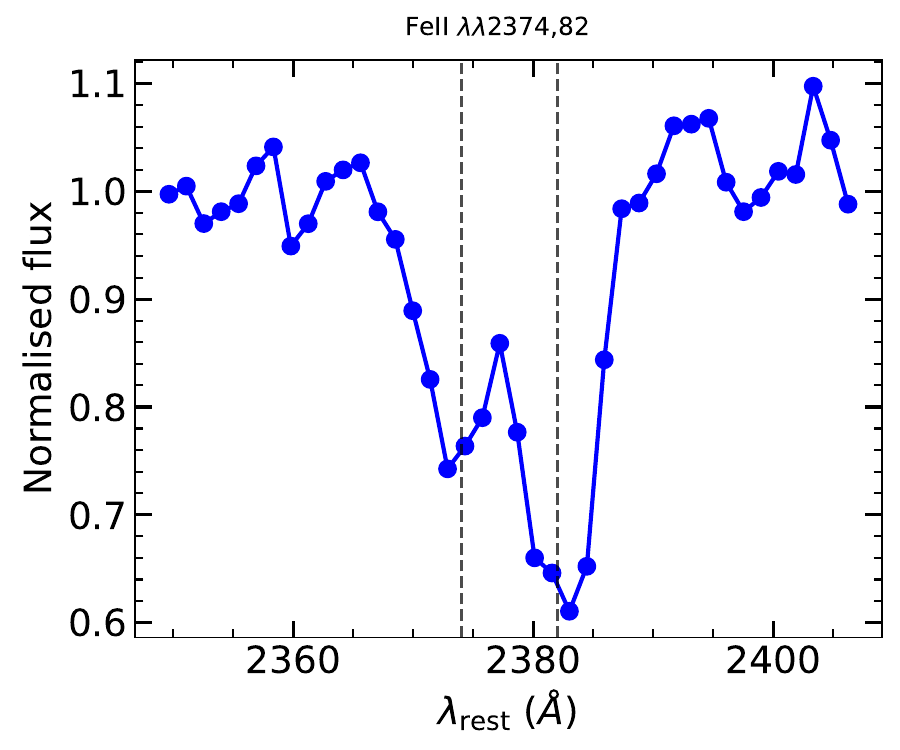}
    \includegraphics[width=0.20\textwidth]{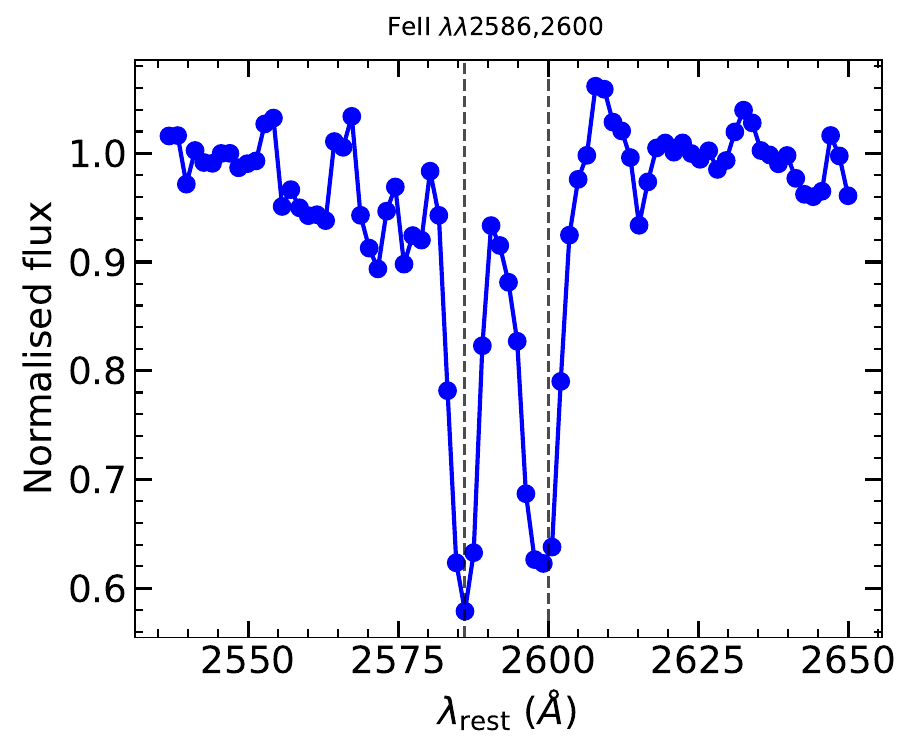}
    \includegraphics[width=0.20\textwidth]{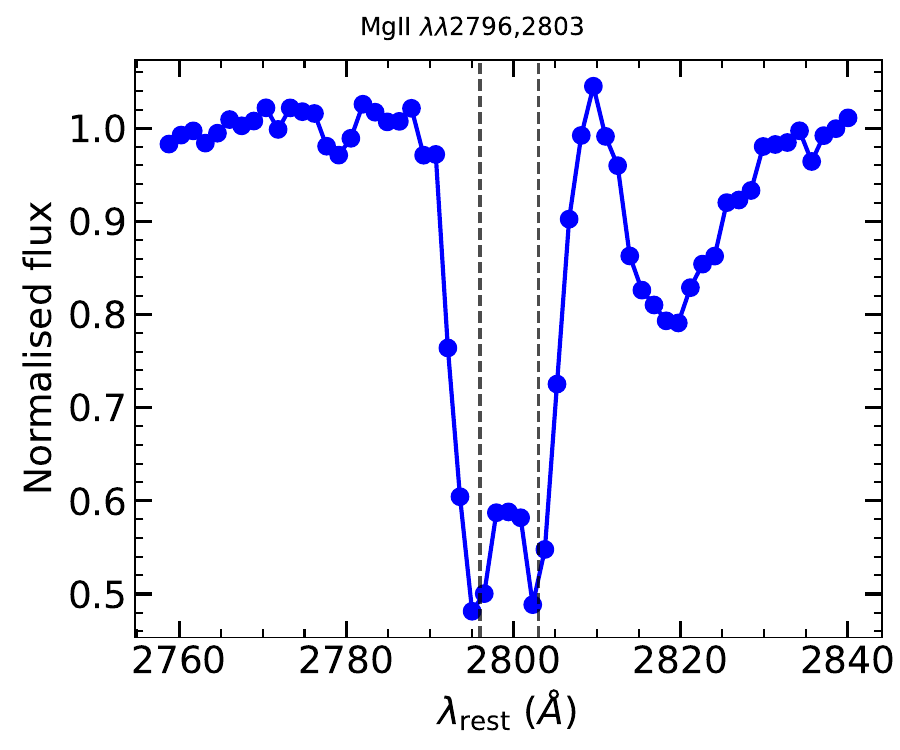}
  \end{center}
  \caption{Same as Fig. \ref{fig:cut_out_SED_all} but for object ID 206679.}
  \label{fig:cut_out_SED_all20}
\end{figure*}

\end{appendix}

\end{document}